\documentclass[twocolumn,twocolappendix,appendixfloats]{aastex631}
\usepackage{apjfonts}
\usepackage{amsmath}
\usepackage{scalerel}
\usepackage{multirow}

\definecolor{gold}{HTML}{FFA500}
\definecolor{pinkred}{HTML}{FFC0CB}

\DeclareSymbolFont{matha}{OML}{txmi}{m}{it}
\DeclareMathSymbol{\varv}{\mathord}{matha}{29}
\DeclareMathOperator{\diag}{diag}

\newcommand\msol{$M_{\odot}$}%

\newcommand{\fex}{\ion{Fe}{2}\protect\scaleto{$/III$}{1.4ex}}%

\newcommand{\caii}{\ion{Ca}{2}}%
\newcommand{\tiii}{\ion{Ti}{2}}%
\newcommand{\nai}{\ion{Na}{1}}%
\newcommand{\mgx}{\ion{Mg}{1}\protect\scaleto{$/II$}{1.4ex}}%

\newcommand{\nii}{\ion{N}{2}}%
\newcommand{\oii}{\ion{O}{2}}%
\newcommand{\siii}{\ion{Si}{2}}%
\def\simgt{\lower.5ex\hbox{$\; \buildrel > \over \sim \;$}}%
\def\simlt{\lower.5ex\hbox{$\; \buildrel < \over \sim \;$}}%
\def\sni{SN Ia}%
\def\he1{${\rm{^{1}He}}$}%
\def\ni56{${\rm{^{56}Ni}}$}%
\def\sni58{${\rm{^{58}Ni}}$}%
\def\co56{${\rm{^{56}Co}}$}%
\def\fer56{${\rm{^{56}Fe}}$}%
\def\fe52{${\rm{^{52}Fe}}$}%
\def\chr48{${\rm{^{48}Cr}}$}%
\def\dm15{$\Delta M_{15}(B)$}%
\def\tas{Type Ia SN}%
\def\tase{Type Ia SNe}%
\def\sn1a{SNe Ia}%
\def\sbv{$s_{BV}$}%
\def\t0{$t_{\rm 0}$}%
\def\chisqr{$\chi^2_{\rm R}$}%
\def\d6s{D$^{\wedge}$6}%
\def\vi{\mbox{$V\!-\!i$}}%

\def\scl{1.2}%

\shorttitle{KSP Infant Type Ia SN Sample}
\shortauthors{Ni et al.}

\graphicspath{{./}{./figures}}

\begin{document}

\title{Infant Type Ia Supernovae from the KMTNet I. Multi-Color Evolution and Populations}

\author[0000-0003-3656-5268]{Yuan Qi Ni}
\affiliation{David A. Dunlap Department of Astronomy and Astrophysics, University of Toronto, 50 St. George Street, Toronto, ON M5S 3H4, Canada}

\author[0000-0003-4200-5064]{Dae-Sik Moon}
\affiliation{David A. Dunlap Department of Astronomy and Astrophysics, University of Toronto, 50 St. George Street, Toronto, ON M5S 3H4, Canada}

\author[0000-0001-7081-0082]{Maria R. Drout}
\affiliation{David A. Dunlap Department of Astronomy and Astrophysics, University of Toronto, 50 St. George Street, Toronto, ON M5S 3H4, Canada}

\author[0000-0002-6261-1531]{Youngdae Lee}
\affiliation{Department of Astronomy and Space Science, Chungnam National University, Daejeon 34134, Republic of Korea}

\author[0009-0002-5988-9898]{Patrick Sandoval}
\affiliation{David A. Dunlap Department of Astronomy and Astrophysics, University of Toronto, 50 St. George Street, Toronto, ON M5S 3H4, Canada}

\author[0000-0002-9914-3129]{Jeehye Shin}
\affiliation{Korea Astronomy and Space Science Institute, 776, Daedeokdae-ro, Yuseong-gu, Daejeon 34055, Republic of Korea}
\affiliation{Korea University of Science and Technology (UST), 217 Gajeong-ro, Yuseong-gu, Daejeon 34113, Republic of Korea}

\author[0000-0002-3505-3036]{Hong Soo Park}
\affiliation{Korea Astronomy and Space Science Institute, 776, Daedeokdae-ro, Yuseong-gu, Daejeon 34055, Republic of Korea}
\affiliation{Korea University of Science and Technology (UST), 217 Gajeong-ro, Yuseong-gu, Daejeon 34113, Republic of Korea}

\author[0000-0001-9670-1546]{Sang Chul Kim}
\affiliation{Korea Astronomy and Space Science Institute, 776, Daedeokdae-ro, Yuseong-gu, Daejeon 34055, Republic of Korea}
\affiliation{Korea University of Science and Technology (UST), 217 Gajeong-ro, Yuseong-gu, Daejeon 34113, Republic of Korea}

\author[0000-0002-5037-951X]{Kyuseok Oh}
\affiliation{Korea Astronomy and Space Science Institute, 776, Daedeokdae-ro, Yuseong-gu, Daejeon 34055, Republic of Korea}

\correspondingauthor{Yuan Qi Ni}
\email{chris.ni@mail.utoronto.ca}

\begin{abstract}
We conduct a systematic analysis of the early multi-band light curves and colors of 19 Type~Ia~Supernovae (SNe) from the Korea Microlensing Telescope Network SN Program, including 16 previously unpublished events.
Seven are detected $\lesssim$~1~day since the estimated epoch of first light and the rest within $\lesssim$~3~days.
Some show excess emission within $<$~0.5~days to $\sim$~2~days, but most show pure power-law rises.
The colors are initially diverse before $\sim$~5~days, but converge to a similar color at $\sim$~10~days.
We identify at least three populations based on 2--5-day color evolution: 
(1) ``early-blues'' exhibit slowly-evolving colors consistent with a $\sim$~17,000~K blackbody; (2) ``early-reds'' have initially blue \bv\ and red \vi\ colors that cannot simultaneously be fit with a blackbody---likely due to suppression of $B$- and $i$-band flux by \fex\ and \caii---and evolve more rapidly; and (3) ``early-yellows'' evolve blueward, consistent with thermal heating from $\sim$~8,000 to 13,000~K.
The distributions of early-blue and early-red colors are compatible with them being either distinct populations---with early-reds comprising (60~$\pm$~15)\% of them---or extreme ends of one continuous population; whereas the early-yellow population identified here is clearly distinct.
Compared to the other populations, early-blues in our sample differ by exhibiting excess emission 
within 1--2 days, nearly constant peak brightness regardless of \dm15\ after standardization, 
and shallower \siii\ features. 
Early-blues also prefer star-forming host environments, while early-yellows and, to a lesser extent, early-reds prefer quiescent ones. These preferences appear to indicate at least two \tas\ production channels based on stellar population age, while early-reds and early-blues may still share a common origin.
\end{abstract}

\keywords{Binary stars (154), Supernovae (1668), Type Ia supernovae (1728), White dwarf stars (1799), Transient sources (1851), Time domain astronomy (2109), Galaxy ages (576)}

\section{Introduction} \label{sec:intro}

Type Ia Supernovae (SNe) are thermonuclear explosions of white dwarf (WD) stars \citep[e.g.,][]{Nugent2011nat}, thought to be triggered as a result of mass transfer in binary systems.
They are the primary producers of iron-peak elements in the Universe \citep[e.g.,][]{Matteucci2012book} and one of the most mature tools for measuring extragalactic distances \citep[e.g.,][]{Albrecht2006}, used to discover the accelerated cosmological expansion and dark energy \citep{Riess1998aj, perlmutter1999apj}.
Despite their importance and extensive efforts to understand their origins, the explosion mechanisms and nature of their companion stars remain poorly understood.
This is especially true for the majority class of ``normal'' \tase\
(comprising $>$ 63\% of them; \citealt{Morrell2024apj})
that are most widely used for cosmological distance measurements due to their similar light curve properties near peak.
For instance, whether the dominant production channel is ``single-degenerate'' \citep[where the companion is a main-sequence, red giant, or helium star;][]{Whelan&Iben1973apj} or ``double-degenerate'' \citep[where it is another WD;][]{Iben&Tutukov1984apjs}, is still debated \citep[see][for reviews]{Maoz2014araa, Ruiter2020iaus}.
In addition, while normal \tase\ have long been theorized to be ignited in the core of a WD when binary accretion or merger causes its mass to approach the critical Chandrasekhar limit \citep[$\sim$ 1.4 \msol;][]{Mazzali2007sci}, recent theoretical studies suggest that the detonation of a thin helium layer on the WD surface can subsequently ignite the core, producing normal \tase\ significantly below the Chandrasekhar mass \citep[or ``sub-Chandrasekhar-mass'' explosions; e.g.,][]{Townsley2019apj}.

The observed population of \tase\ may, in principle, originate from a mixture of multiple explosion mechanisms and progenitor channels that can produce different light curve properties. 
In this case, there is the question of whether the
ratio of different channels correlates with environment variables---e.g., age of star formation environment \citep{Ruiter2009apj, Ruiter2014mnras}---in ways that can bias distance measurements.
The spectroscopic heterogeneity of \tase\ near peak provides some evidence for this, with at least two identifiable sub-populations of normal \tase---Core-Normal/Normal-Velocity (CN/HV); and Broad-Line/High-Velocity \citep[BL/HV; see][for a review of \tas\ subtypes]{Parrent2014apss}---that have been used to refine distance measurements \citep[e.g.,][]{Wang2009apj}.
Two progenitor channels from young and old stellar environments have also been proposed to explain the observed host galaxy mass-dependent bias of \tas\ distance measurements \citep[sometimes called ``mass step'';][]{Childress2014mnras, Jones2023apj}. This points to a potential bias that depends on global age and therefore redshift \citep{Zhang2021mnras}.
Though other explanations have been proposed for the mass step \citep[e.g., dust environment;][]{Brout2021apj}, 
understanding the progenitor/explosion channels of \tase\ and their mixture ratios has clear importance for the reliability of their distance measurements, especially for next-generation SN cosmology based on extrapolating local \tas\ properties to extreme redshifts \citep[e.g, see][]{Albrecht2006}.

Historically, most \tase\ have been discovered and monitored near the peak of their light curves ($\sim$2--3 weeks post explosion) when the emission is dominated by the radioactive decay of \ni56\ in the core. However, the early light curves obtainable within hours to a few days post-explosion contain unique natal information about their origins.
Multiple physical processes related to the progenitor and explosion mechanism have been theorized to produce excess emission and/or short-lived spectroscopic features that fade rapidly in later epochs as the ejecta expands. This includes collision with either the companion star \citep{Kasen2010apj} or circumstellar material \citep{Piro&Morozova2016apj}, which can heat the ejecta surface; as well as subsonic mixing \citep{Maeda2010apj} or surface nuclear burning \citep{Polin2019apj}, both of which can enrich the outer ejecta with radioactive iron-peak elements.

A variety of excess emissions have been recently reported in normal \tase\ observed within a few days post-explosion \citep[e.g.,][]{Marion2016apj, Hosseinzadeh2017apj, Dimitriadis2019apj, Ni2023bapj, Ni2023apj, Wang2024apj}, though the vast majority of such early light curves match simple power-law profiles \citep[e.g.,][]{Nugent2011nat, Foley2012apj, Olling2015nat, Cartier2017mnras, Holmbo2019aa, Miller2020apj, Moon2021apj}, consistent with a centrally-concentrated and monotonic distributions of radioactive \ni56\ below the ejecta surface and disfavouring the presence of giant companion stars.
The nature of the excess emission remains unknown in the majority of cases, though several possible interpretations including the physical processes listed above have been debated \citep{Shappee2018apj, Sand2018apj, Levanon2019apj}.

The colors of \tase\ within hours to a few days post-explosion add critical information about the temperature and the distribution of elements in the photosphere, and observations within $\lesssim$ 3 days post-explosion appear to reveal significant heterogeneity within the population.
\citet[][SET18 hereafter]{Stritzinger2018apj} identify at least two distinct sub-populations based on \bv\ color evolution in this phase. However, more recent studies have debated whether these sub-populations are distinct or extreme ends of a continuous population \citep{Han2020apj, Bulla2020apj}, and the physical origins of the color evolutions themselves remain uncertain.
Part of the difficulty in distinguishing and interpreting early \tase\ color evolution can be attributed to the degeneracy between temperature and spectral line features, which cannot be broken using a single color \citep[e.g., only \bv\ or \mbox{$g\!-\!r$}, as in SET18;][]{Han2020apj, Bulla2020apj}.
For example, two colors were required in the recent case of the normal \tas~2018aoz to deduce the presence of iron-peak elements concentrated near the ejecta surface based on the rapid redward evolution of its \bv\ color compared to its \vi\ color within 1--13 hours since first light \citep{Ni2022natas}.
The relationship between the color evolution of \tase\ and their origins can therefore be better understood with a sufficiently large sample of early ``multi-color'' light curves consisting of observations in at least three filters.

Here, we present our systematic analysis of 19 infant/early \tase\ discovered by the Korea Microlensing Telescope Network (KMTNet) SN Program within a few hours to a few days post-explosion.
Each of them are monitored in three filters ($BVI$-band) from the beginning, providing a unique opportunity to investigate how \tase\ explode and test for the existence of \tas\ sub-populations with distinct explosion processes.
In Section~\ref{sec:obs}, we describe our observations---including of 16 previously unpublished events discovered between 2016 and 2022---and the standardization process of rest-frame $BVi$-band light curves.
Our results are presented in Section~\ref{sec:res}, including a census of early excess emissions in the sample and identification of three distinct sub-populations of \tase\ based on early multi-color evolution, followed by characterization of their relative rates and properties based on their near-peak light curves, spectroscopic features, and host galaxy environments.
Physical processes leading to the observed multi-color evolutions are briefly discussed in Section~\ref{sec:disc}, while a follow-up paper (Ni et al. 2024 in preparation) will explore the implications for the \tas\ explosion mechanisms and progenitor systems in detail.
We summarize and conclude in Section~\ref{sec:conc}.

\begin{deluxetable*}{l|llcc}
\tabletypesize{\footnotesize}
\tablecolumns{9} 
\tablewidth{0.99\textwidth}
 \tablecaption{Type Ia SN discoveries in KSP-monitored fields.}
 \tablehead{
 \colhead{KSP field$\rm ^a$} & \colhead{KSP name$\rm ^b$} & \colhead{IAU Name} & \colhead{Equatorial Coordinates ($\alpha$, $\delta$)} & \colhead{Epoch of First Detection ($t_{\rm det}$)}\\
 \colhead{} & \colhead{} & \colhead{} & \colhead{[J2000]} & \colhead{[UTC; MJD]}
 } 
\startdata 
ESO~149-G003 & KSP-SN-2017gp &  & ($\rm 23^h54^m1^s.129$, $-52\degr48\arcmin30\farcs488$) & 13h02m on Oct. 16, 2017; MJD 58042.543 \\
\hline
ESO~489-G035 & KSP-SN-2016M & AT~2016igg [4] & ($\rm 6^h14^m45^s.309$, $-24\degr29\arcmin55\farcs063$) & 16h10m on Oct. 22, 2016; MJD 57683.674 \\
& KSP-SN-2016ad & SN~2016iew [5,6] & ($\rm 6^h13^m11^s.532$, $-24\degr16\arcmin18\farcs843$) & 15h25m on Nov. 4, 2016; MJD 57696.642 \\
& KSP-SN-2017iw &  & ($\rm 6^h17^m59^s.294$, $-23\degr48\arcmin52\farcs153$) & 01h59m on Mar. 1, 2017; MJD 57813.083 \\
\hline
NGC~2292 & KSP-SN-2016bo &  & ($\rm 6^h44^m31^s.948$, $-25\degr34\arcmin8\farcs543$) & 07h19m on Nov. 15, 2016; MJD 57707.305 \\
\hline
NGC~247 & KSP-SN-2017cv &  & ($\rm 0^h50^m15^s.898$, $-19\degr19\arcmin57\farcs811$) & 17h41m on Jul. 15, 2017; MJD 57949.737 \\
\hline
NGC~2997 & KSP-SN-2018oh & AT~2018ahb [7] & ($\rm 9^h44^m29^s.597$, $-31\degr43\arcmin41\farcs610$) & 03h07m on Mar. 7, 2018; MJD 58184.130 \\
\hline
NGC~300 & KSP-OT-201509b\ \ \ [1] &  &  &  \\
& KSP-SN-2017cz &  & ($\rm 0^h52^m41^s.104$, $-38\degr4\arcmin45\farcs337$) & 19h16m on Jun. 25, 2017; MJD 57929.803 \\
& KSP-SN-2017fo &  & ($\rm 0^h57^m1^s.248$, $-37\degr2\arcmin36\farcs943$) & 01h07m on Oct. 30, 2017; MJD 58056.047 \\
\hline
NGC~3511 & KSP-SN-2019bl & SN~2019bxi [8,9] & ($\rm 11^h6^m28^s.496$, $-22\degr44\arcmin26\farcs525$) & 01h27m on Mar. 6, 2019; MJD 58548.060 \\
\hline
NGC~3717 & KSP-SN-2018ng &  & ($\rm 11^h29^m15^s.446$, $-30\degr31\arcmin5\farcs983$) & 10h07m on May 6, 2018; MJD 58244.422 \\
& KSP-OT-201903ah &  & ($\rm 11^h30^m37^s.202$, $-30\degr14\arcmin9\farcs954$) & 03h30m on Mar. 15, 2019; MJD 58192.146 \\
\hline
NGC~3923 & KSP-SN-2018ku\ \ \ \ [2] & SN~2018aoz &  &   \\
& KSP-SN-2019dz &  & ($\rm 11^h51^m4^s.127$, $-28\degr57\arcmin29\farcs183$) & 05h58m on Feb. 23, 2019; MJD 58537.249 \\
\hline
NGC~59 & KSP-SN-2016bu &  & ($\rm 0^h12^m56^s.061$, $-20\degr44\arcmin17\farcs647$) & 02h27m on Oct. 26, 2016; MJD 57687.102 \\
\hline
NGC~1553 & KSP-SN-2021V\ \ \ \ \ [3] & SN~2021aefx &  &  \\
\hline
ESO~265-G007 & KSP-SN-2021iq &  & ($\rm 11^h4^m7^s.865$, $-46\degr24\arcmin36\farcs696$) & 02h39m on Mar. 25, 2021; MJD 59298.110 \\
\hline
NGC~988 & KSP-SN-202112D & SN~2022zz [10,11] & ($\rm 2^h38^m31^s.546$, $-8\degr15\arcmin45\farcs472$) & 12h00m on Dec. 20, 2021; MJD 59568.500 \\
\enddata
\tablenotetext{{\rm a}}{Each field is selected to contain at least one nearby ($<$ 20~Mpc) galaxy which it is named after.}
\tablenotetext{{\rm b}}{Events with KSP names starting with KSP-SN- are spectroscopically identified as \tase\ (see Section~\ref{sec:spec}), while those starting with KSP-OT- are identified based on light curve features (Section~\ref{subsec:nonearly}) and template fitting (Appendix~\ref{sec:standard}).}
\tablecomments{[1] \citet{Moon2021apj}; [2] \citet{Ni2022natas}; [3] \citet{Ni2023bapj}; [4] \citet{Chambers2016tns}; [5,6] \citet{Tonry2016tns, Takats2016tns}; [7] \citet{Tonry2018tns} ; [8,9] \citet{Nordin2019tns, Fremling2019tns}; [10,11] \citet{Hodgkin2022tns, Gromadzki2022tns}}
\end{deluxetable*} 
\label{tab:survey}

\section{Observations and Data Analysis}\label{sec:obs}

The KMTNet SN Program \citep[KSP;][]{Moon2016spie} discovered 19 infant/early \tase\ between 2015 and 2022. We define an ``infant/early'' event as having detections within $\lesssim$ 3 days of the estimated ``epoch of first light'' (see Section~\ref{sec:excess}), with ``infant'' SNe referring to only those detected within $\lesssim$ 1 day. 
These 19 SNe were discovered in 13 KSP fields (Table~\ref{tab:survey}), each of which is 4 degree$^2$ in size, containing at least one nearby ($<$ 20 Mpc) galaxy: ESO~149-G003, ESO~489-G035, NGC~2292, NGC~247, NGC~2997, NGC~300, NGC~3511, NGC~3717, NGC~3923,
NGC~59, NGC~1553, ESO~265-G007, and NGC~988.
Most of the infant/early \tase\ exploded elsewhere in the field---typically from a higher redshift ($z$ $\sim$ 0.05--0.10)---rather than within the nearby galaxies,
though two cases of KSP-SN-2018ku and 2021V were from the nearby galaxies NGC~2923 and NGC~1566, respectively.
Thirteen of the SNe were classified as Type Ia with at least one spectrum from near peak brightness (Section~\ref{sec:spec}), while the rest were classified based on near-peak light curve features (Section~\ref{subsec:nonearly}).

The KSP fields had typically been monitored in $BVI$ every 4 to 8 hours continuously 
for a few years with gaps caused by the Sun and also by other ongoing programs.
The 17 SNe in Table~\ref{tab:survey} with KSP names starting with ``KSP-SN'' are spectroscopically classified (Section~\ref{sec:spec}) while the two starting with ``KSP-OT'' are classified based on light curve features (Section~\ref{subsec:nonearly}) and \tas\ light curve template fits (Appendix~\ref{sec:standard}). Both naming conventions are sometimes abbreviated as ``KSN'' (= KSP \tas) throughout this paper for convenience. 
Seven events have been reported previously on the Transient Name Server (TNS\footnote{\url{https://www.wis-tns.org}}) and assigned International Astronomical Union (IAU) names, 3 of which were subsequently classified as \tase\ (see references in Table~\ref{tab:survey}). 
We have already reported on our observations of KSN-201509b \citep{Moon2021apj}, KSN-2018ku \citep[= SN~2018aoz;][]{Ni2022natas, Ni2023apj}, and KSN-2021V \citep[= SN~2021aefx;][]{Ni2023bapj} previously, and therefore, their observations are not described again in this section.

\subsection{Photometry} \label{sec:phot}

The KSP uses three 1.6m telescopes of the KMTNet \citep{Kim2016jkas} in Chile, South Africa, and Australia, capable of performing 24-hr continuous monitoring. 
Each telescope of the network is equipped with an identical wide-field CCD camera with 4 degree field-of-view and multiple filters in the visible band.
Images with 60-s exposures were obtained of the fields every 4--8 hours in each of the $B$, $V$, and $I$ bands. 
The $B$, $V$, and $I$ bands are observed in sequence, with an average time difference of $\sim$ 2 minutes between adjacent filters for a given epoch. 
The typical limiting magnitude for a point source in these images ranges from $\sim$ 22 mag during new moon to $\sim$ 21 mag during full moon at an S/N of 3.
This continuous monitoring with a high cadence and multiple colors provides 
an excellent opportunity to discover and investigate optical transients, 
such as SNe, from their infant phases \citep[e.g.,][]{Ni2022natas, Lee2024apj} as well as low-surface brightness objects, such as dwarf galaxies, when hundreds/thousands
of the obtained images are stacked \citep[e.g.,][]{Fan2023mnras}.
Each KSP \tas\ reported in this work was detected in $>$ 200 images with S/N $>$ 3.

\subsubsection{Initial Data Reduction Process} \label{sec:photreduce}

The KSP real-time data processing pipeline first performs the bias subtraction, 
cross-talk removal, and flat-fielding of obtained images. 
The astrometric solution in a given field is obtained by the \texttt{SCAMP}\footnote{\url{http://www.astromatic.net/software/scamp}} \citep{Bertin2006aspc} package
using more than hundreds of unsaturated stars from the second Hubble guide star catalog \citep{Lasker2008aj}, usually resulting in an astrometric precision of $\sim$ 0\farcs12.
For the fields of KSNe-2017gp and 2017fo, where not enough numbers of reference stars are available
from the catalog in the vicinity of the SNe, additional nearby reference stars from the AAVSO Photometric All-Sky Survey (APASS\footnote{\url{https://www.aavso.org/apass}}) database are used to obtain
astrometric solutions of a similar precision.

\subsubsection{Photometric Flux Measurements} \label{sec:photflux}

Point-spread function (PSF) photometry on KSP images was performed using the SuperNova Analysis Package \citep[SNAP;][]{Ni2022zndo}, a custom python-based pipeline for SN photometry and analysis that
we have developed. 
A local PSF was obtained by fitting a Moffat function \citep{Moffat1969aap, Trujillo2001mnras} to nearby reference stars after subtracting sky background determined by fitting a first-order polynomial function to an annulus around the star.
For most SNe (KSNe-2017gp, 2016ad, 2016M, 2017iw, 2016bo, 2017cv, 2018oh, 2019dz, and 2016bu), we applied a forced ``multi-object photometry'' (MOP) method to measure fluxes, similar to the scene modelling methods that have been used for photometry of Cepheids in crowded fields \citep{Riess2016apj}.
The profiles of nearby sources (e.g., host galaxies or overlapping background sources) 
that affect the photometry of the SN were measured in advance from deep 
pre-SN images created by stacking many images taken before the SN explosions with the \texttt{SWARP\footnote{\url{https://www.astromatic.net/software/swarp/}}} \citep{Bertin2002aspc} package.
The fluxes of the SN and nearby sources were measured simultaneously by fitting them together (see Appendix~\ref{sec:mop} for the details).

For the fields of KSNe-2017cz, 2017fo, 2019bl, 2018ng, 2021iq, and 202112D, we found
the MOP method unsuitable due to the difficulty of obtaining the profiles 
of the nearby sources. 
This happens because of the intrinsic shape of host galaxies that cannot be 
modeled by a S\'ersic profile \citep{Caon1993mnras} or positional degeneracy between the host galaxy nucleus and the SN.
In these cases, we performed image subtraction with the \texttt{HOTPANTS\footnote{\url{http://www.astro.washington.edu/users/becker/v2.0/hotpants.html}}} \citep{Becker2015ascl} package.
Deep pre-SN images were subtracted from target images after matching the seeing, leaving a subtracted image containing the SN. 
The SN flux was then measured by fitting the PSF after removal of the background with a first-order polynomial function fitted to an annulus.
Image subtraction is known to produce extreme photometric outliers (Appendix~\ref{subsec:cophot}), which we mitigated by excluding the images that produced obviously poor subtraction results due to bad seeing conditions and data points with $\gtrsim$ 4-$\sigma$ deviation from the smoothed light curve obtained with a Gaussian process model (Appendix~\ref{sec:gpint}).
Note that several images are typically acquired every night in our high-cadence observations which makes it possible to identify the photometric outliers for removal.

Finally, we applied aperture photometry in one case of KSN-201903ah.
In this case, MOP was inapplicable due to the positional degeneracy of the point-like host galaxy with the SN and image subtraction (which tends to add noise and produce outliers) was avoidable due to its small size.
We determined aperture radii containing 90\% of the PSF flux and used circular apertures to measure the combined flux of the SN and host after sky background removal.
We then subtract the flux of the host measured from pre-SN images.  

\begin{deluxetable*}{ccccccccc}
\tabletypesize{\footnotesize}
\tablecolumns{9} 
\tablewidth{0.99\textwidth}
 \tablecaption{KSP photometry of \tase.}
 \tablehead{
 \colhead{KSN} & \colhead{Date [MJD]} & \colhead{1-$\sigma$ [days]} & \colhead{Band} & \colhead{App. Mag.$\rm ^a$} & \colhead{1-$\sigma\rm ^b$} & \colhead{S/N} & \colhead{Phase$\rm ^c$} & \colhead{Abs. Mag.$\rm ^d$}
 } 
\startdata 
2016M           & 	  57683.674 & 	  0.001	 & $B$ & 	  20.81 & 	  0.27 & 	    4.0 & 	  -13.680 & 	  -17.12 \\
2016M           & 	  57684.689 & 	  0.024	 & $B$ & 	  20.82 & 	  0.13 & 	    8.2 & 	  -12.735 & 	  -17.12 \\
2016M           & 	  57684.691 & 	  0.024	 & $V$ & 	  20.67 & 	  0.10 & 	   10.3 & 	  -12.734 & 	  -17.27 \\
2016M           & 	  57684.692 & 	  0.024	 & $i$ & 	  20.92 & 	  0.14 & 	    7.9 & 	  -12.732 & 	  -16.77 \\
2016M           & 	  57685.691 & 	  0.022	 & $B$ & 	  20.38 & 	  0.08 & 	   12.9 & 	  -11.804 & 	  -17.57 \\
2016M           & 	  57685.692 & 	  0.022	 & $V$ & 	  20.38 & 	  0.07 & 	   15.0 & 	  -11.802 & 	  -17.55 \\
2016M           & 	  57685.694 & 	  0.022	 & $i$ & 	  20.72 & 	  0.11 & 	   10.1 & 	  -11.801 & 	  -16.97 \\
2016M           & 	  57686.305 & 	  0.025	 & $B$ & 	  20.19 & 	  0.04 & 	   26.1 & 	  -11.232 & 	  -17.76 \\
2016M           & 	  57686.307 & 	  0.025	 & $V$ & 	  20.22 & 	  0.05 & 	   21.4 & 	  -11.231 & 	  -17.72 \\
2016M           & 	  57686.308 & 	  0.025	 & $i$ & 	  20.49 & 	  0.07 & 	   16.7 & 	  -11.229 & 	  -17.20 \\
2016M           & 	  57686.692 & 	  0.022	 & $B$ & 	  20.02 & 	  0.07 & 	   15.9 & 	  -10.873 & 	  -17.93 \\
2016M           & 	  57686.694 & 	  0.022	 & $V$ & 	  19.97 & 	  0.06 & 	   18.5 & 	  -10.871 & 	  -17.96 \\
2016M           & 	  57686.695 & 	  0.022	 & $i$ & 	  20.53 & 	  0.11 & 	   10.3 & 	  -10.870 & 	  -17.17 \\
2016bu          & 	  57687.102 & 	  0.060	 & $i$ & 	  22.52 & 	  0.22 & 	    4.9 & 	  -14.450 & 	  -15.79 \\
2016M           & 	  57687.217 & 	       	 & $B$ & 	  19.90 & 	  0.05 & 	   20.1 & 	  -10.384 & 	  -18.05 \\
2016M           & 	  57687.219 & 	       	 & $V$ & 	  19.96 & 	  0.09 & 	   12.6 & 	  -10.382 & 	  -17.97 \\
2016M           & 	  57687.220 & 	       	 & $i$ & 	  20.18 & 	  0.18 & 	    6.0 & 	  -10.381 & 	  -17.52 \\
2016M           & 	  57687.312 & 	  0.039	 & $B$ & 	  19.79 & 	  0.03 & 	   35.2 & 	  -10.296 & 	  -18.16 \\
2016M           & 	  57687.313 & 	  0.048	 & $V$ & 	  19.87 & 	  0.04 & 	   27.5 & 	  -10.295 & 	  -18.06 \\
2016M           & 	  57687.315 & 	  0.039	 & $i$ & 	  20.18 & 	  0.05 & 	   22.8 & 	  -10.293 & 	  -17.52 \\
2016M           & 	  57687.667 & 	       	 & $B$ & 	  19.64 & 	  0.06 & 	   17.8 & 	   -9.965 & 	  -18.31 \\
2016M           & 	  57687.669 & 	       	 & $V$ & 	  19.78 & 	  0.07 & 	   16.5 & 	   -9.964 & 	  -18.15 \\
2016M           & 	  57687.670 & 	       	 & $i$ & 	  19.97 & 	  0.09 & 	   12.4 & 	   -9.963 & 	  -17.74 \\
2016bu          & 	  57687.753 & 	  0.181	 & $V$ & 	  22.38 & 	  0.22 & 	    4.9 & 	  -13.867 & 	  -16.25 \\
2016bu          & 	  57687.756 & 	  0.180	 & $i$ & 	  22.31 & 	  0.28 & 	    3.9 & 	  -13.863 & 	  -15.99 \\
\enddata
\tablenotetext{{\rm a}}{The $BV$-band apparent magnitudes are in the Vega system,
while the $i$-band magnitudes are in the AB system (see Section~\ref{sec:photcal} text).}
\tablenotetext{{\rm b}}{1-$\sigma$ error of apparent magnitudes includes detection S/N and flux calibration error (Section~\ref{sec:phot}).}
\tablenotetext{{\rm c}}{Phase is rest-frame days since $B$-band maximum ($t_{\rm max}$; Table~\ref{tab:snparam}).}
\tablenotetext{{\rm d}}{Dereddened and K--corrected rest-frame absolute magnitudes (see Sections~\ref{sec:restlc} text).}
\tablecomments{The entire observed magnitudes are available in the electronic edition.}
\end{deluxetable*} 
\label{tab:lc}

\subsubsection{Photometric Flux Calibration} \label{sec:photcal}

Photometric flux calibration was performed against 5--21 nearby standard reference stars
from the APASS database whose apparent magnitudes are in the range of 15--16 mag.
The photometry of the APASS reference stars are given in the Johnson $BV$ and Sloan $i$ bands; 
therefore, our photometry is calibrated to the $BVi$ bands.
We correct the $B$-band instrumental magnitudes of the reference stars and SNe before calibration 
due to the known difference between the filter response functions of the KMTNet and Johnson systems \citep{Park2017apj, Park2019apj}.
This filter difference produces a systematic color-dependence in the observed instrumental magnitudes of APASS reference stars, while no such color-dependence has been found for the $V$ and $i$ bands. 

Our $B$-band photometry correction was made in two steps as follows. 
We first corrected the reference star instrumental magnitudes based on their \bv\ colors following the procedure detailed in \citet{Park2017apj}. 
For SNe,
whose spectra differ substantially from stars and evolve with time,
we apply spectrophotometric (S)--corrections \citep{Stritzinger2002aj} to their instrumental magnitudes
by performing synthetic photometry on \tas\ spectral templates \citep{Hsiao2007apj} fitted to the $V$- and $i$-band light curves \citep[as done in the case of KSN-201509b;][also see Appendix~\ref{sec:standard}]{Moon2021apj}.
The calibrated and S--corrected photometric detections are presented in Table~\ref{tab:lc}.
Note that clusters of 2--10 images taken within the same night in the early light curves were sometimes binned (using \texttt{SWARP}) to maximize detection S/N, in which case the 1-$\sigma$ range of the epochs for the binned images is provided in column 3 of the table.

\begin{deluxetable*}{llcccccccc}
\tabletypesize{\footnotesize}
\tablecolumns{10} 
\tablewidth{0.99\textwidth}
 \tablecaption{Spectroscopic data of KSP Type Ia SNe.}
 \tablehead{
 \colhead{KSN} & \colhead{Date$\rm ^a$} & \colhead{Phase$\rm ^b$} & \colhead{Telescope} & \colhead{Instrument} & \colhead{R} & \colhead{Wavelength} & \colhead{Redshifts$\rm ^c$} & \colhead{\nai~D$\rm ^d$} & \colhead{\siii~$\lambda$6355~\AA$\rm ^e$}\\
 \colhead{} & \colhead{[UT]} & \colhead{} & \colhead{} & \colhead{} & \colhead{} & \colhead{[\AA]} & \colhead{} & \colhead{[mag]} & \colhead{[10$^3$~km~s$^{-1}$]}
 } 
\startdata 
2017gp & 2017-11-15.07 & 10.33 & du Pont & WFCCD & 800 & 3600--9200 & 0.135 $\pm$ 0.005 & not seen & $-$10.71 $\pm$ 0.40 \\
 & 2024-01-27.04 & host & Gemini S & GMOS & 1690 & 3700--10000 & \textbf{0.1364 $\pm$ 0.0002} & \\
 \hline
2016M & 2016-11-9.32 & 2.74 & Gemini S & GMOS & 1690 & 5300--10000 & 0.071 $\pm$ 0.005 & not seen & $-$12.60 $\pm$ 0.37 \\
 & 2024-01-23.18 & host & SOAR & Goodman & 930 & 5000--9000 & \textbf{0.0752 $\pm$ 0.0002} & \\
 \hline
 2016ad & 2016-11-16.27 & $-$5.46 & Gemini S & GMOS & 1690 & 3700--10000 & 0.058 $\pm$ 0.005 & not seen & $-$11.63 $\pm$ 0.02 \\
 & 2016-11-18.32 & $-$3.52 & NTT & EFOSC2 & 355 & 3650--9250 & 0.058\ \ \ \ \ [1] & not seen & $-$11.43 $\pm$ 0.09 \\
& 2017-05-1.01 & host & Magellan-Clay & LDSS-3 & 860 & 4250--10000 & \textbf{0.0613 $\pm$ 0.0002} \\
\hline
2017iw & 2017-03-28.02 & 12.16 & du Pont & WFCCD & 800 & 3600--9200 & 0.063 $\pm$ 0.006 & not seen & $-$9.87 $\pm$ 0.21 \\
 & 2024-01-23.10 & host & SOAR & Goodman & 930 & 5000--9000 & \textbf{0.0617 $\pm$ 0.0008} \\
\hline
2016bo & 2018-01-6.53 & host & Gemini S & GMOS & 1690 & 3700--7000 & \textbf{0.0563 $\pm$ 0.0001} & \\
& NED & host & UK Schmidt & 6dF & 1000 & 4000--7500 & 0.0565 $\pm$ 0.0002 & [4]\ \ \ \ \ \ \ \ \ \ \ \ \ \ \ \ \ \ \ \ \\
\hline
2017cv & NED & host & UK Schmidt & 6dF & 1000 & 4000-7500 & \textbf{0.0866 $\pm$ 0.0002} & [4]\ \ \ \ \ \ \ \ \ \ \ \ \ \ \ \ \ \ \ \ \\
\hline
2018oh & 2018-03-18.06 & $-$3.15 & Gemini S & GMOS & 1690 & 3700--10000 & & not seen & $-$12.53 $\pm$ 0.24 \\
& 2018-03-19.15 & host & du Pont & WFCCD & 800 & 3600--9200 & \textbf{0.0566 $\pm$ 0.0001} \\
& 2018-03-20.17 & $-$1.15 & du Pont & WFCCD & 800 & 3600--9200 & 0.054 $\pm$ 0.002 & not seen &  $-$12.18 $\pm$ 0.46 \\
& 2018-04-5.08 & 13.91 & Gemini S & GMOS & 1690 & 3700--7000 & & not seen & $-$10.80 $\pm$ 0.08 \\
& 2018-05-10.96 & 47.87 & Gemini S & GMOS & 1690 & 5000--10000 & & not seen & not seen \\
& NED & host & 2MASS & ANN$z$ &  &  & 0.051 $\pm$ 0.015 & [5]\ \ \ \ \ \ \ \ \ \ \ \ \ \ \ \ \ \ \ \ \\
\hline
2017cz & 2017-11-15.14 & host & du Pont & WFCCD & 800 & 3600--9200 & \textbf{0.108 $\pm$ 0.001} & \\
\hline
2017fo & 2016-08-11 & host & Magellan-Clay & LDSS-3 & 860 & 4250--10000 & \textbf{0.167}\ \ \ \ \ [2] \\
& 2017-11-14.03 & $-$0.19 & Gemini S & GMOS & 1690 & 3700--10000 & 0.168 $\pm$ 0.006 & blended &  $-$10.87 $\pm$ 0.14 \\
\hline
2019bl & 2019-03-27.18 & 6.31 & Gemini S & GMOS & 1690 & 3700--10000 & 0.062 $\pm$ 0.006 & blended & $-$10.86 $\pm$ 0.07 \\
& NED & host & UK Schmidt & 6dF & 1000 & 4000-7500 & \textbf{0.0637 $\pm$ 0.0001} & [4]\ \ \ \ \ \ \ \ \ \ \ \ \ \ \ \ \ \ \ \ \\
\hline
2018ng & 2018-05-19.05 & $-$0.62 & Gemini S & GMOS & 1690 & 5000--10000 & (0.073 $\pm$ 0.007) & blended & $-$11.50 $\pm$ 0.42 \\
 & 2018-06-14.97 & 24.42 & Gemini S & GMOS & 1690 & 5000--10000 & & blended & not seen \\
& $\hookrightarrow$$\rm ^f$ & host & & & & & \textbf{0.0748 $\pm$ 0.0002} \\
& NED & host & 2MASS & ANN$z$ &  &  & 0.073 $\pm$ 0.015 & [5]\ \ \ \ \ \ \ \ \ \ \ \ \ \ \ \ \ \ \ \ \\
\hline
2019dz & 2019-03-27.21 & 14.41 & Gemini S & GMOS & 1690 & 5000--10000 & \textbf{0.140 $\pm$ 0.004} & not seen & $-$10.18 $\pm$ 0.05 \\
& NED & host & 2MASS & ANN$z$ &  &  & 0.101 $\pm$ 0.015 & [5]\ \ \ \ \ \ \ \ \ \ \ \ \ \ \ \ \ \ \ \ \\
\hline
2016bu & 2017-01-28.04 & host & du Pont & WFCCD & 800 & 3600--9200 & \textbf{0.115 $\pm$ 0.002} & \\
& NED & host & 2MASS & ANN$z$ &  &  & 0.102 $\pm$ 0.015 & [5]\ \ \ \ \ \ \ \ \ \ \ \ \ \ \ \ \ \ \ \ \\
\hline
2021iq & 2021-04-19.04 & 8.61 & Gemini S & GMOS & 1690 & 3700--10000 & 0.098 $\pm$ 0.006 & 0.021 $\pm$ 0.014 & $-$9.36 $\pm$ 0.23 \\
& 2021-07-1.96 & host & Gemini S & GMOS & 1690 & 5000--10000 & \textbf{0.0927 $\pm$ 0.0003} & \\
\hline
202112D & 2022-01-3.10 & $-$3.26 & Gemini S & GMOS & 1690 & 3700--10000 & 0.048 $\pm$ 0.006 & 0.023 $\pm$ 0.016 & $-$12.21 $\pm$ 0.117 \\
& 2022-01-22.80 & 17.33 & SALT & RSS & 1000 & 3700--8300 & [3] &  & $-$10.14 $\pm$ 0.77 \\
& $\hookrightarrow$$\rm ^f$ & host &  &  &  &  & \textbf{0.0531 $\pm$ 0.0003} \\
& NED & host & Sloan & SPEC2 & 2000 & 3600-10000 & 0.05285 $\pm$ 0.00001 & [6]\ \ \ \ \ \ \ \ \ \ \ \ \ \ \ \ \ \ \ \ \\
\enddata
\tablenotetext{{\rm a}}{Date of acquisition; except for the NED results, which are obtained by querying the listed ($\alpha$, $\delta$) (J2000) in Table~\ref{tab:hosts}.}
\tablenotetext{{\rm b}}{Phase is rest-frame days since $B$-band maximum ($t_{\rm max}$; Table~\ref{tab:snparam}).}
\tablenotetext{{\rm c}}{Bold font indicates our adopted values of the host galaxy redshift ($z_{\rm host}$) throughout the paper.}
\tablenotetext{{\rm d}}{Visibility of \nai~D feature at the redshift of the SN ($z_{\rm host}$ from Table~\ref{tab:snparam}).
The measured host $E(\bv)$ [mag] is provided when a resolved doublet is visible.}
\tablenotetext{{\rm e}}{Visibility of \siii~$\lambda$6355~\AA\ feature. The measured velocity [10$^3$~km~s$^{-1}$] of the \siii~$\lambda$6355~\AA\ PVF (see Section~\ref{sec:sivel}) is provided when visible.}
\tablenotetext{{\rm f}}{The explosion spectra of these SNe contain visible emission lines from their host galaxies which are used for host galaxy redshift estimation.}
\tablecomments{[1] \citet{Takats2016tns}; [2] \citet{Moon2021apj}; [3] \citet{Gromadzki2022tns}; [4] \citet{Jones2004mnras}; [5] Photometric redshift obtained with Artificial Neural Network \citep[ANN$z$;][]{Bilicki2014apjs} trained on data from spectroscopic redshifts surveys, including 6dF and SDSS; [6] \citet{Albareti2017apjs}}
\end{deluxetable*} 
\label{tab:spec}

\begin{figure*}[hbtp]
\epsscale{\scl}
\begin{center}
\includegraphics[width=1.00\textwidth]{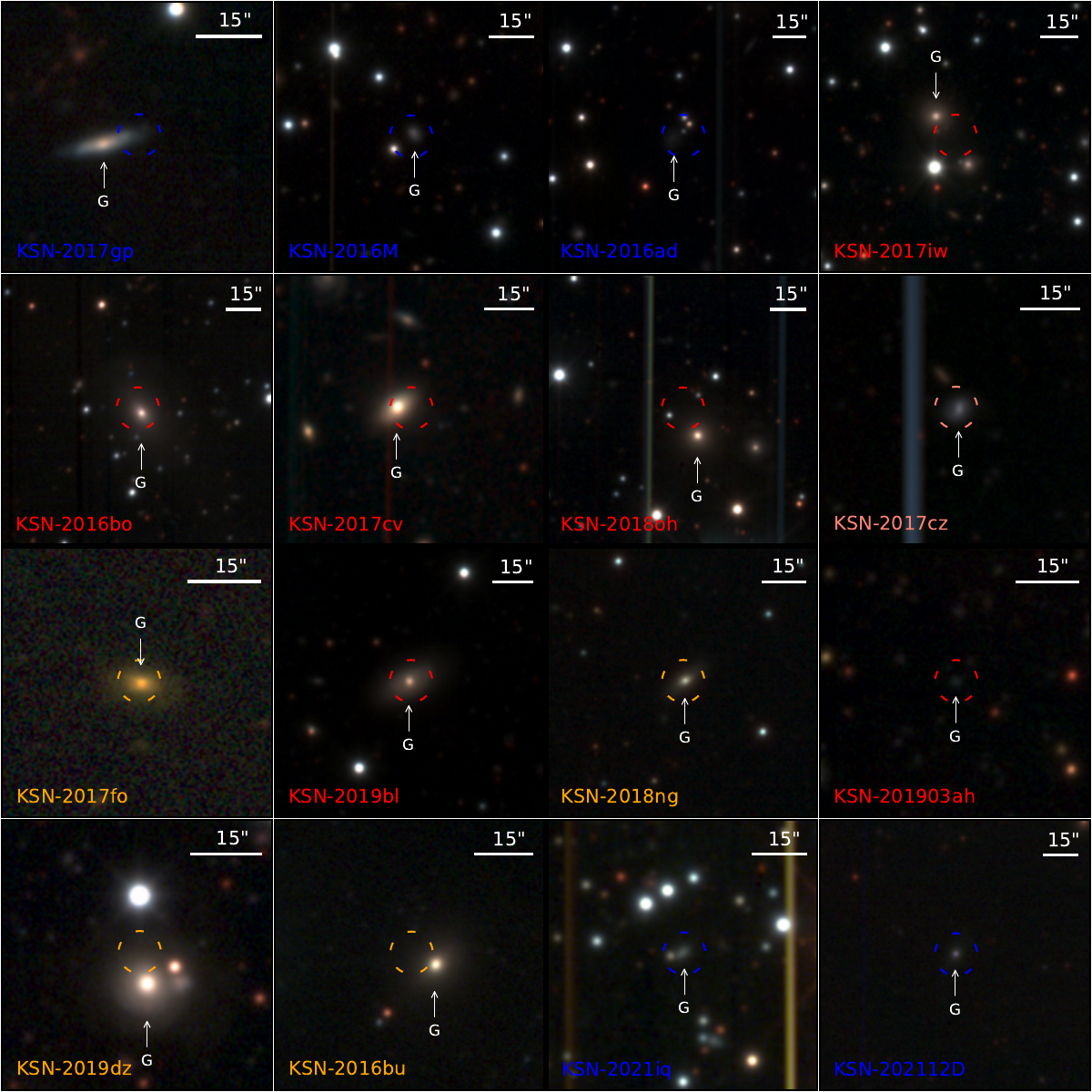}
\end{center}
\caption{Deep $BVi$ pre-explosion stack image stamps in RGB format for each \tas\ from the KSP, whose positions are indicated with dashed circles (colored by population; Section~\ref{subsec:popid}), showing their host galaxies (indicated with white arrows labelled ``G''; Table~\ref{tab:hosts}).
To effectively reveal faint morphological features of the galaxies, the intensity scale of the images is proportional to the square of the observed flux with minima and maxima spanning from the 1-$\sigma$ noise floor below the median flux to 110\% of the central flux of each image at SN peak.
The RGB colors are relative to that of the Sun (= 5778~K blackbody) in the host galaxy rest frame.
Up is North and East is to the left,
and the width of every image stamp is equal to a physical 
scale of 150 kpc in the rest frame.
(This figure is available as an animation in the online journal showing the evolution of each \tas---from $-$10 days since first light to $+$40 days at 0.1-day intervals---in stamps of the same images but with equal angular size of 1\farcm33$\times$1\farcm33.
In the animation, synthetic point-sources matching the image PSFs and smoothed $BVi$-band light curves are injected at the SN positions. The real-time duration of the animation is 50 seconds---i.e., 1 second = 1 day.)}
\label{fig:hosts}
\end{figure*}

\subsection{Spectroscopy} \label{sec:spec}

We conducted spectroscopic observations of 16 \tase\ from the KSP, 
and present 15 explosion and 10 host galaxy spectra of 14 previously unpublished events
(see Sections~\ref{sec:siclass}, \ref{sec:sivel} and \ref{subsec:hostppxf} below for the spectra and analysis).
Note that spectra of two SNe (KSNe-2018ku and 2021V) were already published in \citet{Ni2022natas, Ni2023bapj, Ni2023apj}.
We used a combination of the Gemini Multi-Object Spectrograph \citep[GMOS;][]{Hook2004} on the 8.1m Gemini-South telescope and Goodman high-throughput spectrograph \citep{Clemens2004spie} on the 4.1m Southern Astrophysical Research (SOAR) telescope at Cerro Pach\'on; as well as the Wide-Field CCD \citep[WFCCD;][]{Weymann2001apj} on the 2.5m du Pont telescope and Low Dispersion Survey Spectrograph-3 \citep[LDSS-3;][]{Allington-Smith2002pasp} on the 6.5m Magellan-Clay telescope at Las Campanas Observatory to obtain the spectra.
We supplement these data with published spectra of KSNe-2016ad \citep{Takats2016tns} and 202112D \citep{Gromadzki2022tns} from TNS; 
the host galaxy of KSN-2017fo obtained by \citet[][source ``G'' in the paper]{Moon2021apj} 
in the study of KSN-201509b;
and spectroscopic data of some of the host galaxies from the NASA/IPAC Extragalactic Database (NED\footnote{\url{https://ned.ipac.caltech.edu}}).
The entire set of spectroscopic data used in this work (excluding those of KSNe-2018ku and 2021V already published) is summarized in Table~\ref{tab:spec}.

Spectral data reduction consists of bias and flat-field corrections,
spectral extraction, wavelength calibration against spectra from calibration lamps
obtained immediately after target observations, as well as flux calibration
using spectrophotometric standards observed in the same setup obtained
in the same night or sometimes in the same semester. 
GMOS and Goodman spectra were reduced using the custom \texttt{gmos} suite of \texttt{IRAF\footnote{IRAF is distributed by the National Optical Astronomy Observatory, which is operated by the Association for Research in Astronomy, Inc. under cooperative agreement with the National Science Foundation.}} 
\citep{Tody1993aspc} tasks and the Goodman Data-Reduction Pipeline\footnote{\url{https://soardocs.readthedocs.io/projects/goodman-pipeline}}, respectively. 
Flux calibration for GMOS and Goodman was performed using \texttt{IRAF}.
Spectra from the du Pont and Magellan telescopes were also reduced using standard tasks within \texttt{IRAF}, while their flux calibration and telluric corrections were performed 
with a set of custom IDL scripts \citep{Matheson2008, Blondin2012aj}. 
The full set of reduced spectra will be made available on WISeREP \citep{Yaron2012pasp}.

We spectroscopically classify the 11 events with explosion spectra by cross-correlating the spectrum nearest $B$-band maximum with SN templates using \texttt{SNID}\footnote{\url{https://people.lam.fr/blondin.stephane/software/snid/index.html}}.
For each event, we obtain the best-matching template as well as the redshift ($z_{\rm SN}$) that maximizes the quality of the match, which is quantified by the product $r \times lap$ of the ``height-to-noise ratio'' of the correlation function at its maximum (= $r$) and the overlap between the wavelengths (from $\lambda_1$ to $\lambda_2$) of the SN spectrum and the template in $\ln(\lambda)$ space---i.e., $lap$ = $\ln(\lambda_1/\lambda_2)$.
Note that $r \times lap >$ 5 typically constitutes an acceptable match \citep{Blondin2007apj}.
\texttt{SNID} classifies all 11 events as \tase, which is based on 80--100\% of more than 50 acceptably matching templates being \tase, except in the case of KSN-2018ng where only 2 matches with $r \times lap >$ 5 were found (both with \tas\ templates) due to blending with its host spectrum.
Except for KSNe-2018ng, the matching \tas\ templates had $r \times lap$ values that were well-within the acceptable range---attaining maxima between 6 and 30 in all cases---and comprised mostly (70--100\%) of normal \tase, favouring normal Type Ia sub-classification (see Section~\ref{sec:siclass} for more detailed spectral sub-classification).
For all 11 near-peak spectra, Table~\ref{tab:spec} (column 8) provides the median and standard deviation of $z_{\rm SN}$ for the matching \tas\ templates, including less reliable ones obtained for KSN-2018ng in parentheses.

\begin{deluxetable*}{lccccccc|cccr}
\tabletypesize{\footnotesize}
\tablecolumns{9} 
\tablewidth{0.99\textwidth}
 \tablecaption{Host galaxy properties of KSP Type Ia SNe.}
 \tablehead{
 \colhead{KSN} & \colhead{Host galaxy ($\alpha$, $\delta$) (J2000)} & \colhead{Morph} & \colhead{SN location} & $B$ & $V$ & \bv\ & \colhead{$M_{\star}$} & \colhead{SFR} & \colhead{sSFR} & \colhead{Age} & \colhead{[M/H]}
 } 
\startdata
\textcolor{blue}{2017gp$^b$} & ($\rm 23^h54^m2^s.038,\ -52\degr48\arcmin32\farcs735$) & S & N.W. (edge) & 18.39 $\pm$ 0.10 & 17.42 $\pm$ 0.07 & 0.97 & 15.4 & 0.84 & 5.51 & 0.11 & $-$1.21 \\
\textcolor{blue}{2016M$^b$} & ($\rm 6^h14^m45^s.213,\ -24\degr29\arcmin54\farcs946$) & Irr & S.E. (edge) & 19.63 $\pm$ 0.01 & 18.84 $\pm$ 0.02 & 0.78 & 0.86 & 0.05 & 6.02 & 8.45 & $-$0.20 \\
\textcolor{blue}{2016ad$^b$} & ($\rm 6^h13^m11^s.865,\ -24\degr16\arcmin21\farcs166$) & dIrr & N.W. (edge) & 19.66 $\pm$ 0.01 & 18.95 $\pm$ 0.02 & 0.71 & 0.28 & 0.07 & 24.3 & 3.63 & $-$0.51 \\
\textcolor{red}{2017iw$^r$} & ($\rm 6^h17^m59^s.880,\ -23\degr48\arcmin44\farcs500$) & E & S.W. (halo) & 17.97 $\pm$ 0.02 & 16.94 $\pm$ 0.03 & 1.02 & 7.09 &  &  & 0.34 & $-$0.27 \\
\textcolor{red}{2016bo$^r$} & ($\rm 6^h44^m31^s.830,\ -25\degr34\arcmin10\farcs999$) & E & N.E. (core) & 16.18 $\pm$ 0.02 & 15.14 $\pm$ 0.02 & 1.04 & 31.2 &  &  & 6.67 & 0.12 \\
\textcolor{red}{2017cv$^r$} & ($\rm 0^h50^m16^s.230,\ -19\degr19\arcmin57\farcs500$) & E & N.W. (edge) & 17.51 $\pm$ 0.03 & 16.19 $\pm$ 0.07 & 1.32 & 64.3 &  &  & 5.90 & 0.02 \\
\textcolor{red}{2018oh$^r$} & ($\rm 9^h44^m29^s.100,\ -31\degr43\arcmin53\farcs800$) & E & N.E. (halo) & 17.24 $\pm$ 0.06 & 16.01 $\pm$ 0.03 & 1.23 & 28.5 &  &  & 13.5 & $-$0.02\\
\textcolor{pinkred}{2017cz$^{(r)}$} & ($\rm 0^h52^m41^s.050,\ -38\degr4\arcmin46\farcs495$) & Irr & E. (core) & 19.43 $\pm$ 0.03 & 18.75 $\pm$ 0.03 & 0.68 & 1.37 & 0.29 & 21.4 & 1.11 & $-$0.07 \\
\textcolor{gold}{2017fo$^y$} & ($\rm 0^h57^m1^s.216,\ -37\degr2\arcmin37\farcs907$) & S0 & Nuclear & 19.16 $\pm$ 0.04 & 17.97 $\pm$ 0.05 & 1.18 & 23.0 &  &  & 4.55 & $-$0.37 \\
\textcolor{red}{2019bl$^r$} & ($\rm 11^h6^m28^s.561,\ -22\degr44\arcmin27\farcs161$) & S & Nuclear & 17.18 $\pm$ 0.03 & 16.05 $\pm$ 0.03 & 1.13 & 26.1 & 0.20 & 0.77 & 6.70 & 0.02 \\
\textcolor{gold}{2018ng$^y$} & ($\rm 11^h29^m15^s.400,\ -30\degr31\arcmin5\farcs902$) & S0 & Nuclear & 18.23 $\pm$ 0.02 & 17.05 $\pm$ 0.02 & 1.18 & 14.2 \\
\textcolor{red}{201903ah$^r$} & ($\rm 11^h30^m37^s.211,\ -30\degr14\arcmin9\farcs970$) & dE & Coincident & 23.16 $\pm$ 0.01 & 22.57 $\pm$ 0.01 & 0.58 & 0.03 \\
\textcolor{gold}{2019dz$^y$} & ($\rm 11^h51^m4^s.007, -28\degr57\arcmin36\farcs068$) & E & N. (edge) & 17.87 $\pm$ 0.02 & 16.66 $\pm$ 0.01 & 1.21 & 77.9 \\
\textcolor{gold}{2016bu$^y$} & ($\rm 0^h12^m55^s.630,\ -20\degr44\arcmin20\farcs900$) & E & N.E. (halo) & 17.84 $\pm$ 0.04 & 16.68 $\pm$ 0.03 & 1.16 & 39.4 &  & & 3.64 & 0.22 \\
\textcolor{blue}{2021iq$^b$} & ($\rm 11^h4^m7^s.865$,\ $-46\degr24\arcmin36\farcs696$) & Irr & Coincident & 20.97 $\pm$ 0.03 & 20.06 $\pm$ 0.05 & 0.91 & 0.66 &  &  & 1.50 & $-$1.54\\
\textcolor{blue}{202112D$^b$} & ($\rm 2^h38^m31^s.569,\ -8\degr15\arcmin45\farcs583$) & S & Coincident & 18.83 $\pm$ 0.02 & 18.16 $\pm$ 0.04 & 0.67 & 0.62 & 0.27 & 43.6 & 0.98 & $-$1.40
\enddata
\tablecomments{Columns include: (1) KSP name (colored by population in the superscript; Section~\ref{subsec:popid}); (2) host galaxy equatorial coordinate; (3) morphological type; (4) location of the SN in the host galaxy; (5-7) apparent magnitudes and colors [mag]; (8) stellar mass [10$^{10}$~\msol]; (9) star formation rate [\msol~year$^{-1}$] from NUV luminosities (Section~\ref{subsec:hostsfr}); (10) specific star formation rate (= SFR/$M_{\star}$) [10$^{-12}$~year$^{-1}$]; (11--12) luminosity-weighted age [Gyr] and metallicity [dex]. For (11) and (12), $\log_{10}({\rm Age})$ and [M/H] both have conservative 1-$\sigma$ uncertainties of $\lesssim$ 0.3 dex \citep[see Figure 7 of][]{Lee2023mnras}.}
\end{deluxetable*} 
\label{tab:hosts}

\subsection{Host Galaxy and Redshift}\label{sec:host}

We identify host galaxies for the SNe and measure their redshifts using several methods as follows. 
First, we identified the host galaxies for each SN by examining the pre-SN stacked images shown in Figure~\ref{fig:hosts}.
For all but one SN (KSN-201903ah), we found either: (i) one resolved source in the vicinity of the SN, which we identify as the host galaxy; or (ii) multiple resolved sources, but only one that has a comparable redshift to $z_{\rm SN}$ (see Section~\ref{sec:spec}), consistent with it being the host.
We obtained redshifts for these host galaxies (Table~\ref{tab:spec}, column 8) through a combination of acquiring spectra and cross-matching the candidates against NED. 

For each of our spectra, we fit emission/absorption lines---typically H$\alpha$, H$\beta$, [\nii], [\oii], and/or the \nai~Doublet (\nai~D)---independently with Voigt profiles \citep{Armstrong1967}.
The host redshift and its uncertainty are the mean and standard deviation of the multiple line redshifts.
These spectroscopic redshifts for the host galaxies agree with $z_{\rm SN}$ to within the 3-$\sigma$ level, as do the NED redshifts when they are available.
Our adopted value for the host galaxy ($z_{\rm host}$; bold in Table~\ref{tab:spec}) is 
determined in the following way: 
(1) for 11 of the SNe, we use the redshift obtained from our own spectra mostly observed with higher resolution;
(2) for 3 cases where we do not have our own spectra, we adopt the NED spectroscopic redshifts;
(3) for one remaining case (KSN-2019dz) where only a NED photometric redshift is available, we adopt the spectroscopic redshift of the SN ($z_{\rm SN}$).

In the case of KSN-201903ah for which we have no explosion or host spectra, we estimated $z_{\rm SN}$ = 0.132 $\pm$ 0.003 by fitting \tas\ templates to the light curve (see Appendix~\ref{sec:standard}), assuming the velocity of the SN rest frame is $\lesssim$ 500~km~s$^{-1}$ with respect to the Hubble flow following the method detailed in \citet{Moon2021apj}.
There are no resolved sources visible in the pre-SN stacked image.
However, we identified an isolated point source coincident with the SN position; and considering it is the brightest source
within 25~kpc (Figure~\ref{fig:hosts}), it is most likely the host.
No spectrum was obtained for the host due to its faintness, so we adopt the template-fitted $z_{\rm SN}$ as its redshift (= $z_{\rm host}$).

Finally, we carry out photometry for the host galaxies with a standardized process as follows. First, objects around each host galaxy are masked using Source Extractor \citep{Bertin&Arnouts1996aas}. 
Then, fluxes of ellipse apertures around the host galaxy are measured using the \texttt{IRAF} task \texttt{ellipse}. The magnitude of the host galaxy is determined from an optimized aperture and a background level using a growth curve. 
We conduct flux calibration using standard stars in APASS as described in Section~\ref{sec:photcal}, correcting the $B$-band magnitudes of the host for \bv\ color dependence \citep[for more details, see][]{Park2017apj, Park2019apj}.
Table~\ref{tab:hosts} compiles the location and properties of the host galaxy for each \tas.

\subsection{Transformation to Rest-Frame Light Curve}\label{sec:restlc}

We transform the apparent magnitudes ($m_{\lambda}$) of the SNe in each observed filter (= $BVi$) $\lambda$ to absolute magnitudes ($M_{\lambda_0}$) in the same rest-frame filters $\lambda_0$ with the equation
\begin{equation}
    M_{\lambda_0} = m_{\lambda} - A_{\lambda, {\rm MW}} - DM - K_{\lambda_0,\lambda} -  A_{\lambda_0, {\rm host}}
    \label{eq:standard}
\end{equation}
where $A_{\lambda, {\rm MW}}$ and $A_{\lambda_0, {\rm host}}$ represent extinction from the Milky Way (MW) and host galaxy, respectively, in each filter, $DM$ is the distance modulus, and $K_{\lambda_0,\lambda}$ represents K--correction between the observed and rest-frame filters \citep{Hogg2002, Oke1968apj}.
All of our SNe suffer from relatively little Galactic extinction since our target fields are away from the Galactic plane, with $A_{V, {\rm MW}}$ ranging in 0.03--0.3 mag obtained from the extinction model of \citet{Schlafly&Finkbeiner2011apj}.
Similar to S--corrections (Section~\ref{sec:photcal}), we estimate K--corrections by performing synthetic photometry on fitted \tas\ templates, simultaneously obtaining $DM$ and $A_{\lambda_0, {\rm host}}$ as best-fit parameters (see Appendix~\ref{sec:standard} for the details).
Table~\ref{tab:snparam} summarizes the key measured parameters of each SN, including $z_{\rm host}$, $DM$, and extinction.

We assess the validity of the template-fitted $DM$ and $A_{\lambda_0, {\rm host}}$ using the following independent estimation methods. 
First, we estimate the expected Hubble flow distance modulus, $DM_{\rm cosmo}(z_{\rm host})$, in the cosmology of \citet{Riess2016apj}, with corrections for peculiar velocities due to the Virgo Supercluster, Great Attractor, and Shapley Supercluster \citep{Mould2000apj}, adopting $H_0$ uncertainty of $\pm$5~km~s$^{-1}$~Mpc$^{-1}$ following the identical method used by NED.
We find broad agreement between $DM_{\rm cosmo}(z_{\rm host})$ and the template-fitted $DM$, with the differences between them ranging in $\pm$3-$\sigma$ of their combined uncertainties.
Throughout the paper, we adopt the template-fitted $DM$ since they are less dependent on the choice of cosmological parameters, and more precise than the Hubble flow ones (Table~\ref{tab:snparam}).
From $DM$, we then calculated the luminosity distance as $D_L = 10^{1+DM/5}$~pc and the angular distance \citep{Hogg1999} as $D_A = D_L/(1+z_{\rm cosmo}^2)$, where $z_{\rm cosmo}$ is the redshift attributable to cosmological expansion obtained by solving $DM = DM_{\rm cosmo}(z_{\rm cosmo})$.

Secondly, we assess the host galaxy extinction ($A_{\lambda_0, {\rm host}}$) by searching for \nai~D features in the $z_{\rm host}$ rest-frame for the 11 SNe with explosion spectra (Table~\ref{tab:spec}).
Resolved doublets seen in two cases of KSNe-2021iq and 202112D were fitted with Voigt doublet profiles to measure their equivalent widths.
We then estimate $E(\bv)_{\rm host}\sim$ 0.021 $\pm$ 0.14 and 0.023 $\pm$ 0.16 mag for them, respectively, by assuming a Milky Way–like correlation between \nai~D equivalent width and dust extinction \citep{Poznanski2012mnras}, the reddening law of \citet{Fitzpatrick1999pasp} with $R_V$ = 3.1, and uncertainty of 68\% for $A_V$ estimated using this method \citep{Phillips2013apj}.
These values are consistent with the template-fitted $E(\bv)_{\rm host}$ of the two SNe (= 0.0 and 0.03, respectively).
Strongly blended doublets were also seen for the three SNe that were coincident with their host galaxy nuclei (KSNe~2017fo, 2019bl, and 2018ng; Table~\ref{tab:hosts}).
However, for these SNe, the \nai~D features are unlikely to be entirely from foreground absorption since the spectral flux is largely dominated by background emission from the host galaxy, and thus, they were not used for $E(\bv)_{\rm host}$ measurements.
All of the \nai~D search results are shown in Table~\ref{tab:spec} (column 9) where we note that most events had non-detections of \nai~D in their spectra, consistent with the small template-fitted values of $A_{\lambda_0, {\rm host}}$ ranging in $\sim$ 0--0.1 mag (Table~\ref{tab:snparam}). 

\begin{figure*}[t!]
\epsscale{\scl}
\begin{center}
\includegraphics[width=0.9\textwidth]{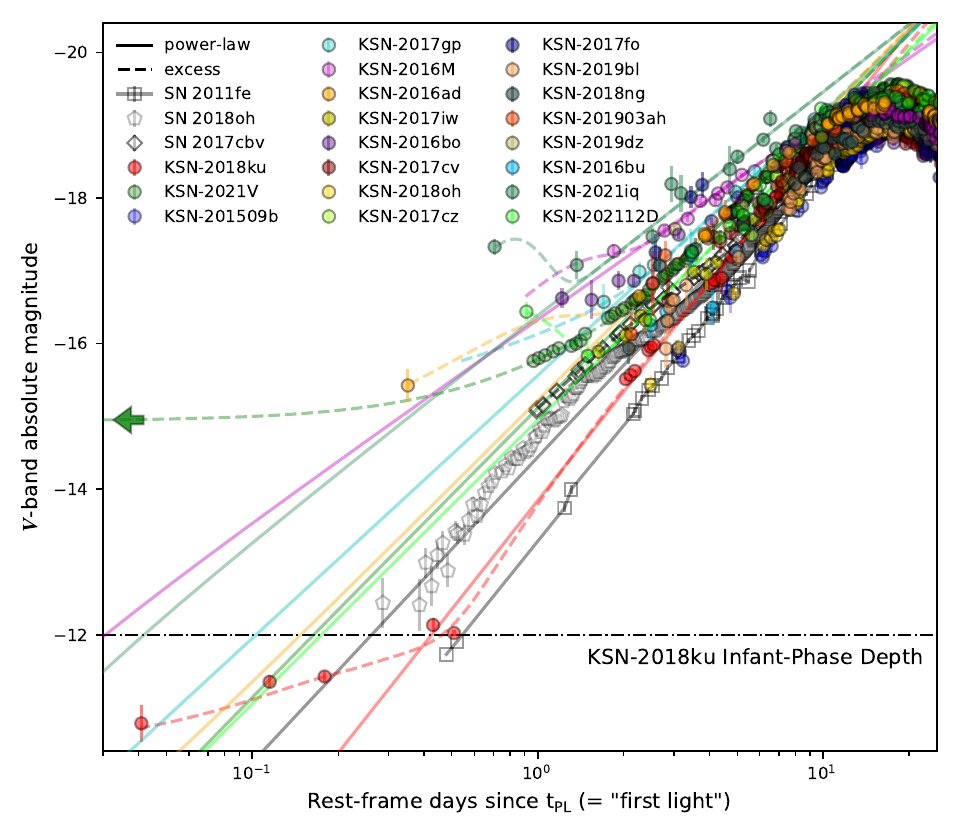}
\end{center}
\caption{Rest-frame $V$-band early light curves of KSP infant/early \tase\ (colored data points) compared to other prototypical normal \tase\ in their nearest-to-$V$ bands, SNe~2011fe \citep[$g/V$][]{Nugent2011nat}, 2018oh \citep[$K2$][]{Dimitriadis2019apj}, and 2017cbv \citep[$V$][]{Hosseinzadeh2017apj}, as well as the following models (Appendix~\ref{sec:reggaus}): (1) pure power-law fits (solid lines) used to estimate $t_{\rm PL}$ (often called the epoch of ``first light''); and (2) power-law $+$ Gaussian fits representing early excess emission (dashed curves). The errorbars represent the 1-$\sigma$ uncertainty level throughout the paper, unless otherwise indicated. The horizontal dot-dashed line delineates the ``infant-phase depth'' $>$ $-$12~mag, where the evolution of \tase\ was revealed by KSN-2018ku \citep[= SN~2018aoz;][]{Ni2022natas, Ni2023apj} but has remained otherwise unexplored to date. The arrow represents a detection from prior to $t_{\rm PL}$, first identified in KSN-2021V \citep[= SN~2021aefx;][]{Ni2023bapj}.}
\label{fig:ksplc}
\end{figure*}

\section{Results}\label{sec:res}

We examine the early multi-band light curves and colors of the full sample of 19 infant/early 
Type Ia SNe from the KSP, including the three previously studied events 
KSNe-201509b \citet{Moon2016spie}, 2018ku \citep[= SN~2018aoz;][]{Ni2022natas, Ni2023apj}, 
and 2021V \citep[= SN~2021aefx;][]{Ni2023bapj} for comparison and population analysis.

\subsection{Early Light Curves and Excess Emission} \label{sec:excess}

Figure~\ref{fig:ksplc} shows the $V$-band light curves (colored circles) of the KSP \tase\ in rest-frame.
Typically, the onset of the light curve in \tase\ is estimated by fitting a power-law model (e.g, colored solid lines) to the rising early part up to $\sim$ 40\% of maximum light \citep[e.g.,][see Appendix~\ref{sec:reggaus} for the details]{Olling2015nat, Moon2021apj, Ni2022natas}.
This onset epoch ($t_{\rm PL}$) has often been referred to as the epoch of
``first light''.
We note, however, that since the power-law rise of \tase\ is thought to be driven by radioactive emission from the centrally-concentrated main distribution of \ni56\ in the ejecta \citep{Piro&Nakar2013apj, Piro&Nakar2014apj}, the precise quantity estimated by $t_{\rm PL}$ is the onset of the central \ni56-driven power-law rise. 
Thus, the epoch of ``first light'' estimated this way typically follows the epoch of explosion by a ``dark phase'' of $\lesssim$ 1 day \citep[see][]{Piro&Nakar2014apj}, depending on the depth of the main \ni56\ distribution and photon diffusion \citep[e.g., 0.4 days in KSN-2018ku and 0.5 days in KSN-2021V;][]{Ni2023bapj, Ni2023apj}.
If another source of emission were present (e.g. an over-density of \ni56\ near the ejecta surface or collision with the companion star), it would, in principle, be possible to observe emission prior to the epoch of ``first light'' \citep[e.g., as found in KSN-2021V;][]{Ni2023bapj}.
Note that Table~\ref{tab:snparam} summarizes the observed and measured properties, including $t_{\rm PL}$, for the 16 previously unpublished \tase.

Of the 19 KSP \tase, 7 are detected within $\lesssim$ 1 day since ``first light'' and the rest within $\lesssim$ 3 days (Table~\ref{tab:snparam}).
As seen in Figure~\ref{fig:ksplc}, the SN light curves undergo similar evolution near peak, which is a known characteristic of \tase\ used to standardize their peak luminosities for distance measurements \citep{Phillips1999aj}.
However, they are apparently much more diverse at earlier times, especially within the so-called ```infant phase'' of $\lesssim$ 1 day. For instance, the observed luminosities during 0.5--0.7 days span at least 5~mag, which is a hundred-fold difference in brightness.

For the 16 previously unpublished events, we examine whether the light curves accommodate ``early excess emission'' sometimes found in \tase, including normal ones \citep[e.g.,][]{Dimitriadis2019apj, Ni2023bapj, Ni2023apj, Wang2024apj}, using the method summarized below (see Appendix~\ref{sec:reggaus} for additional details).
We first iteratively fit power-laws to the early $BVi$-band light curves over time intervals with a range of start times (and fixed end times) to determine if the data points in any early time interval deviate from pure power-law rise---i.e., if the inclusion of the data points worsens the fit quality.
If such a deviation is found, then we examine those data points for consistency with typical early excess emissions that have been observed in \tase\ based on: (1) positive S/N above the underlying power-law rise; (2) reasonable power-law fit parameters when the data points are excluded; and (3) reasonable excess emission properties, as measured by a power-law $+$ Gaussian fit (see the example below) wherein the Gaussian parameters are regularized based on measured ones from a few prototypical cases of \tase\ with very well-sampled early excess emissions.

If no deviation consistent with early excess emission is found, then we consider the light curves as showing pure power-law rise.
Note that for cases accommodating early excess emission, $t_{\rm PL}$ is determined from a power-law fit that excludes the data points containing excess emission.
We caution that early excess emission in \tase\ can be as faint as $-$11$^{\rm th}$ mag (see Figure~\ref{fig:ksplc}), and it is therefore possible that the SNe showing pure power-law rise in our data could reveal excess emission with earlier/deeper observations.

\begin{figure}[t!]
\epsscale{\scl}
\plotone{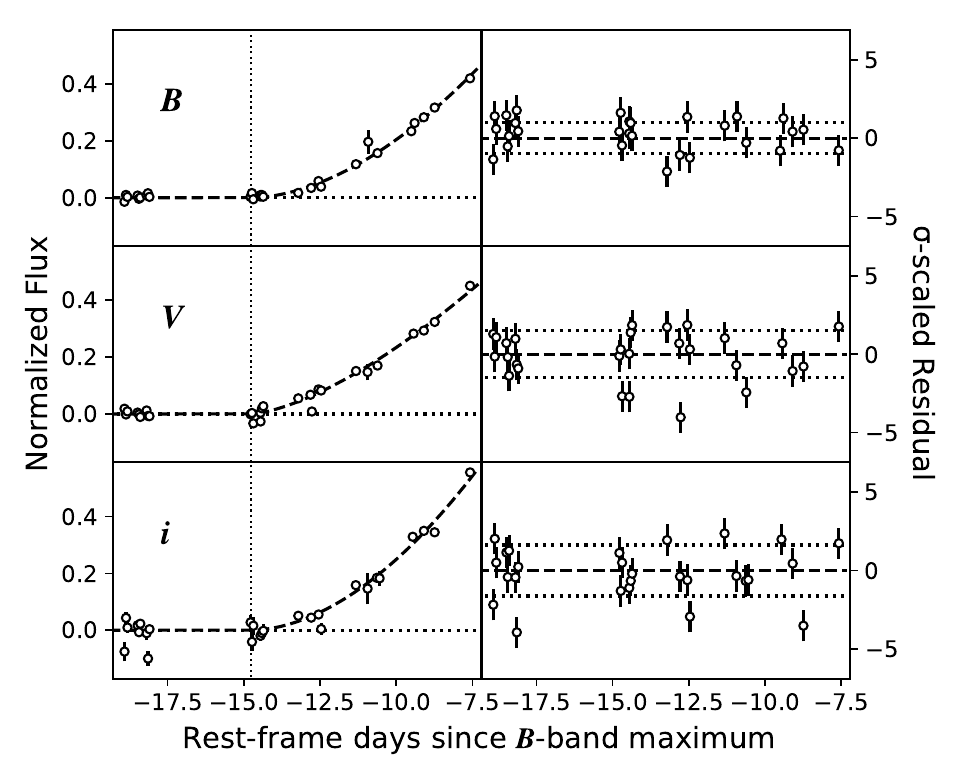}
\caption{(Left) Normalized early rest-frame $BVi$-band (top to bottom) light curves of KSN-2017iw up to $\sim$ 40\% of $B$-band peak flux compared to the best-fit power-law model (dashed curves; Equation~\ref{eq:pow}) with \chisqr\ = 2.2 (Appendix~\ref{sec:reggaus}).
The vertical and horizontal dotted lines represent $t_{\rm PL}$ of the fit (= ``first light'') and zero flux, respectively.
(Right) Fit residuals divided by the photometric errorbars. The horizontal dashed line represents zero residual while the dotted ones represent the 1-$\sigma$ region of the residual data points.
\label{fig:powfit}}
\end{figure}

\begin{figure}[t!]
\epsscale{\scl}
\plotone{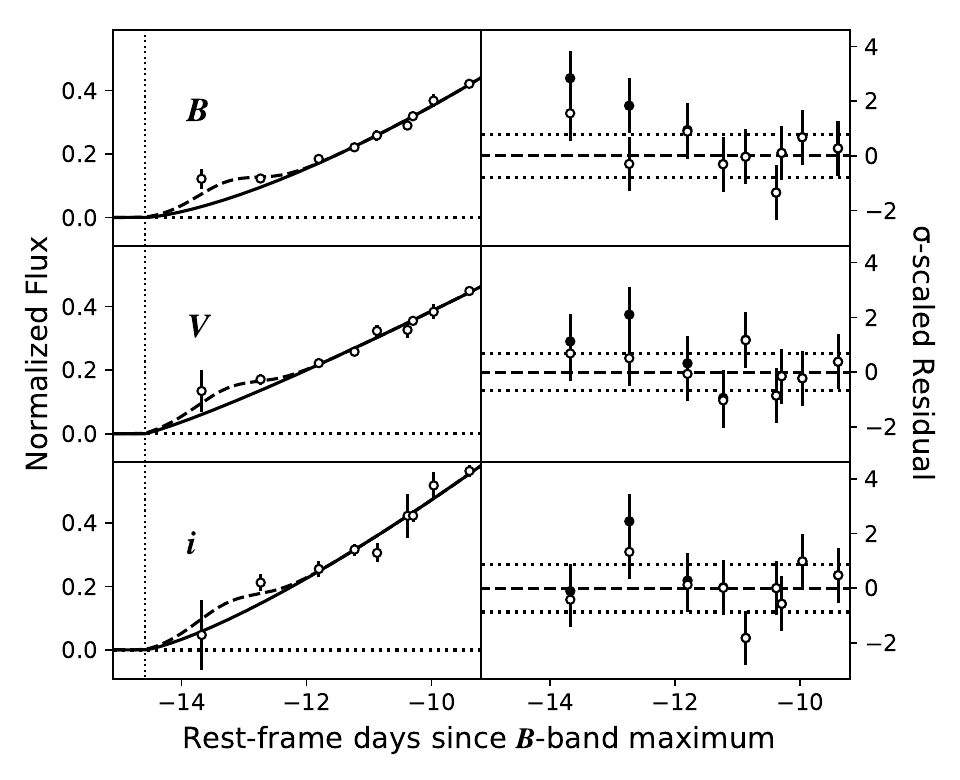}
\caption{Left panels are the same as Figure~\ref{fig:powfit}, but comparing the rest-frame $BVi$-band light curves of KSN-2016M to the best-fit power-law $+$ Gaussian model (dashed curves; Equations~\ref{eq:pow} + \ref{eq:gex}) with \chisqr\ = 1.1 and its power-law component (solid curves). Right panels compare the residuals of the power-law $+$ Gaussian (open circles) to those of its power-law component (filled circles).
\label{fig:reggausfit}}
\end{figure}

Figures~\ref{fig:powfit} and \ref{fig:reggausfit} show the $BVi$-band light curve fits (left panels) and residuals (right panels) for two example cases, KSN-2017iw and KSN-2016M. These are typical examples of pure power-law rise and early excess emission, respectively.
The fit residuals for KSN-2017iw are consistent with noise, indicating that the pure power-law model adequately explains its early light curve.
For KSN-2016M, comparison of the residuals of the power-law $+$ Gaussian (open circles) to those of the power-law component alone (filled circles) reveals excess flux (i.e., positive residual for the power-law component) in the earliest detections, which is better fit by including the Gaussian ``excess emission'' component. 
(Note that a pure power-law fit to the data of KSN-2016M is inconvergent, favouring an onset epoch of $t_{\rm PL} \lesssim -$45 days and power-law indices $\gtrsim$ 10 in all three $BVi$ bands).

Of the full sample of 19 KSP \tase, 12 show pure power-law rise.
The remaining 7, KSNe-2017gp, 2016M, 2016ad, 2021iq, 202112D, 2018ku and 2021V,
accommodate early excess emission, where the early excess emissions in the latter 2 events were previously reported.
The $V$-band light curves of the power-law $+$ Gaussian fits used to characterize the excess emission properties are shown in Figure~\ref{fig:ksplc} with the correspondingly-colored dashed curves, while the power-law fits used to infer their epochs of first light (see above) are shown with solid curves.
($B$- and $i$-band light curve fits are shown in the Appendix; Figures~\ref{fig:Bsplit}--\ref{fig:Isplit}).

Figure~\ref{fig:ksplc} also shows three non-KSP \tase\ in their nearest-to-$V$ bands that provide prototypical examples of \tase\ with pure power-law rise \citep[SN~2011fe;][]{Nugent2011nat}---which appears as a straight line in the log-log plot---and well-sampled early excess emissions \citep[SNe~2017cbv and 2018oh;][]{Hosseinzadeh2017apj, Dimitriadis2019apj}.
Early excess emissions are visible across a large range of luminosities spanning $\sim$ $-$11$^{\rm th}$ to $-$17.5$^{\rm th}$~mag, contributing to the aforementioned diversity of \tas\ light curves in the infant phase.
They fade on a variety of timescales, though most are still visible shortly after infant phase, during 1--2 days after first light, except in the case of KSN-2018ku where the light curve appears consistent with power-law rise after only 0.5 days.

\subsection{Early Multi-Color Evolution} \label{sec:early}

Table~\ref{tab:snparam} lists the first detection of color ($t_{\rm col}$) for the 16 previously unpublished \tase\ in our sample.
Note that the default $BVi$ sequence observational mode of KSP 
can provide almost immediate color coverage from the first detection 
except for the cases where only a single filter has sufficient S/N.
Color detections were obtained within $\lesssim$ 3 days after first light for all but one event (KSN-2017cz), providing an opportunity to examine the multi-color evolution of \tase\ at early phases with a large, uniform sample.

To characterize the early color evolution, we first smoothly interpolate the light curves during the first 10 days since first light using a ``2-D'' Gaussian process model, which simultaneously fits all three $BVi$-band light curves as a function of wavelength and time (see Appendix~\ref{sec:gpint} for the details).
For all 19 KSP \tase, we visually confirm that the model predictions capture the light curve and color evolution in all filters during this phase, while
their uncertainties are typically less than the photometric errorbars of individual data points (see Figures~\ref{fig:Bsplit}--\ref{fig:Isplit}).
For one case of KSN-2018ku, we note that the model predictions during 0.5--2.0 days---where we have an observational data gap (see Figure~\ref{fig:ksplc})---have relatively large uncertainties due to the rapid (0.5-day) timescale of light curve variations just prior to it; however, this is not seen in the other events with slower evolution compared to the size of gaps that can be found in the KSP data (Appendix~\ref{sec:gpint}).
The model-predicted \bv\ and \vi\ colors and their uncertainties contained in the covariance matrix are therefore used as smooth proxies for the intrinsic color evolution of the KSP \tase.

\begin{figure}[t!]
\epsscale{\scl}
\plotone{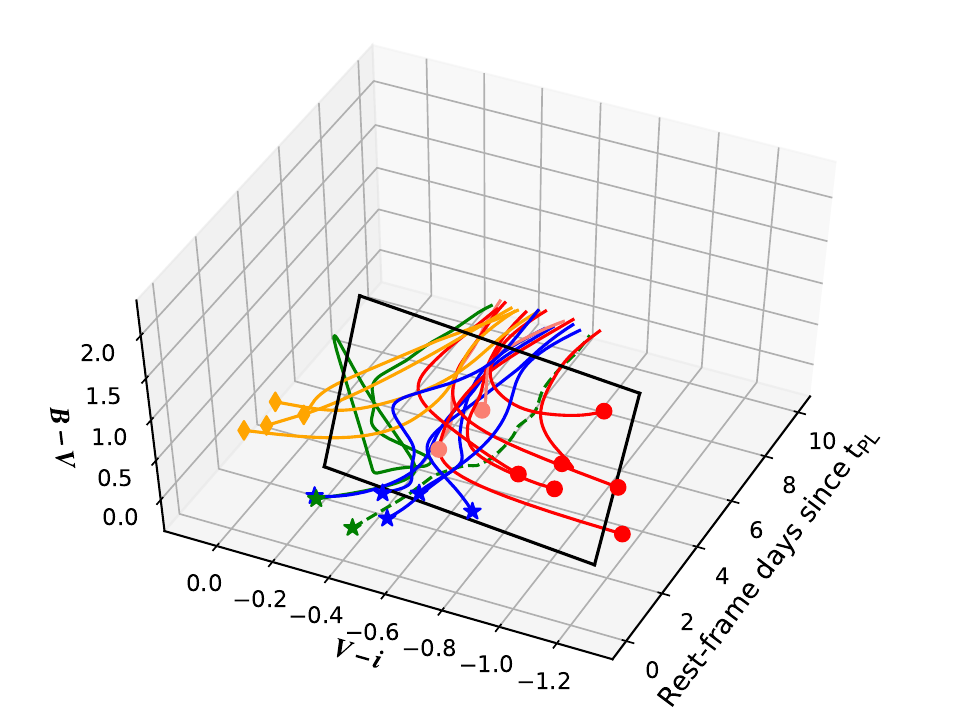}
\caption{\bv\ (z-axis) and \vi\ (y-axis) multi-color evolution of the KSP \tase\ 
along the time since $t_{\rm PL}$ (= ``first light''; x-axis)
represented by smoothed interpolants (colored curves with symbols representing the first color detection), 
showing at least three populations with distinct behaviors---early-red events (red/pink with circles), early-blue events (blue with stars), 
and early-yellow events (gold with diamonds)---as well as SNe~2018aoz (solid green with star) and 2021aefx (dashed green with star).
The events with star symbols accommodate excess emission in their early light curves, while those with circles and diamonds are consistent with pure power-law rise. The black rectangle represents the oblique slice (Equation~\ref{eq:slice}) that separates the early-red and early-blue populations (Section~\ref{subsec:popsep} and Figure~\ref{fig:slice}). (This figure is available as an interactive figure in the online journal where the axes can be rotated).
\label{fig:gp3d}}
\end{figure}

\begin{figure}[t!]
\epsscale{\scl}
\plotone{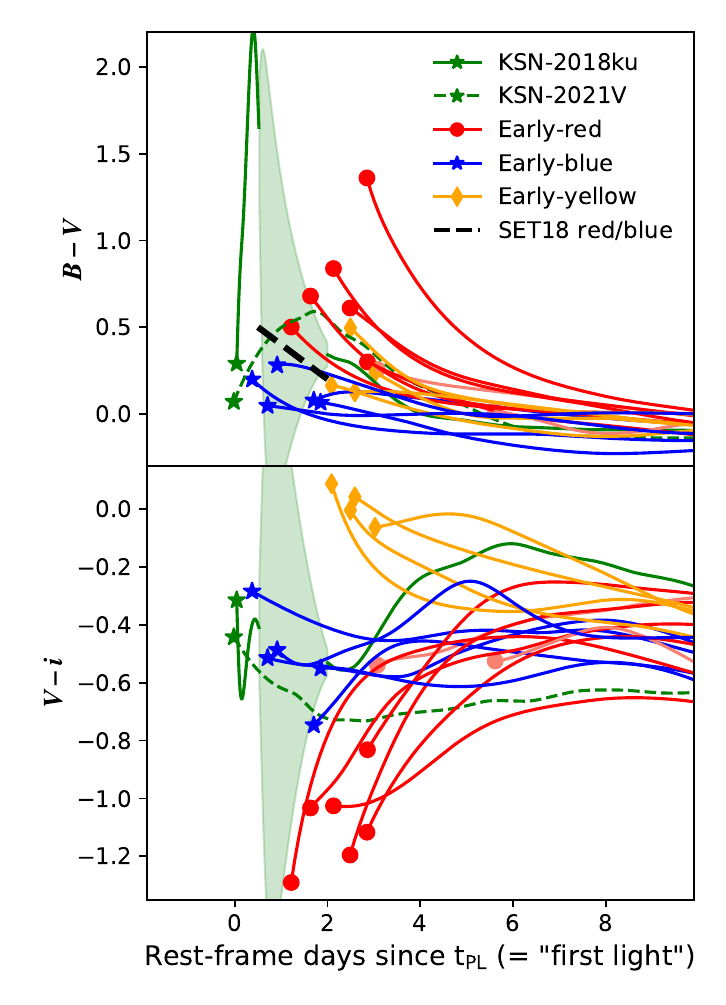}
\caption{Projection of the color evolution in Figure~\ref{fig:gp3d} 
onto \bv\ (top) and \vi\ (bottom) axes.
The black dashed line in the top panel separates red and blue objects of SET18 based on \bv\ color, which approximately correspond to early-red and early-blue in this work.
The green shaded region represents the gap in the observations of KSN-2018ku from $\sim$ 0.5 to 2.0 days, showing the 1-$\sigma$ uncertainties of its interpolated colors which are particularly large due to its rapid color evolution in infancy prior to the gap within $\lesssim$ 0.5 days.
\label{fig:gpcolor}}
\end{figure}

Figures~\ref{fig:gp3d} and \ref{fig:gpcolor} show the multi-color evolution of the 19 KSP \tase\ in color-color space (\bv\ versus \vi) over time (in days since first light) starting from their first detections of color:
Figure~\ref{fig:gp3d} for a 3-D view, and 
Figure~\ref{fig:gpcolor} for 2-D projections of \bv\ and \vi\ versus time.
The multi-color evolutions of the SNe converge to a similar color at \bv\ = $-$0.07 $\pm$ 0.06 mag and \vi\ = $-$0.45 $\pm$ 0.12 mag by 10 days.
However, at earlier phases, the colors apparently diversify sharply, with standard deviations of \bv\ and \vi\ color increasing to 0.26 and 0.28 mag, respectively, at 3 days.

Notably, the color evolutions of the SNe accommodating early excess emission
represented by the curves with star symbols
begin from a similar initial color \bv\ = 0.15 $\pm$ 0.10 mag and \vi\ = $-$0.48 $\pm$ 0.14 mag, which is based on their first color detections within $\lesssim$ 2 days.
The initial \vi\ color is consistent with the color that all of the SNe in our sample converge to at late times ($\sim$ 10 days).
The green solid and dashed curves represent KSNe-2018ku and 2021V, respectively. 
They show clearly different color evolution from the rest of our sample, which may be partially attributable to earlier detection as discussed in Section~\ref{subsec:popspec}.
Except for these two cases, the colors of the SNe that accommodate early excess emission are slower-evolving overall than those without excess emission, showing little change between 3 and 10 days with $\Delta$(\bv) = $-$0.13 mag and $\Delta$(\vi) = $-$0.01 mag, which is comparable to the standard deviations of their initial \bv\ and \vi\ colors: 0.08 and 0.06 mag, respectively.

\subsection{Early Multi-Color Populations} \label{sec:pop}

\subsubsection{Identification of Three \tas\ Sub-Populations}\label{subsec:popid}

We identify three distinct populations of \tase\ in the KSP sample based on similar behaviors in their early multi-color evolution, shown with different colored curves and starting symbols in Figures~\ref{fig:gp3d} and \ref{fig:gpcolor} as follows:
\begin{enumerate}
    \item \textbf{Early-blue population ($\bigstar$ symbol $+$ blue curve):} These events show little \bv\ evolution between 1--10 days since first light, being confined to blue \bv\ colors within $\pm$ 0.3 mag. \vi\ is also approximately stationary in this phase at a blue \vi\ color of $\sim$ $-$0.51 $\pm$ 0.06 mag. As mentioned above (Section~\ref{sec:early}), all of them accommodate early excess emission in their light curves, which provides a likely explanation for their blue color at early phases (see Section~\ref{sec:disc}).
    
    \item \textbf{Early-red population ({\Large $\bullet$} symbol $+$ red/pink curve):} These events show opposing evolution blueward and redward in \bv\ and \vi\ respectively. They appear to start from \bv\ $\gtrsim$ 0.5 mag at 1 day and \vi\ $\lesssim$ $-$1.0 mag at 2 days.
    All of their light curves are consistent with pure power-law rise.
    In addition, we tentatively associate two events detected at later phases ($\gtrsim$ 3 days; pink curves) with this population since they show the opposing evolution in \bv\ and \vi\ and pure power-law rise; however, we note that their first color detections broadly overlap with both the early-red and early-blue populations, and earlier observations may reveal excess emission and/or only weak color evolution more consistent with early-blue events.
    
    \item \textbf{Early-yellow population ($\blacklozenge$ symbol $+$ gold curve):}
    These events are defined by blueward color evolution during 1--10 days in both \bv\ and \vi, where the former is less than what is seen in early-red events but more than in early-blue events.
    Straddling the boundary between early-red and early-blue when viewed in \bv\ color alone, they can appear to be compatible with either population. However, their early \vi\ colors starting from $>$ $-$0.2 mag at 2 days clearly separate them from both early-red and early-blue events.
\end{enumerate}

\subsubsection{Separation of Early-Red and Early-Blue Populations}\label{subsec:popsep}

SET18 previously identified two groups of \tase\ based on early \bv\ color evolution alone: (1) ``red'' objects exhibiting \bv\ $\gtrsim$ 0.2 mag at 0.2 days and as much as $\sim$ 0.5 mag in earlier phases---represented by the black dashed line in the top panel of Figure~\ref{fig:gpcolor}; 
and (2) ``blue'' objects with \bv\ colors within $\sim$ $-$0.2 to 0.05 mag that evolve relatively slowly compared to the red objects.
These groups approximately correspond to the early-red and early-blue populations defined here, though as mentioned above, there is debate over whether they are separate or extreme ends of a single continuous population \citep[e.g., see][]{Han2020apj, Bulla2020apj}.
Part of the difficulty in separating the two groups lies in the presence of the events on the boundary between the two, which, within the KSP sample, appears to comprise largely of early-yellow events with clearly distinct behavior from both groups in \vi\ color (bottom panel).

We examine whether the early-red and early-blue populations excluding the overlapping early-yellows are separable with multi-color information as follows.
First, we determine the ``oblique slice'' in the 3-D space of \bv, \vi, and time (shown in Figure~\ref{fig:gp3d}) that maximizes the separation (defined below) between the two populations. Such an oblique slice can be described by the equation
\begin{equation}
    t = t_0 + \alpha [(\bv) - (\bv)_0] + \beta [(\vi) - (\vi)_0]\ \ {\rm days}
\end{equation}
adopting (\bv)$_0$ = $-$0.07 mag and (\vi)$_0$ = $-$0.45 mag to match the late-time color that the entire sample converges to at 10 days since first light (see Section~\ref{sec:early}).

We fit for ($t_0$, $\alpha$, $\beta$) to maximize the separation between the colors of early-red and early-blue events that intersect with the slice, or ``intersection colors'' denoted (\bv)$_{\rm slice}$ and (\vi)$_{\rm slice}$.
Note that since each SN is discovered at different times after ``first light'', not all of them may intersect with an arbitrary plane through this 3-D space.
Multi-color curves are considered ``nearly intersecting'' the slice if the time delay between the slice and the beginning of the curve is less than the 1-$\sigma$ uncertainty of the estimated epoch of ``first light''; in which case, we include the color from the nearest observed epoch after the slice.
Including later colors for these nearly-intersecting cases conservatively estimates the early-red/blue separation due to the divergence of their colors in earlier epochs as mentioned above (Section~\ref{sec:early}).

When maximizing the separation between the early-red (= $r$) and early-blue (= $b$) events, we define ``separation'' as the $Z$-score (or Mahalanobis distance) between the distributions of their intersection colors in the oblique slice
\begin{equation}
    Z = \sqrt{(\Vec{\mu}_r - \Vec{\mu}_b)^T(\Sigma_r + \Sigma_b)^{-1}(\Vec{\mu}_r - \Vec{\mu}_b)}
\end{equation}
whose means and variances are $\mu_{(r,b)}$ and $\Sigma_{(r,b)}$ estimated by fitting a hierarchical Gaussian class-conditional model to both the observed colors and errors (see Appendix~\ref{sec:hgaus} for the details).
We use the Nelder-Mead algorithm\footnote{\texttt{scipy.optimize.minimize} (\url{https://scipy.org})} to obtain a maximized $Z$-score of 1.46 with the following slice:
\begin{equation}
    t = 1.74 + 1.43[(\bv) + 0.07] + 0.79[(\vi) + 0.45]\ \ {\rm days}
    \label{eq:slice}
\end{equation}
as shown with the black rectangle in Figure~\ref{fig:gp3d}.
This $Z$-score corresponds to an approximate p-value of 13.7\% under Hotelling's $T^2$ test, indicating the distributions of early-red and early-blue colors are marginally separated.
The intersection colors of the 11 events are provided in Table~\ref{tab:slice}.

\begin{deluxetable}{lcc}
\tabletypesize{\footnotesize}
\tablecolumns{9} 
\tablewidth{0.99\textwidth}
 \tablecaption{Intersection colors of early-red and early-blue events with the oblique slice shown in Figure~\ref{fig:gp3d} (black rectangle).}
 \tablehead{
 \colhead{KSN} & \colhead{(\bv)$_{\rm slice}$ [mag]} & \colhead{(\vi)$_{\rm slice}$ [mag]}
 } 
\startdata
\textcolor{blue}{2017gp$^b$} & 0.08 $\pm$ 0.18 & $-$0.74 $\pm$ 0.24 \\
\textcolor{blue}{2016M$^b$} & 0.07 $\pm$ 0.12 & $-$0.55 $\pm$ 0.14 \\
\textcolor{blue}{2016ad$^b$} & $-$0.02 $\pm$ 0.13 & $-$0.40 $\pm$ 0.14 \\
\textcolor{red}{2017iw$^r$} & 0.50 $\pm$ 0.12 & -0.94 $\pm$ 0.17 \\
\textcolor{red}{2016bo$^r$} & 0.31 $\pm$ 0.12 & $-$0.78 $\pm$ 0.17 \\
\textcolor{red}{2017cv$^r$} & 0.30 $\pm$ 0.23 & $-$0.83 $\pm$ 0.38 \\
\textcolor{red}{2018oh$^r$} & 0.61 $\pm$ 0.24 & $-$1.20 $\pm$ 0.38 \\
\textcolor{red}{2019bl$^r$} & 1.18 $\pm$ 0.33 & $-$1.04 $\pm$ 0.64 \\
\textcolor{red}{201903ah$^r$} & 0.72 $\pm$ 0.24 & $-$1.03 $\pm$ 0.33 \\
\textcolor{blue}{2021iq$^b$} & 0.01 $\pm$ 0.13 & $-$0.54 $\pm$ 0.15 \\
\textcolor{blue}{202112D$^b$} & 0.22 $\pm$ 0.26 & $-$0.51 $\pm$ 0.30 
\enddata
\tablecomments{KSP names are colored according to population (Section~\ref{sec:pop}) in the superscript.}
\end{deluxetable} 
\label{tab:slice}

\begin{figure}[t!]
\epsscale{\scl}
\plotone{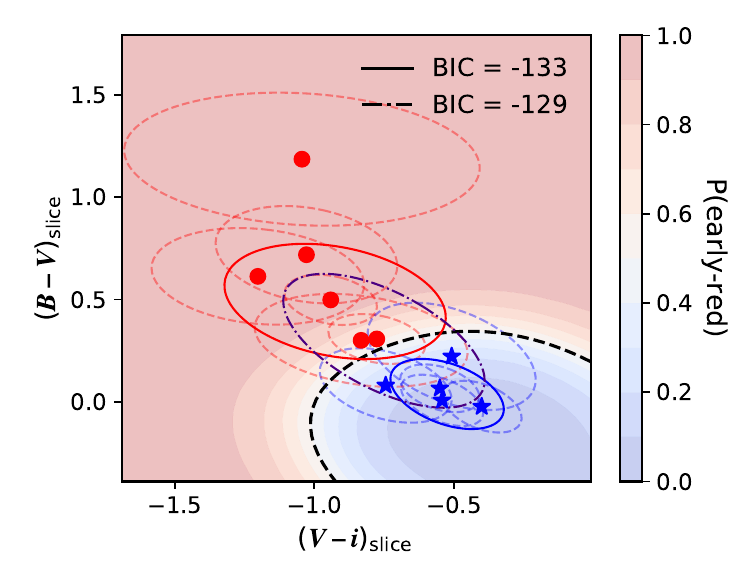}
\caption{Intersections of the color curves of early-red (red filled circles) and early-blue (blue filled stars) events from Figure~\ref{fig:gp3d} with the oblique slice in the 3-D space of \bv, \vi, and time defined by Equation~\ref{eq:slice} (black rectangle in Figure~\ref{fig:gp3d}).
The red/blue dashed covariance ellipses represent the 1-$\sigma$ uncertainties of their colors.
The distributions of the colors of the early-red and early-blue events as two distinct populations (separated by $Z$-score $\sim$ 1.5)
and that obtained for them as one population are compared using solid red/blue (for two populations) and dot-dashed indigo (for one population) 1-$\sigma$ covariance ellipses, respectively.
The color distributions and 1-$\sigma$ covariance ellipses are computed from best-fit hierarchical Gaussian class-conditional models (Appendix~\ref{sec:hgaus}) whose BIC are shown in the legend.
The black dashed curve is the decision boundary between the early-red and early-blue distributions,
while the red/blue-shaded contour regions map the classification probability for early-red (= 1) vs early-blue (= 0) as defined by the colorbar.
\label{fig:slice}}
\end{figure}

Next, we investigate whether the early-red and early-blue events in this slice favour single or separate distributions for them by comparing best-fit hierarchical Gaussian class-conditional models with one and two classes.
Figure~\ref{fig:slice} shows the intersection colors of the slice and the 1-$\sigma$ ellipses of the best-fit two-class (solid red and blue) and one-class (dot-dashed indigo) distributions. 
The early-red and early-blue events are clearly separable in the slice, but also seem compatible with being from either extreme ends of a single continuous distribution or two distinct distributions.
We quantify the model comparison by estimating the Bayesian information criteria:
\begin{equation}
    {\rm BIC} = p\ln{n} - \ln{\hat{\mathcal{L}}}
\end{equation}
where $p$ and $n$ are the number of parameters and observations, respectively, and $\hat{\mathcal{L}}$ is the optimized model likelihood.
The two-component and one-component model have similar BIC of $-$133 and $-$129, respectively, again indicating that the populations are compatible with either model.

We obtain a decision boundary for classifying early-red and early-blue events based on their intersection colors with the slice by drawing 10$^6$ Monte Carlo samples from the best-fit two-class model above to approximate their 2-D probability distribution functions (PDFs) $P(\Vec{x}|r)$ and $P(\Vec{x}|b)$, respectively.
The PDFs are smoothed by fitting a Gaussian kernel density estimator\footnote{\texttt{sklearn.neighbors.KernelDensity} (\url{https://scikit-learn.org})} with bandwidth of 0.1 mag to the samples.
We then calculate the posterior early-red classification probability $P(r | \Vec{x})$ for the observed colors $\Vec{x}$ = (\vi, \bv) using Bayes rule
\begin{equation}
    P(r | \Vec{x}) = \frac{P(\Vec{x}| r) P(r)}{P(\Vec{x}| r) P(r) + P(\Vec{x}| b) P(b)}
\end{equation}
adopting the observed mixture probabilities $P(r)$ = 6/11 and $P(b)$ = 5/11 of the intersecting early-red and early-blue events in the slice, respectively, as priors. 
The decision boundary is the level set $\{\Vec{x}\,|\,P(r | \Vec{x}) = 0.5\}$ shown with the black dashed curve in Figure~\ref{fig:slice}.
As seen in the figure, the boundary successfully separates the two observed populations.

\subsubsection{The Earliest Cases of KSNe-2018ku and 2021V}\label{subsec:popspec}

In addition to the three populations with nearly monotonically evolving colors described above, we highlight KSNe-2018ku and 2021V. These events show initially redward evolution in \bv\ and blueward evolution in \vi, which is not observed in either of the three populations.
However, both SNe become similar (especially KSN-2018ku; see Figure~\ref{fig:gpcolor}) 
to early-red events with the opposite color evolution---blueward and redward in \bv\ and \vi, respectively---starting from $\sim$ 2 days.
In addition, the consistency between the color evolutions of KSN-2018ku and early-red events may start from as early as $\sim$ 0.5 days due to the observational data gap during 0.5--2.0 days where the Gaussian process predictions have large uncertainties (see Section~\ref{sec:early}).
(Note that such data gaps do not produce uncertain Gaussian process predictions for the early-blue, early-red, and early-yellow events since the color evolutions that characterize them occur on longer timescales than the gaps).

The discovery of KSN-2018ku at an earlier phase (1 hour after first light) and fainter luminosity ($-$10.5$^{\rm th}$ mag; Figure~\ref{fig:ksplc}) than any previously observed \tas\ and compatibility of its light curve and color evolution with early-red events in later phases points to the possibility that KSN-2018ku is simply a member of the early-red population that was discovered exceptionally early.
In particular, similar short-lived features as the redward color evolution and excess emission observed in KSN-2018ku before $\sim$ 0.5 days cannot be firmly ruled out in any of the other early-red SNe in the KSP sample (or red events in SET18) since their color detections all come from $\gtrsim$ 1 day. This highlights the limited state of exploration for the infant-phase multi-color evolution of early-red events with pure power-law rise which comprise the majority of early \tase\ (see Section~\ref{sec:zpop}).

On the other hand, the early light curve of KSN-2021V is featured with a bright ($\sim$ $-$15$^{\rm th}$ mag) and long-lasting ($\sim$ 2--3 days) excess emission (see Figure~\ref{fig:ksplc}) that is incompatible with the power-law rise of early-red events, while its red \bv\ color ($\gtrsim$ 0.5 mag) at $\sim$ 1.5 days and blue \vi\ color ($\lesssim$ $-$0.6 mag) starting from $\sim$ 2 days make it incompatible with early-blue and early-yellow ones, respectively (Figure~\ref{fig:gpcolor}).
Thus, the early features of KSN-2021V appear to be unique compared to the other 18 \tase\ in the KSP sample.

\subsection{Near-Peak Light Curve Properties} \label{sec:peak}

\subsubsection{Light Curve Evolution After the Early Phase} \label{subsec:nonearly}

The KSP \tas\ light curves all show a gradual ascent to peak brightness over $\sim$ 2 weeks, increasing by $>$ 2 mag and up to 8.8 mag in brightness from their first detections.
The peak is followed by a slow decline, initially by 0.8--1.6 mag over the first 15 days post-peak in $B$ band and then further decelerating to a much slower decline rate, consistent with \tase\ powered by \ni56\ and \co56\ radioactive decay \citep[e.g., see][]{Hamuy1996aj}.
The $i$-band light curves reach a primary peak before the $B$ and $V$ bands, followed by a secondary $i$-band peak associated with the recombination of iron group elements in the ejecta \citep{Kasen2006apj} $\sim$ 20--35 days later, which is typically seen in normal \tase\ \citep[e.g., see][]{Burns2014apj}.
In the absence of explosion spectroscopy (see Section~\ref{sec:spec}), these light curve features provide photometric confirmation of \tas\ classification for 6 cases of KSNe-2016bo, 2017cv, 2017cz, 201903ah, and 2016bu, as well as 201509b \citep[see][]{Moon2021apj}.

By fitting the rest-frame light curves from $\sim$ $-$10 to 30 days since peak with a high-order polynomial function, we measure the peak epochs $t_{{\rm max}, (B,V,i)}$, peak absolute magnitudes $M_{B,V,i}$(peak), and $B$-band decline rate over the first 15 days post-peak \dm15\ \citep[or ``Phillips parameter'';][]{Phillips1999aj}. See Table~\ref{tab:snparam} for the measured parameters of the 16 previously unpublished KSP \tase, and references for those of KSNe-201509b, 2018ku, and 2021V \citep[][]{Moon2021apj, Ni2022natas, Ni2023bapj}.
The total \dm15\ range of 0.8--1.6 mag for the sample falls within the normal range of decline rates for \tase\ on the Phillips relation \citep[0.8--1.7 mag;][]{Prieto2006apj}. 
The rise times from first light to $B$-band maximum range in 13--18 days, consistent with ranges seen in larger samples of \tase\ with similar \dm15---e.g., 12--19 days for the ZTF sample with \dm15\ in 0.8--1.5 mag \citep{Miller2020apj}.

\begin{deluxetable*}{|l|c|c|c|c|c|c|c|c|c|c|c|c|c|c|c|c|}[ht!]
\tabletypesize{\footnotesize}
\tablecolumns{17} 
\tablewidth{0pt}
 \tablecaption{KSP \tase\ basic properties.}
 \rotate
 \tablehead{
 \colhead{KSN} & \colhead{$z_{\rm host}$} & \colhead{$DM$} & \colhead{($DM_{\rm cosmo}$)} & \colhead{$A_{\lambda, {\rm MW}}$} & \colhead{$A_{\lambda, {\rm host}}$} & \colhead{$t_{{\rm max}, B}$} & \colhead{$t_{{\rm max}, V}$} & \colhead{$t_{{\rm max}, i}$} & \colhead{$M_{{\rm max}, B}$} & \colhead{$M_{{\rm max}, V}$} & \colhead{$M_{{\rm max}, i}$} & \colhead{$\Delta M_{15, B}$} & \colhead{$t_{\rm PL}$} & \colhead{$\alpha_{B,V,i}$}  & \colhead{$t_{\rm det} - t_{\rm PL}$} & \colhead{$t_{\rm col} - t_{\rm PL}$}\\
 \colhead{} & \colhead{} & \colhead{[mag]} & \colhead{[mag]} & \colhead{[mag]} & \colhead{[mag]} & \colhead{[MJD]} & \colhead{[MJD]} & \colhead{[MJD]} & \colhead{[mag]} & \colhead{[mag]} & \colhead{[mag]} & \colhead{[mag]} & \colhead{[MJD]} & \colhead{} & \colhead{[days]} & \colhead{[days]}
 }
\startdata
\textcolor{blue}{2017gp$^b$} & 0.1364 & 39.01 & (38.94 & 0.042 & 0.000 & 58060.33 & 58061.61 & 58057.40 & -19.44 & -19.41 & -18.50 & 0.84 & 58041.93 & 1.72, 1.44, 1.59 & 0.54 & 1.70 \\
(Excess) & $\pm$0.0002 & $\pm$0.08 & $\pm$0.15) &  &  & $\pm$0.28 & $\pm$0.31 & $\pm$0.80 & $\pm$0.02 & $\pm$0.02 & $\pm$0.03 & $\pm$0.04 & $\pm$1.31 &  &  &  \\
\hline
\textcolor{blue}{2016M$^b$} & 0.0752 & 37.91 & (37.56 & 0.089 & 0.000 & 57698.38 & 57700.44 & 57696.20 & -19.40 & -19.19 & -18.45 & 1.22 & 57682.70 & 1.52, 1.10, 1.64 & 0.90 & 1.85 \\
(Excess) & $\pm$0.0002 & $\pm$0.08 & $\pm$0.15) &  &  & $\pm$0.19 & $\pm$0.32 & $\pm$0.58 & $\pm$0.02 & $\pm$0.02 & $\pm$0.04 & $\pm$0.02 & $\pm$3.61 &  &  &  \\
\hline
\textcolor{blue}{2016ad$^b$} & 0.0613 & 37.09 & (37.09 & 0.086 & 0.000 & 57714.06 & 57714.20 & 57710.36 & -19.48 & -19.33 & -18.57 & 0.88 & 57696.27 & 1.59, 1.57, 1.56 & 0.35 & 0.37 \\
(Excess) & $\pm$0.0002 & $\pm$0.08 & $\pm$0.15) &  &  & $\pm$0.17 & $\pm$0.19 & $\pm$0.41 & $\pm$0.01 & $\pm$0.01 & $\pm$0.02 & $\pm$0.03 & $\pm$0.29 &  &  &  \\
\hline
\textcolor{red}{2017iw$^r$} & 0.0617 & 37.48 & (37.11 & 0.130 & 0.148 & 57827.11 & 57829.09 & 57825.66 & -19.06 & -19.03 & -18.40 & 1.44 & 57811.35 & 1.83, 1.53, 1.89 & 1.63 & 1.63 \\
 & $\pm$0.0008 & $\pm$0.08 & $\pm$0.15) &  &  & $\pm$0.13 & $\pm$0.20 & $\pm$0.22 & $\pm$0.01 & $\pm$0.01 & $\pm$0.02 & $\pm$0.03 & $\pm$0.21 &  &  &  \\
\hline
\textcolor{red}{2016bo$^r$} & 0.0563 & 37.23 & (36.90 & 0.296 & 0.000 & 57720.02 & 57721.63 & 57717.95 & -19.06 & -19.00 & -18.44 & 1.41 & 57706.02 & 1.34, 1.03, 1.36 & 1.22 & 1.22 \\
 & $\pm$0.0001 & $\pm$0.08 & $\pm$0.15) &  &  & $\pm$0.20 & $\pm$0.32 & $\pm$0.24 & $\pm$0.02 & $\pm$0.02 & $\pm$0.04 & $\pm$0.03 & $\pm$0.36 &  &  &  \\
\hline
\textcolor{red}{2017cv$^r$} & 0.0866 & 38.17 & (37.86 & 0.057 & 0.094 & 57962.44 & 57964.33 & 57960.29 & -19.11 & -19.02 & -18.52 & 1.43 & 57946.63 & 1.96, 1.47, 2.21 & 2.86 & 2.86 \\
 & $\pm$0.0001 & $\pm$0.08 & $\pm$0.15) &  &  & $\pm$0.08 & $\pm$0.14 & $\pm$0.13 & $\pm$0.01 & $\pm$0.01 & $\pm$0.01 & $\pm$0.02 & $\pm$1.12 &  &  &  \\
\hline
\textcolor{red}{2018oh$^r$} & 0.0566 & 37.36 & (36.96 & 0.286 & 0.000 & 58198.38 & 58200.54 & 58196.22 & -18.97 & -18.99 & -18.37 & 1.48 & 58181.51 & 2.41, 2.07, 2.37 & 2.48 & 2.49 \\
 & $\pm$0.0001 & $\pm$0.08 & $\pm$0.15) &  &  & $\pm$0.09 & $\pm$0.13 & $\pm$0.12 & $\pm$0.01 & $\pm$0.01 & $\pm$0.01 & $\pm$0.02 & $\pm$0.38 &  &  &  \\
\hline
\textcolor{pink}{2017cz$^{(r)}$} & 0.1080 & 38.67 & (38.39 & 0.031 & 0.000 & 57947.27 & 57947.87 & 57941.83 & -19.56 & -19.42 & -18.47 & 1.06 & 57927.17 & 2.98, 2.45, 2.66 & 2.38 & 5.62 \\
 & $\pm$0.0010 & $\pm$0.08 & $\pm$0.15) &  &  & $\pm$0.13 & $\pm$0.16 & $\pm$0.32 & $\pm$0.01 & $\pm$0.01 & $\pm$0.02 & $\pm$0.02 & $\pm$1.31 &  &  &  \\
\hline
\textcolor{gold}{2017fo$^y$} & 0.1670 & 39.29 & (39.42 & 0.034 & 0.306 & 58071.25 & 58071.07 & 58066.98 & -19.45 & -19.26 & -18.63 & 1.36 & 58053.03 & 1.32, 1.31, 1.04 & 2.59 & 2.59 \\
 & $\pm$0.0010 & $\pm$0.09 & $\pm$0.14) &  &  & $\pm$0.51 & $\pm$0.42 & $\pm$8.10 & $\pm$0.05 & $\pm$0.02 & $\pm$0.03 & $\pm$0.18 & $\pm$0.70 &  &  &  \\
\hline
\textcolor{red}{2019bl$^r$} & 0.0637 & 37.65 & (37.24 & 0.160 & 0.000 & 58562.47 & 58564.12 & 58560.04 & -19.16 & -19.06 & -18.30 & 1.50 & 58545.07 & 2.92, 1.69, 2.49 & 2.81 & 2.85 \\
 & $\pm$0.0001 & $\pm$0.08 & $\pm$0.14) &  &  & $\pm$0.11 & $\pm$0.19 & $\pm$0.18 & $\pm$0.01 & $\pm$0.01 & $\pm$0.02 & $\pm$0.03 & $\pm$0.33 &  &  &  \\
\hline
\textcolor{gold}{2018ng$^y$} & 0.0748 & 37.83 & (37.60 & 0.160 & 0.000 & 58257.72 & 58260.73 & 58254.99 & -19.05 & -19.07 & -18.34 & 1.39 & 58242.18 & 1.88, 1.68, 1.22 & 2.08 & 2.09 \\
 & $\pm$0.0002 & $\pm$0.08 & $\pm$0.15) &  &  & $\pm$0.17 & $\pm$0.31 & $\pm$0.38 & $\pm$0.02 & $\pm$0.02 & $\pm$0.02 & $\pm$0.04 & $\pm$1.17 &  &  &  \\
\hline
\textcolor{red}{201903ah$^r$} & 0.1324 & 38.86 & (38.90 & 0.183 & 0.000 & 58574.11 & 58575.74 & 58569.97 & -19.42 & -19.41 & -18.38 & 0.91 & 58554.74 & 1.83, 1.49, 1.79 & 2.13 & 2.13 \\
 & $\pm$0.0030 & $\pm$0.07 & $\pm$0.16) &  &  & $\pm$0.19 & $\pm$0.30 & $\pm$0.47 & $\pm$0.01 & $\pm$0.01 & $\pm$0.03 & $\pm$0.03 & $\pm$1.17 &  &  &  \\
\hline
\textcolor{gold}{2019dz$^y$} & 0.1400 & 39.19 & (39.03 & 0.210 & 0.078 & 58552.79 & 58553.84 & 58546.81 & -19.35 & -19.28 & -18.62 & 1.11 & 58535.79 & 1.63, 1.20, 1.18 & 1.28 & 3.02 \\
 & $\pm$0.0040 & $\pm$0.08 & $\pm$0.17) &  &  & $\pm$0.27 & $\pm$0.24 & $\pm$0.54 & $\pm$0.02 & $\pm$0.02 & $\pm$0.02 & $\pm$0.04 & $\pm$1.05 &  &  &  \\
\hline
\textcolor{gold}{2016bu$^y$} & 0.1150 & 38.65 & (38.53 & 0.054 & 0.000 & 57703.21 & 57706.57 & 57699.38 & -19.29 & -19.32 & -18.44 & 1.15 & 57684.97 & 2.24, 1.83, 1.51 & 1.91 & 2.49 \\
 & $\pm$0.0020 & $\pm$0.08 & $\pm$0.15) &  &  & $\pm$0.24 & $\pm$0.15 & $\pm$0.24 & $\pm$0.02 & $\pm$0.02 & $\pm$0.02 & $\pm$0.02 & $\pm$0.35 &  &  &  \\
\hline
\textcolor{blue}{2021iq$^b$} & 0.0927 & 38.45 & (38.08 & 0.311 & 0.056 & 59313.63 & 59314.74 & 59309.71 & -19.44 & -19.35 & -18.53 & 0.94 & 59297.34 & 1.19, 1.23, 0.89 & 0.71 & 0.71 \\
(Excess) & $\pm$0.0003 & $\pm$0.08 & $\pm$0.15) &  &  & $\pm$0.21 & $\pm$0.30 & $\pm$0.30 & $\pm$0.01 & $\pm$0.02 & $\pm$0.02 & $\pm$0.03 & $\pm$1.22 &  &  &  \\
\hline
\textcolor{blue}{202112D$^b$} & 0.0531 & 37.26 & (36.73 & 0.070 & 0.078 & 59585.54 & 59586.22 & 59581.74 & -19.48 & -19.40 & -18.73 & 1.13 & 59567.54 & 1.85, 1.56, 1.75 & 0.91 & 0.91 \\
(Excess) & $\pm$0.0003 & $\pm$0.09 & $\pm$0.15) &  &  & $\pm$0.15 & $\pm$0.23 & $\pm$0.35 & $\pm$0.02 & $\pm$0.02 & $\pm$0.02 & $\pm$0.02 & $\pm$0.49 &  &  &  \\
\enddata
\tablecomments{Columns include: (1) KSP name colored by population in the superscript (Section~\ref{subsec:popid}) and whether the early light curves accommodate excess emission (Section~\ref{sec:excess}); (2) redshift ($z_{\rm host}$; Section~\ref{sec:host}); (3--4) distance modulus from \tas\ template fitting and Hubble flow ($DM$ and $DM_{\rm cosmo}$; Section~\ref{sec:restlc}); (5--6) Milky Way and host galaxy extinction ($A_{\lambda, {\rm MW}}$ and $A_{\lambda, {\rm host}}$; Section~\ref{sec:restlc}); (7--13) epochs of $B$-, $V$-, and $i$-band maximum ($t_{{\rm max}, B, V, i}$), peak $B$-, $V$-, and $i$-band absolute magnitude ($M_{{\rm max}, B, V, i}$) and Phillips parameter ($\Delta M_{15, B}$; Section~\ref{subsec:nonearly}); (14--15) onset of power-law rise ($t_{\rm PL}$, or epoch of ``first light''; Section~\ref{sec:excess}) and power-law indices ($\alpha_{B,V,i}$; Appendix~\ref{sec:reggaus}); and (16--17) Detection ($t_{\rm det}$; Table~\ref{tab:survey}) and first color detection ($t_{\rm col}$; Section~\ref{sec:early}) since first light in rest-frame days.}
\end{deluxetable*} 
\label{tab:snparam}

\subsubsection{Comparison with the Phillips Relation} \label{subsec:philcomp}

\begin{figure}[t!]
\epsscale{\scl}
\plotone{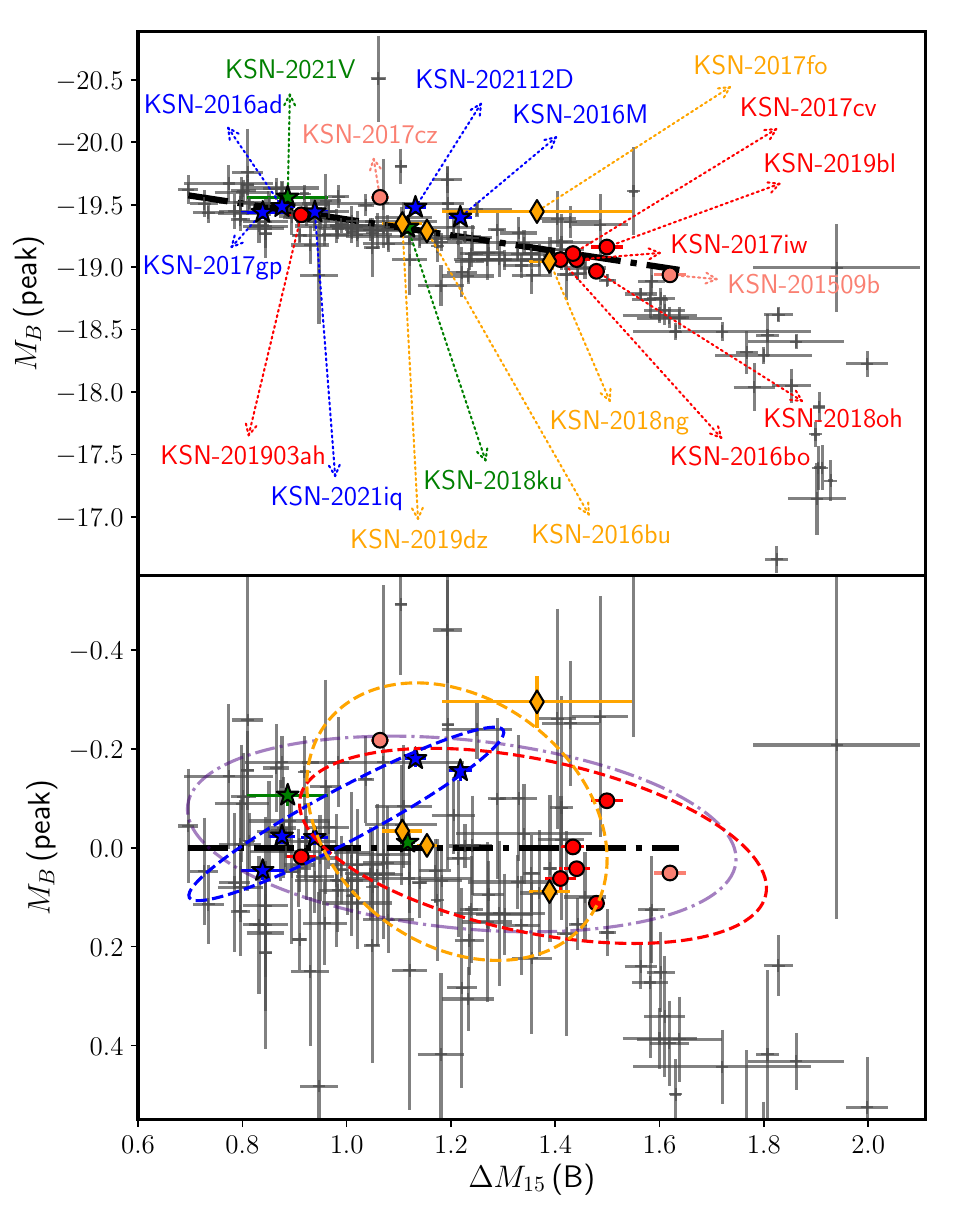}
\caption{(Top) Phillips diagram of $M_B$\,(peak) and \dm15\ of the KSP \tase\ (colored symbols; same as Figure~\ref{fig:gp3d}), as annotated with arrows, compared to the Phillips relation \citep[black dashed line;][]{Prieto2006apj} and other \tase\ \citep[grey crosses;][]{Burns2018apj}. 
(Bottom) Residuals of $M_B$\,(peak) and \dm15\ of \tase\ from the top panel 
after subtracting the Phillips relation.
The three dashed covariance ellipses represent the 2-$\sigma$ ranges,
which are obtained with the best-fit hierarchical Gaussian class-conditional model,
for the early-blue, early-red, and early-yellow populations shown with blue, red, and gold colors,
respectively.
The transparent, indigo, dot-dashed covariance ellipse represents the combined distribution for early-red and early-blue events.
\label{fig:phillips}}
\end{figure}

Figure~\ref{fig:phillips} (top panel) compares $M_B$\,(peak) and \dm15\ of the KSP \tase\ (color-coded by their multi-color populations from Section~\ref{subsec:popid}) to a much larger sample of \tase\ from \citet{Burns2018apj}. Also shown is the Phillips relation, which is used to calibrate \tas\ luminosities for distance measurements \citep[the relation shown is from][]{Prieto2006apj}.
The early-blue events appear tightly clustered around $M_B$\,(peak) = $-$19.45 $\pm$ 0.03 mag, which is on the brighter extreme of peak absolute magnitudes for normal \tase, and confined to the slower-declining end of the Phillips relation with \dm15\ $\lesssim$ 1.2.
On the other hand, the early-red events are found with a large range of brightnesses and decline rates across the Phillips relation. They have \dm15\ values ranging from 0.9 to 1.6 mag, which is nearly the full range of the KSP sample.
This is consistent with the near-peak light curve properties of the red and blue groups identified by SET18, where blue events mainly come from slower-declining normal or 91T/99aa-like \tase\
while red events span the full range of \dm15 for normal \tase.
No clear distinction is seen between the distributions of properties for the early-red and early-yellow events, though we note that the latter appear more tightly clustered around $M_B$\,(peak) $\sim$ $-$19.3 mag and \dm15\ $\sim$ 1.2 mag, which may be at least partially attributable to the smaller number of early-yellow samples.

We further examine the distributions of the near-peak properties for each population and compare them to the Phillips relation by fitting the ``Phillips residuals'' and \dm15\ of the populations with a hierarchical Gaussian class-conditional model. (Note, this is the same type of model that is used to fit the distributions of population colors in Section~\ref{subsec:popsep}). The ``Phillips residuals'' are $M_B$\,(peak) after subtracting the Phillips relation (black dashed line in Figure~\ref{fig:phillips}) to eliminate the known correlation of $M_B$\,(peak) and \dm15\ in \tase\ \citep{Prieto2006apj}.
The model has three classes ($b, r, y$) representing the three populations and is fit to both the data and uncertainties.
The mean ($\Vec{\mu}$), variance ($\Vec{\sigma}^2$), and correlation ($\rho$) of the Phillips residuals and \dm15\ for the three populations are estimated using maximum likelihood.

The bottom panel of Figure~\ref{fig:phillips} compares the Phillips residuals of the populations to correspondingly-colored dashed 2-$\sigma$ ellipses of their distributions estimated from the fit.
Both the early-red and early-yellow distributions seem consistent with the Phillips relation, though the latter is more clustered in \dm15\ as mentioned above.
On the other hand, the Phillips residuals of early-blue events appear to be highly correlated---with best-fit $\rho$ = $-$0.94, versus $\rho$ = 0.44 and 0.27 for early-red and early-yellow events, respectively---which suggests at that they follow a somewhat different $M_B$\,(peak) and \dm15\ relationship than the Phillips relation.
We caution, however, that a high correlation coefficient can also arise by random chance due to the small sample sizes.
For instance, in 10$^6$ Monte Carlo data sets, each with three populations of 5, 8, and 4 samples drawn from a bivariate normal distribution with independent components, $|\rho| >$ 0.94 arises in at least one of the populations 7.7\% of the time.

For the early-blue events, we also estimate a weak correlation of $\rho$ = 0.37 between $M_B$\,(peak) and \dm15, which suggests that they have nearly constant brightness regardless of \dm15, consistent with their tight clustering around $M_{B}$\,(peak) = $-$19.45 $\pm$ 0.03 mag (see above).
However, we note that the standardization of rest-frame light curves for the \tase\ in our sample relies on the assumption that their observer-frame light curves near peak are compatible with a normal \tas\ template under certain transformations---including stretch and S/K-corrections (Appendix~\ref{sec:standard}) as well as reddening and $DM$ that are independently validated with spectroscopic data for all of the early-blue events\footnote{In particular, the Hubble flow and template-fitted $DM$ for all early-blue events agree to within 3-$\sigma$ of their uncertainties, while spectroscopic \nai~D measurements of host galaxy extinction available for two of the early-blue events similarly agree with their template-fitted $E(\bv)_{\rm host}$ (see Table~\ref{tab:snparam}).} (Section~\ref{sec:restlc})---which is also commonly assumed in the standardization of normal \tas\ light curves for cosmological studies \citep[e.g., see][]{Riess2016apj, Burns2018apj}.
Consequently, SNe that match this assumption will necessarily follow the normal \tas\ Phillips relation after standardization, whereas differences between the intrinsic light curve evolution of SNe and the template can lead to deviations from the Phillips relation even when their intrinsic $M_B$\,(peak) and \dm15\ are consistent with the relation.

For early-red and early-yellow events, consistency between the distributions of their Phillips residuals and \dm15\ values (as shown in the bottom panel of Figure~\ref{fig:phillips}) is confirmed by a permutation test as follows.
We first calculate the distance between two populations as the difference between their means---obtained by an errorbar-weighted average of the Phillips residuals and \dm15---after scaling each dimension by the standard deviation of the mean. Note that this assumes the means of the Phillips residuals and \dm15\ are weakly correlated (as determined above).
Next, we consider the null hypothesis that early-red and early-yellow events follow the same distribution of Phillips residuals and \dm15\ and the class labels of their data points in the figure are randomly assigned.
The distribution of possible distances obtainable under the null hypothesis is obtained from 10$^5$ random permutations of the class assignments of early-red and early-yellow events.
Finally, we reject the null hypothesis if the observed distance is ``extreme''---i.e., unlikely under the null hypothesis---compared to the median of the distribution.
We find that the distance is increased 66\% of the time under random permutations of early-red and early-yellow events, indicating that the observed difference between their Phillips residuals and \dm15\ values is near the median of what can be expected under the null hypothesis. This supports that their distributions of Phillips residuals and \dm15\ are indistinct.
On the other hand, the distance is only increased 9\% of the time for permutations of early-red and early-blue events, indicating a lower likelihood for their observed parameters to be from the same distribution.

\subsubsection{Investigation of Continuous Relationships between Peak Magnitudes and Early Colors} \label{subsec:philcont}

\begin{figure}[t!]
\epsscale{\scl}
\plotone{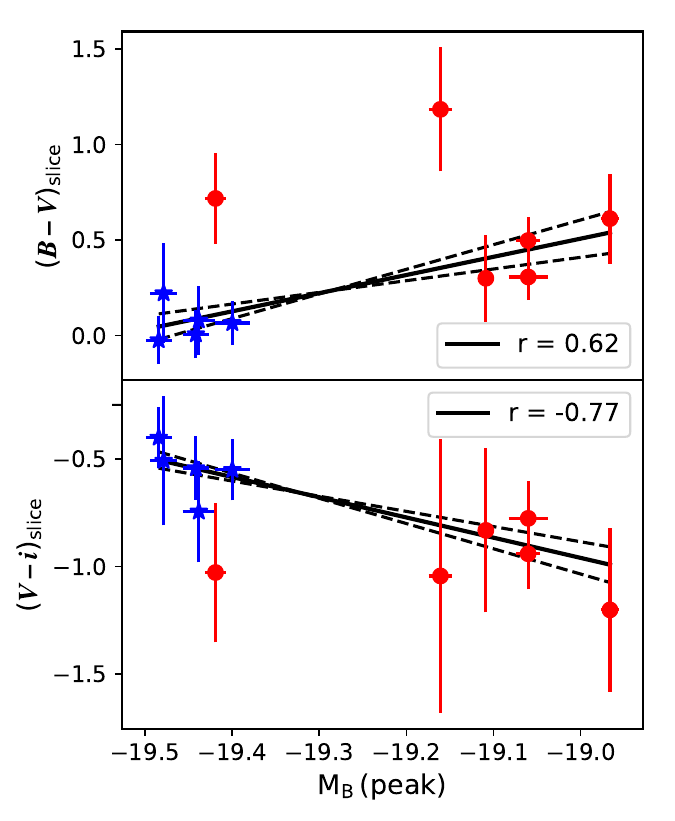}
\caption{Intersection colors of early-red (red circles) and early-blue (blue stars) events with the oblique slice (Figure~\ref{fig:slice}, black rectangle) that maximizes their separation, $(\bv)_{\rm slice}$ (top) and $(\vi)_{\rm slice}$ (bottom) from Table~\ref{tab:slice}, compared to $M_B$\,(peak) of the events. Black solid lines with dashed 1-$\sigma$ error regions are lines of best fit to the data with Pearson correlation coefficients ($r$) in the legend.
\label{fig:slicemag}}
\end{figure}

Figure~\ref{fig:slicemag} compares $M_B$\,(peak) of the early-red and early-blue events to their intersection colors (Table~\ref{tab:slice}) in the slice that maximizes their separation.
As mentioned in Section~\ref{subsec:popsep}, their intersection colors show consistency with being from either two distinct populations or a single continuous population.
In the latter case, $M_B$\,(peak) of the population appears to be correlated and anti-correlated with (\bv)$_{\rm slice}$ and (\vi)$_{\rm slice}$, respectively.
Applying orthogonal distance regression\footnote{\texttt{scipy.odr}} to $M_B$\,(peak) and the colors as well as their uncertainties, we obtain the following lines of best-fit
\begin{align}
    (\bv)_{\rm slice} &= (0.95 \pm 0.34)\times M_B + (18.6 \pm 6.6)\ {\rm mag}\\
    (\vi)_{\rm slice} &= -(0.94 \pm 0.23)\times M_B - (18.8 \pm 4.5)\ {\rm mag}
\end{align}
with Pearson correlation coefficients $r$ = 0.62 and $-$0.77, respectively.
We estimate p-values of 4.2\% and 0.6\% for $|r| >$ 0.62 and 0.77, respectively, under a Student's $t$-test of the null hypothesis (i.e., no relationship).
This suggests, in the case where early-red and early-blue events form a single continuous population, that the peak brightnesses of \tase\ from the population are likely related to the same physical process that determines the colors in infancy.

In this case, the Phillips residuals and \dm15\ of the combined early-red and early-blue population are only weakly correlated, with $\rho$ = 0.25. This indicates that the population $M_B$\,(peak) and \dm15\ are consistent with the Phillips relation overall, as shown in Figure~\ref{fig:phillips} (transparent, indigo, dot-dashed 2-$\sigma$ covariance ellipse). 
However, we emphasize that even in this case, the early-blue and early-red events comprising the extreme ends of the continuous population still show notably distinct properties in our sample, including early light curve features (Section~\ref{sec:excess}) and color evolution (Section~\ref{sec:pop}), as well as near-peak light curve parameters after standardization as found in Section~\ref{subsec:philcomp} above.

\begin{figure*}[t!]
\epsscale{\scl}
\begin{center}
\includegraphics[width=1.0\textwidth]{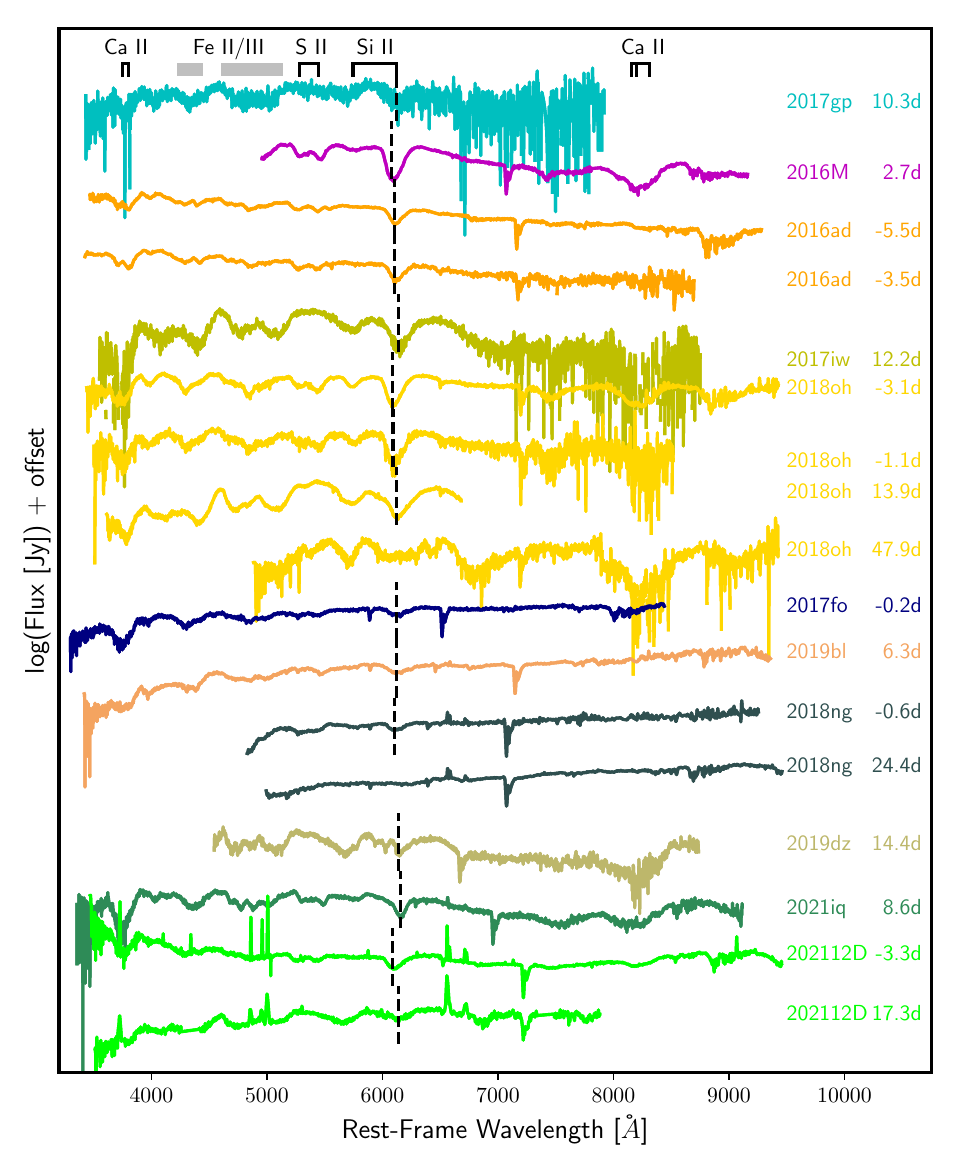}
\end{center}
\vspace{-6mm}
\caption{Explosion spectra of KSP \tase\ (same color as each SN from Figure~\ref{fig:ksplc}) from epochs on the right in rest-frame days since $B$-band maximum. The locations of prominent spectral features are marked at the top of the panel. The vertical dashed lines show the minima of the Si~II~$\lambda$6355~\AA\ absorption feature when it is detected (Table~\ref{tab:spec}, column 10). Note that two events, KSNe-2018ng and 201212D, show narrow emission lines due to blending with the host galaxy (see Tables~\ref{tab:spec} and \ref{tab:hosts}).}
\label{fig:expspec}
\end{figure*}

\subsection{Spectral Sub-Classification} \label{sec:siclass}

Figure~\ref{fig:expspec} presents all 15 explosion spectra obtained by the KSP (Section~\ref{sec:spec}).
We sub-classify four of the previously unpublished KSP \tase\ (KSNe-2016M, 2016ad, 2018oh, and 202112D) by measuring the near-peak pseudo-equivalent widths (pEWs) of their \siii\ $\lambda$6355~\AA\ and 5972~\AA\ absorption features following the classification scheme of \citet{Branch2006pasp}.
(Note that only those four SNe in Table~\ref{tab:spec} had explosion spectra obtained within $\pm$5 days since $B$-band maximum and were not coincident with bright host galaxy nuclei which can affect pEW measurements).

\begin{figure}[t!]
\epsscale{\scl}
\plotone{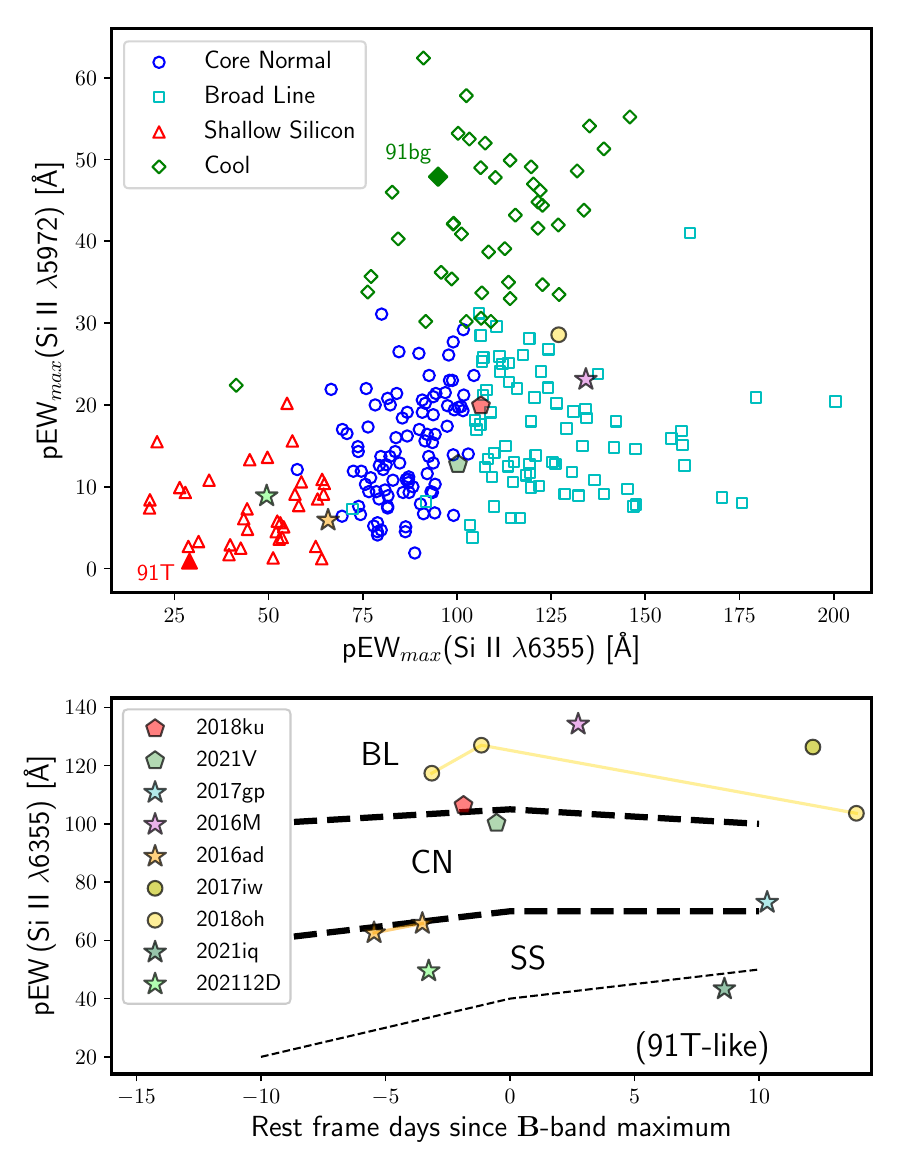}
\caption{\siii\ pEWs of KSP \tase\ (filled symbols with black borders marked by the same color as each SN from Figure~\ref{fig:ksplc}) near peak (top; within $\pm$5 days since $B$-band maximum) and between $-$10 and 10 days since $B$-band maximum versus phase (bottom).
The star, circle, and diamond symbols represent pEWs of early-blue, early-red, and early-yellow events, respectively, while the pentagons represent those of KSNe-2018ku and 2021V.
The top panel compares the pEWs to a sample of \tase\ \citep{Blondin2012aj} from the Core-Normal (CN) and Broad-Line (BL) normal subtypes; and the Shallow Silicon (SS) and Cool (CL) subtypes that contain normal as well as 91T-like and 91bg-like peculiar events, respectively.
Black dashed lines (bottom) show the boundaries of the CN and SS subtypes, as well as 91T-like peculiar events in SS, based on the large sample of \tase\ with pEW evolution measurements shown in Figure~3a of \citet{Phillips2022apj}.
\label{fig:class}}
\end{figure}

Figure~\ref{fig:class} (top panel) compares the near-peak pEWs (pEW$_{\rm max}$) of the four SNe as well as 2018ku and 2021V \citep{Ni2023bapj, Ni2023apj} with those of a large sample of \tase\ \citep{Blondin2012aj} from the CN and BL normal subtypes, as well as the Shallow Silicon (SS) and Cool (CL) subtypes that overlap with normal events and the ``91T-like'' and ``91bg-like'' peculiar subtypes, respectively.
Two early-blue events (KSNe-2016ad and 201212D) are from the SS subtype, while one early-blue (KSN-2016M) and one early-red (KSN-2018oh) event are from the BL subtype.

We also sub-classify three additional SNe (KSNe-2017gp, 2017iw, and 2021iq) that are also not coincident with bright host galaxy nuclei but have later explosion spectra using the following photometric and spectroscopic criteria:
\begin{itemize}
    \item Photometric: First, we rule out the CL/91bg-like subtypes based on their $i$-band light curves \citep[see][and references therein]{Taubenberger2013apj, Dhawan2017aa}---all three events reach a primary $i$-band maximum before the epoch of $B$-band maximum (Table~\ref{tab:snparam}) and show a clear secondary $i$-band maximum.
    \item Spectroscopic: Second, we distinguish between CN and SS subtypes following the method of \citet{Phillips2022apj} using an extended set of pEW (\siii~$\lambda$6355~\AA) observations from $\sim$ $-$10 days to as late as $\sim$ 10 days since $B$-band maximum as detailed below.
\end{itemize}

In the bottom panel of Figure~\ref{fig:class} we compare the evolution of pEW (\siii~$\lambda$6355~\AA) between $\sim$ $\pm$10 days since $B$-band maximum for all 9 KSP \tase---i.e., all the ones that have explosion spectra and are also not coincident with bright host galaxy nuclei---with boundaries for the CN and SS subtypes as well as 91T-like peculiar events from \citet[][see Figure 3a therein]{Phillips2022apj}.
All but one (KSN-2016M) of the 5 early-blue events (star symbols) are consistent with the SS subtype, though two of them (KSNe-2017gp and 2016ad) are on the boundary between the CN and SS subtypes.
The rest are all from the BL and CN subtypes.

These results indicate that the early-red and early-blue events in our sample prefer CN/BL (considering KSN-2018ku and 2021V as tentatively most similar to early-red; see Sections~\ref{subsec:popsep} and \ref{sec:lines-cases}) and CN/SS subtypes, respectively, which concurs with the finding of SET18 that their blue events are all from the SS subtype.

\begin{figure}[t!]
\epsscale{\scl}
\plotone{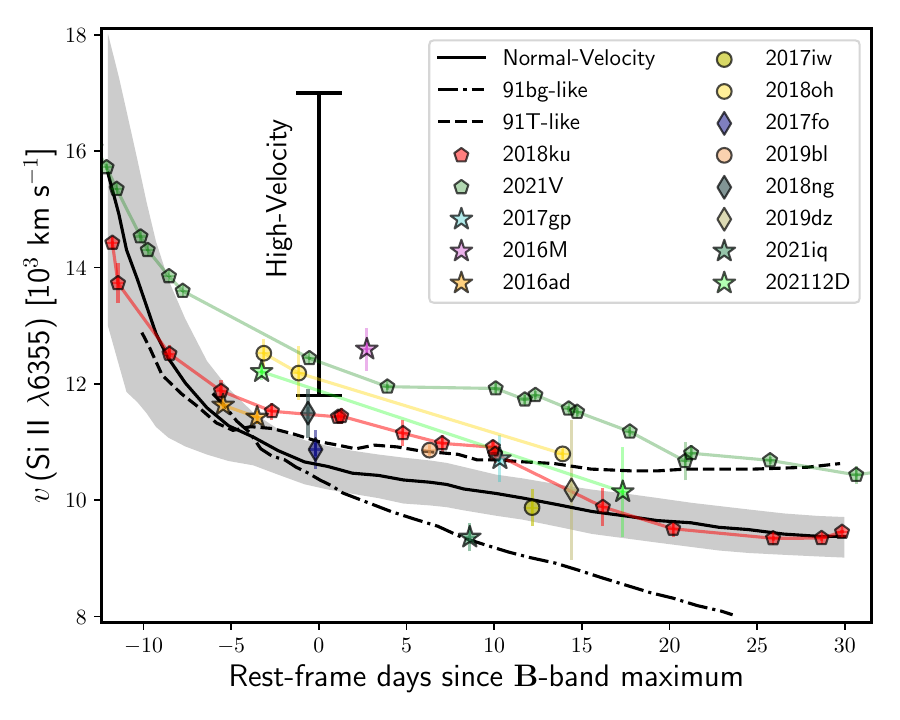}
\caption{\siii\ velocity evolution of KSP \tase\ (connected sets of data points 
marked by the same color as each SN from Figure~\ref{fig:ksplc}) compared 
to that of the Normal-Velocity (NV; black solid curve and shaded 1-$\sigma$ region) and High-Velocity (HV; vertical black velocity range at 0 days) subtypes of normal \tase\ \citep{Wang2009apj} as well as that of 91T/91bg-like peculiar subtypes (black dashed/dot-dashed curves). The star, circle, and diamond symbols represent the velocities of early-blue, early-red, and early-yellow events, respectively, while the pentagons represent those of KSNe-2018ku and 2021V.
The errorbars represent the combined uncertainties of the best-fit \siii\ minimum and of redshift.
\label{fig:wang}}
\end{figure}

\subsection{Photospheric Velocity Evolution} \label{sec:sivel}

We measure the photospheric velocity of the previously unpublished KSP \tase\ with explosion spectra from the minimum of the \siii~$\lambda$6355~\AA\ absorption feature (called ``\siii~velocity'').
Note that the \siii~velocity near peak is often used to estimate the ``characteristic ejecta velocity'' of \tase\ \citep[e.g.,][]{Li2019apj, Zhang2020mnras, Ni2023bapj} and separates normal events into NV and HV subtypes \citep{Wang2009apj}. 
We clearly identify the \siii~$\lambda$6355~\AA\ absorption feature in 13 spectra (vertical dashed lines in Figure~\ref{fig:expspec}) from the photospheric phase, as mentioned above (Section~\ref{sec:siclass}).
For all 13 spectra, we were able to fit the feature with a single skewed Gaussian profile \citep[as done for KSN-2018ku;][]{Ni2022natas}.
(The measured velocities and uncertainties from the fit are presented in column 10 of Table~\ref{tab:spec}).

Figure~\ref{fig:wang} compares the \siii\ velocities for the KSP \tase\ from the three early-multi-color populations
as well as KSNe-2018ku \citet{Ni2022natas} and 2021V \citep{Ni2023bapj} with expectations for the NV and HV subtypes of normal \tase\ as well as 91bg/91T-like peculiar subtypes \citep{Wang2009apj}.
While two of the early-blue events show velocities compatible with the 91T-like subtype---associated with blue events in SET18---we find no strong association overall between the velocities of the five early-blue events in the figure and any particular subtype. (At least one early-blue event shows consistency with each subtype).
The three early-red events in the figure are compatible with the velocity evolution of HV and NV normal subtypes, matching the spectroscopic classifications of KSNe-2018ku and 2021V \citep[see][]{Ni2022natas, Ni2023bapj} as well red events in SET18.
The three early-yellow events in the figure all exhibit slower velocities---two of them below the HV threshold near $B$-band maximum \citep[11,800~km~s$^{-1}$;][]{Wang2009apj} and the third showing consistency with the NV subtype at a later phase---indicating lower ejecta velocities overall or distribution of intermediate-mass elements in deeper ejecta layers.

\subsection{Host Galaxy Properties} \label{sec:hostpop}

\subsubsection{Morphological Classification}\label{subsec:hostmorph}

We conduct morphological classification of the host galaxies for the 16 previously unpublished \tase\ visually\footnote{Classification is done by four of us (S.C.K., H.S.P., Y.L., and J.S.) independently based on the images, and the most voted class is determined as the final morphological type for each host galaxy.} based on the image stamps in Figure~\ref{fig:hosts} (host galaxies labelled ``G'').
The stamps share a common physical scale (width = 100 kpc), RGB color scale\footnote{``Grey'' of the color scale is defined as the $BVi$ color of the Sun de-redshifted to the SN rest-frame and dereddened for MW extinction}, and intensity scale spanning from 1-$\sigma$ below the noise floor to 110\% of the central flux at SN peak. 
The intensity scale of the image is proportional to the square of the observed flux, to enable better visibility and comparison of morphological features.
Stellar masses ($M_{\star}$) of the host
galaxies are measured based on their rest-frame photometry using the stellar mass to $V$-band light ratio and \bv\ color \citep[see Table 7 in][]{Bell2003apjs}.
The results are presented in Table~\ref{tab:hosts} (columns 3 and 8).
Note, that our transformation of the host galaxy photometry to rest frame includes corrections of MW extinction and $DM$, but not $K$-corrections since the galaxy spectra are dominated by continuum flux rather than broad spectral features and all of the $z_{\rm host}$ are relatively small ($<$ 0.2), which leads to little change in the continuum level within any given filter.

The early-red events in our sample are found in a variety of host environments, ranging from elliptical (E/dE) galaxies for the majority of them (KSNe-2017iw, 2016bo, 2017cv, 2018oh, and 201903ah) to irregular (Irr) and spiral (S) galaxies (KSNe-2017cz and 2019bl) or apparently hostless environments \citep[KSN-201509b;][]{Moon2021apj}.
The early-yellow and early-blue events, on the other hand, appear to originate from distinct host environments. All 4 of the early-yellow events are found within massive, red (\bv\ $\sim$ 1.2 mag), early-type elliptical (E) and lenticular (S0) galaxies or their halos, which are associated with old stellar populations and low star formation rates (SFRs). In contrast, the 5 early-blue events in the KSP sample all come from within blue (\bv\ $\lesssim$ 1.0 mag), late-type spiral (S) or irregular (Irr/dIrr) galaxies associated with young stellar populations and high SFRs \citep[e.g., see][]{Poulain2021mnras}.

\subsubsection{Star Formation Rates}\label{subsec:hostsfr}

We estimate the SFRs of the late-type (S, Irr, and dIrr) host galaxies by cross-referencing their coordinates (Table~\ref{tab:hosts}) with near-ultraviolet (NUV) observations of the Galaxy Evolution Explorer \citep[GALEX\footnote{\url{https://galex.stsci.edu/GR6/}};][]{Bianchi2017apjs}.
For 6 out of 7 of them, a detected source is found at their coordinate and we measure SFR using dereddened NUV luminosity (adopting $DM$ from Table~\ref{tab:snparam}) and Equation 1 of \citet{Kennicutt1998araa}.
We note that although some early-type galaxies (E and S0) are also detected in NUV, we do not calculate their SFRs because NUV fluxes from those galaxies are not considered to be originated from star formation \citep{Yi1999apj, Peng2009apjl}.
Table~\ref{tab:hosts} (column 9 and 10) provides the estimated SFR for the 6 NUV-detected late-type galaxies, as well as ``specific'' SFR (sSFR = SFR/$M_{\star}$).
We confirm the presence of star formation in 5 of them---including 4 galaxies hosting early-blue events and 1 hosting a possibly early-red event---with sSFR $\sim$ 10$^{-11}$~yr$^{-1}$ or more, but the remaining galaxy (hosting an early-red event) is found to be quiescent with sSFR $<$ 10$^{-12}$~yr$^{-1}$ \citep{Houston2023mnras} which is at least one order of magnitude lower than the other sSFRs.

\subsubsection{Stellar Population Age and Metallicity}\label{subsec:hostppxf}

\begin{figure}[t!]
\epsscale{\scl}
\plotone{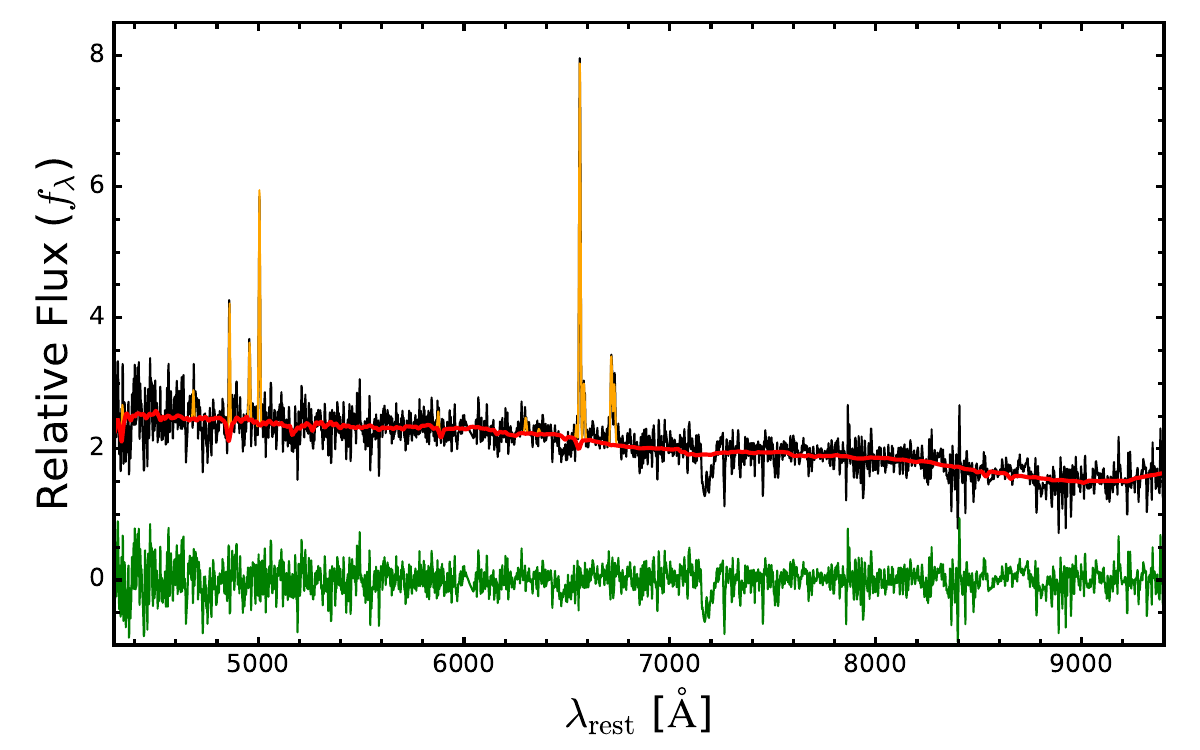}
\caption{Example of the pPXF fit for the host galaxy spectrum of KSN-2016ad (black) with \chisqr\ = 0.9. The red and orange curves represent the stellar spectrum and emission line components of the model, respectively. The green spectrum is the fit residual.
\label{fig:ppxf}}
\end{figure}

We investigate the age/metallicity environment of the 13 host galaxies for which spectra are available (either from our observations or NED; see Table~\ref{tab:spec} for the details) by carrying out a full spectrum fitting process using penalized pixel-fitting \citep[pPXF;][]{Cappellari2017mnras}.
Figure~\ref{fig:ppxf} shows an example of the fit for the host of KSN-2016ad.
In the fitting, we use the E-MILES stellar population models \citep{Vazdekis2016mnras}.
For each spectrum, we perform an initial fit while masking emission lines (Section~\ref{sec:host}) and some telluric lines to determine stellar kinematics. Then, we perform a second pPXF fit with fixed stellar kinematics to derive the emission line kinematics and luminosity-weighted age and metallicity [M/H] of the stellar populations (Table~\ref{tab:hosts}; columns 11 and 12).
We found excellent matches between the best-fit pPXF models and the spectra in all cases, with \chisqr\ ranging in 0.9--1.3.
The results confirm that the stellar populations of the morphologically early-type host galaxies of our sample---with mean luminosity-weighted age and [M/H] of 5.8 $\pm$ 4.0 Gyr and $-$0.05 $\pm$ 0.21 dex, respectively---are indeed older \citep[and, therefore, more metal-rich;][]{Gallazzi2005mnras} than the late-type ones, with corresponding mean age and [M/H] of 3.2 $\pm$ 3.0 Gyr and $-$0.70 $\pm$ 0.62 dex, respectively.

\begin{figure}[t!]
\epsscale{\scl}
\plotone{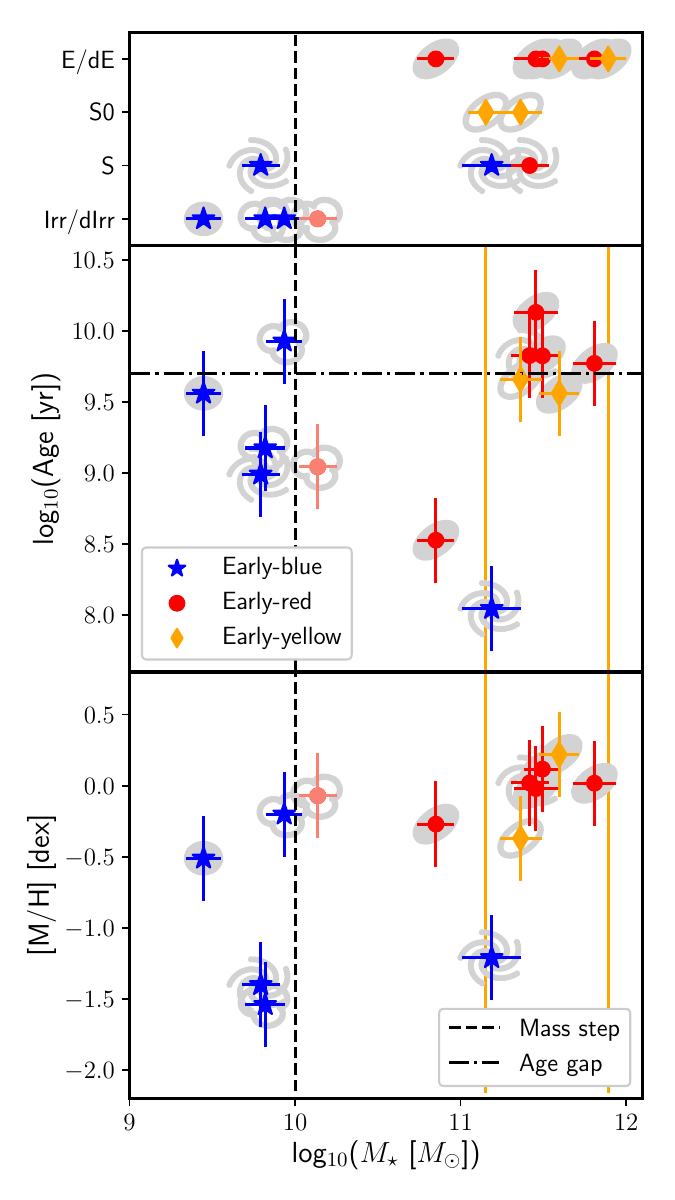}
\caption{Morphological type (top), age (middle), and metallicity ([M/H]; bottom) as a function of stellar mass ($M_{\star}$) for KSP \tas\ host galaxies (colored symbols; same as Figure~\ref{fig:gp3d}), highlighting differences between the hosts of early-blue events and those of early-red and early-yellow events. 
Grey symbols correspond to morphological type: E (filled ellipse), S0 (open ellipse), S (spiral), Irr (open trefoil) and dIrr (filled circle). 
Note that age and metallicity are only available for the 13 galaxies with spectra (see Table~\ref{tab:spec}).
The black vertical dashed line and horizontal dot-dashed line (middle panel) represent
the ``mass step'' boundary of 10$^{10}$~\msol\ \citep{Jones2023apj} and 
age gap in the bimodal distribution of stellar population ages at $\sim$ 6~Gyr \citep{Chung2023apj},
respectively, for \tas\ host galaxies.
The two gold vertical solid lines in the middle and bottom panels indicate $M_{\star}$ 
for the two early-yellow events of KSNe-2018ng and 2019dz for which no host spectra are available.
The $M_{\star}$ for the two early-red events of KSNe-201903ah and 201509b are 
\emph{below the lower limit of the shown x-axis}.
\label{fig:hostmass}}
\end{figure}

Figure~\ref{fig:hostmass} compares the morphological type (top), luminosity-weighted stellar population ages (middle), and metallicities (bottom) as a function of host stellar mass for the three early multi-color populations.
(Note that two early-red events with very low-mass hosts, KSN-201903ah with $\log_{10}(M_{\star})$ = 8.5 and KSN-201509b which is either hostless or hosted by a dwarf galaxy [\citealt{Moon2021apj}], are not shown).
The distribution of early-red events confirms that they come from a variety of host galaxies with stellar masses spanning an extreme range from the lowest (below 8.5 dex) to one of the highest (11.8 dex) among our sample, though most come from more massive ($>$ 10$^{10}$~\msol) hosts with older, more metal-rich stellar populations.
The early-blue and early-yellow events in our sample appear to prefer opposite extremes of host galaxy masses and stellar population properties: the former come from lower-mass galaxies and/or younger, lower-metallicity stellar populations towards the bottom and left sides of the panels, consistent with them being from active star-forming environments as mentioned above; while all of the latter originate from early-type galaxies at the upper end of the stellar mass distribution between 10$^{11}$ and 10$^{12}$~\msol\ that tend to harbor old, metal-rich stellar populations, consistent with quiescent environments.

\subsubsection{Comparison to Stritzinger et al. Results and the Host Galaxy ``Mass Step'' for Distance Measurements}\label{subsec:hostcomp}

These results found above, showing apparent preferences of the early multi-color populations in the KSP sample for different host environments, differ from the findings of SET18, which found no obvious trends between early color type and host properties and/or the locations of the SNe relative to their hosts.
However, we note that the 13 red and blue objects of the SET18 sample are all from spiral galaxies at extremely low redshifts ($z <$ 0.025), so a comparison cannot be made for SNe hosted by early-type galaxies.

Considering that the early-blue events exhibit different near-peak light curve properties (after standardization with normal \tas\ templates; see Section~\ref{subsec:philcomp}) and host environment preferences from the early-red and early-yellow events, it is possible that the differences between these populations contribute to the 
the observed host galaxy mass-dependent bias of \tas\ distance measurements (called ``mass step'', see Section~\ref{sec:intro}).
This possibility calls for further future investigation with a larger sample size, though we briefly examine the consistency between the mass step and the host galaxy properties of \tase\ from the early multi-color populations in our sample below.

The dashed vertical line in Figure~\ref{fig:hostmass} shows the mass step boundary of $M_{\star}$ = 10$^{10}$~\msol\ \citep{Childress2013apj, Jones2023apj}, where we find all but one of the early-blue events in our sample on the low-mass side of the step while all of the early-yellow events and all but two (i.e., KSNe-201903ah and 201509b with $M_{\star} < 10^9$~\msol; Section~\ref{subsec:hostppxf}) of the early-red events are found on the high-mass side.
One explanation that has been considered for the origin of the mass step is dependency of the near-peak properties of \tase\ on stellar population age, either with two progenitor channels \citep{Childress2014mnras} or with the bimodal distribution of host galaxy stellar population ages (i.e., old ``red sequence'' and young ``blue cloud''). 
The horizontal dot-dashed line in the top panel shows a proposed separation of \tase\ to explain the mass step based two groups of young and old host galaxies separated by an age gap at $\sim$ 6 Gyr \citep{Chung2023apj}.
All but one of the early-blue events are found on the side of the younger galaxies.
Thus, separation of host galaxies based on the mass step or age gap could conceivably produce a difference in \tas\ standardization due to the mass and age preferences of early-blue events.

\subsection{Redshift Distribution and Relative Rates of Events in the Early Multi-Color Populations}\label{sec:zpop}

\begin{figure}[t!]
\epsscale{\scl}
\plotone{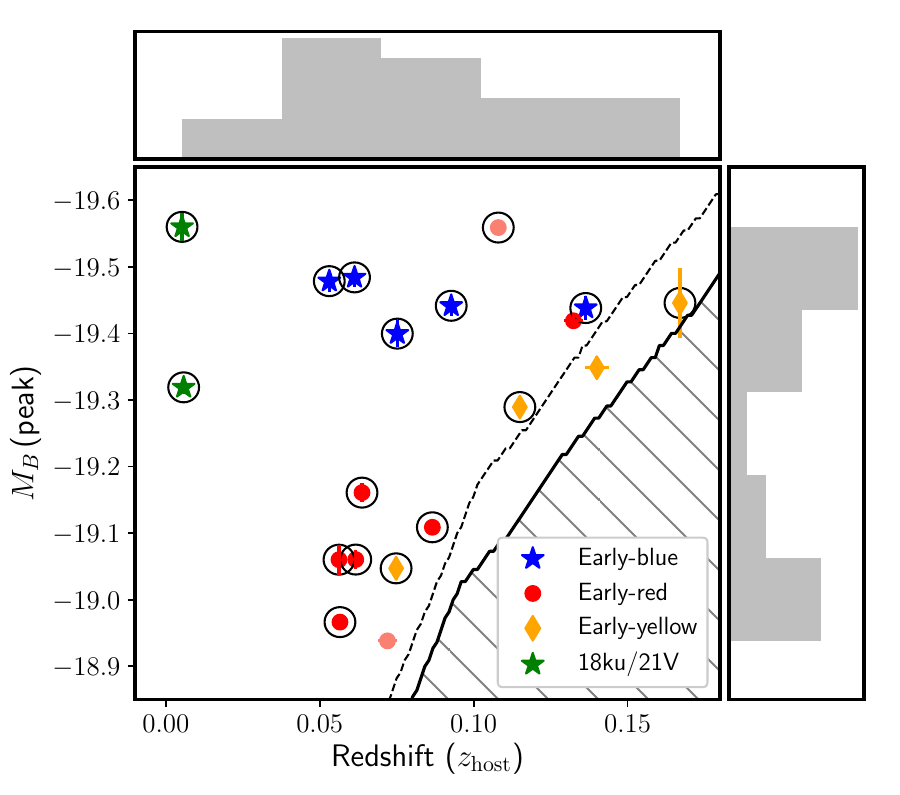}
\caption{Distribution of KSP \tas\ (same colored symbols as in Figure~\ref{fig:gp3d}) peak $B$-band absolute magnitude and redshift with marginal distributions along the top and right axes.
The circles indicate SNe with very reliable redshifts measured from host galaxy spectra. 
The solid and dashed curves represent the redshift upper limits of a \tas\ 
detectable in KSP with the detection limit of 21 mag for at least 30 and 35 days, respectively,
whereas the hatched region does the redshift range where a \tas\ at a given magnitude 
would be detectable for less than 30 days.
\label{fig:zmag}}
\end{figure}

Figure~\ref{fig:zmag} shows the distribution of $M_B$\,(peak) versus host galaxy redshift for the KSP \tase.
Most of them are from $z\sim$ 0.07, as shown by the top marginal distribution.
The full KSP sample is clearly not volume limited, with an upper limit for the redshift of detectable early \tase\ that depends on $M_B$\,(peak).
We estimate this upper limit by adopting the synthetic stretch-corrected KSP light curve of SN~2011fe (see Appendix~\ref{sec:synph}) as representing typical \tase\ with various $M_B$\,(peak). With this, we determine the redshift at which \tase\ with different $M_B$\,(peak) would be detectable for only 30 and 35 days with typical KSP detection limits of $>$ 21 mag. These are shown as solid and dashed lines in Figure~\ref{fig:zmag}. Note that the light curve analyses of \tase\ included in the KSP sample typically requires detections spanning $\sim$ $\pm$15 days around the peak. Thus, redshifts in the grey hatched region would be excluded.

The KSP sample of \tase\ with early multi-color light curves spans the complete luminosity range for normal \tase\ on the Phillips relation out to $z \sim$ 0.08 (Figure~\ref{fig:zmag}), 
where early-red and early-blue events are found with 5:3 ratio,
similar to the reported proportion of red and blue events (7:5)
from $z<$ 0.025 in SET18.
Accounting for the small sample size, we estimate the intrinsic relative rate of early-red events based on the observed KSP events within $z \lesssim$ 0.08 with Bayes rule as follows.
We consider the sampling of early-red events from a population comprised of early-red and early-blue events as Bernoulli trials with probability $P_r$ of drawing an early-red event. 
The posterior PDF of $P_r$ is Beta($n_r+1$, $n_b+1$) for a uniform prior on $P_r \in [0,1]$, where $n_r$ = 5 and $n_b$ = 3 are the observed numbers of early-red and early-blue events.
The expected value and variance are
\begin{align}
    \mathbb{E}(P_r) &= \frac{n_r+1}{n_r+n_b+2}\\
    Var(P_r) &= \frac{(n_r+1)(n_b+1)}{(n_r+n_b+2)^2 (n_r+n_b+3)}
\end{align}
which estimate $P_r \sim$ (60 $\pm$ 15)\% for the relative rate of early-red events to early-reds and early-blues.
This result is consistent with $P_r$ that can be estimated using the observed numbers of red and blue events within $z \lesssim$ 0.025 from SET18, which results in $P_r \sim$ (57 $\pm$ 13)\%, though this does not account for possible early-yellow events in their sample.

While early-blue events are found out to $z \sim$ 0.14 in the KSP sample,
it is unclear whether the early-red to early-blue ratio will continue to be seen at higher $z$ without deeper observations capable of detecting more distant events at the fainter end of normal \tase.
Early-yellow events, on the other hand, are apparently missing at low $z$ with only one of them being found 
within $z\sim$ 0.08, indicating that they are intrinsically rarer compared to their early-red and early-blue counterparts.
We estimate $P_y$ = (18 $\pm$ 11)\% as the 
relative rate of early-yellow events (to early-blues, reds, and yellows) based on the one sample within $z\lesssim$ 0.08.  

Beyond $z\gtrsim$ 0.1, the early-yellow events appear to 
comprise a larger fraction of the bright \tase\ in our sample, 
which suggests either a possible transition in intrinsic \tase\ properties at $z\sim$ 0.1, or that the observed rates of early-yellow events in our sample are significantly skewed by the small sample size.
Since no observational effects or theoretical mechanisms predict a transition 
in intrinsic \tas\ properties at $z\sim$ 0.1 as far as we are aware, 
the latter explanation for the redshift distribution of early-yellow events seems more likely.
Similarly, we note that under-representation of SNe with intermediate $M_B$\,(peak) appears to be present in our sample as well 
(Figure~\ref{fig:zmag}, see the marginal distribution on the right axis), but this pattern likely also arises from our small sample size since \tas\ peak brightnesses are not known to be bimodal \citep[e.g., see][]{Ashall2016mnras}.

\section{Discussion}\label{sec:disc}

Here, we discuss the basic physical mechanisms that can produce the multi-color evolutions
seen in the three \tas\ sub-populations that we identify in this work alongside those of
KSNe-2018ku and 2021V. 
(Note that our forthcoming paper [Ni et al. 2024 in prep.] will focus
on their ejecta properties, explosion mechanisms, and progenitor systems
in the context of \tas\ origin scenarios which we attempt to remain agnostic towards in this paper).

The spectral energy distributions (SEDs) of \tase\ in the infant to early phase of 1--3 days are first formed within an optically thick photosphere in the ejecta---or ``color depth'' since it is the origin of the color temperature of the SED \citep{Piro&Nakar2014apj}---and then affected by broad absorption features of various elements that form line photospheres in the outer ejecta \citep[for a review of these features in \tase, see][]{Parrent2014apss}.
As a result, color evolution is mainly driven by a combination of thermal and line effects.

\begin{figure}[t!]
\epsscale{\scl}
\plotone{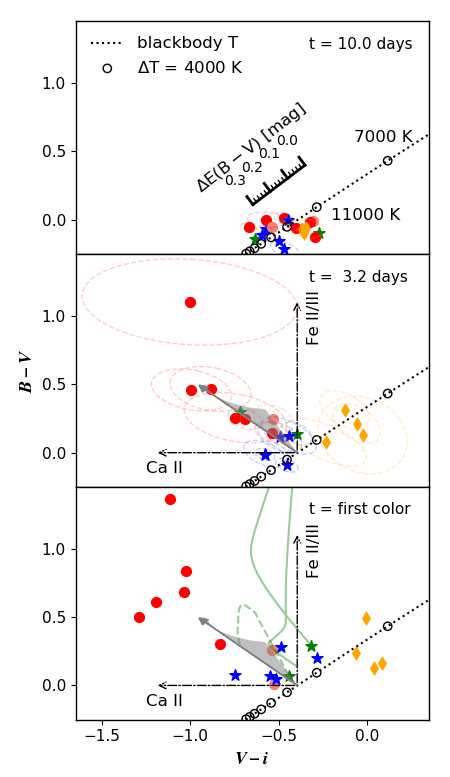}
\caption{Color-color (\bv\ vs. \vi) diagrams of KSP \tase\ (same colored symbols as in Figure~\ref{fig:gp3d}, 
with 1-$\sigma$ uncertainty ellipses) in rest frame 
at three phases: (top) 10 days since first light; (middle) 3.2 days; and (bottom) the first color detection ($t_{\rm col}$; Section~\ref{sec:early}). 
Dotted lines represent the colors of blackbodies for temperatures marked by open circles in 4000~K increments starting from 7000~K on the right.
Dot-dashed arrows (middle and bottom) 
indicate the directions of color change caused by \fex\ and \caii\ flux suppression 
in the $B$ and $i$ bands on a $\sim$ 13,000~K blackbody, respectively.
The gray arrow is the direction of the combined color change caused by \fex\ and \caii\
for SN~2011fe at 3.5 days (which we adopt as a template; Appendix~\ref{sec:synph}), 
with the shaded area representing its uncertainty.
Green curves (bottom) show the color evolution of KSN-2018ku (solid) 
and 2021V (dashed) from $t_{\rm col}$ (green stars) 
to 3.2 days at the other terminus.
(Note that KSN-2018ku reaches above the limit of the panel; see Figure~\ref{fig:gpcolor}).
While interstellar extinction can produce a similar color 
change---with a ruler (top) showing its direction---as a blackbody, this is not
responsible for the observed color evolution of the SNe (see text).
\label{fig:gpphys}}
\end{figure}

Figure~\ref{fig:gpphys} shows color-color (\bv\ versus \vi) diagrams of the KSP \tase\ with dotted ellipses for their 1-$\sigma$ uncertainties at three epochs (top to bottom panels): (1) 10~days after first light when all of the SNe have homogeneized in their evolution to a similar \bv\ and \vi\ color (Section~\ref{sec:early}); (2) 3.2 days, an earlier phase when the colors of the SNe are more divergent and we have coverage for all but one of the SNe in our sample (KSN-2017cz with color detection at $>$ 5 days); 
and (3) the earliest color detections for all SNe.
Note that while the first two panels show cross-sections in time, the colors in the final panel are not obtained at the same phase for all \tase\ but are illustrative of the maximum divergence of colors for our sample.
We investigate the possible origins for the color evolution of \tase\ in the figure in terms of thermal and line mechanisms below.

\subsection{Origin of Color Evolution I: Thermal Energy Sources} \label{sec:thermal}

The \bv\ and \vi\ colors of thermal blackbodies at various temperatures are shown as the dotted line in Figure~\ref{fig:gpphys}, with open circles along the line marking 4,000~K increments from 7,000 to 35,000~K.
As seen in the top panel, the KSP \tase\ appear clustered around a similar color temperature of $\sim$ 17,000~K by 10 days.
The small differences of the clustered SN colors at this phase may 
be partly attributable to intrinsic temperature differences of their photospheres 
and/or uncertainties in interstellar dust extinction correction.
We note that extinction in principle can produce somewhat similar color changes  
as blackbody temperature---as shown with the black ruler in the panel 
for a range of $\Delta E(\bv)$---but its color change is expected to be constant.
Thus,
the color evolutions of the infant/early \tase\ seen in Figure~\ref{fig:gpphys} 
are not due to uncertainties from extinction correction.

The light curves of \tase\ are mainly powered by the radioactive heating of \ni56\ produced by the explosion. 
The distribution of \ni56\ is primarily concentrated towards the ejecta core \citep[e.g.,][]{Arnett1982apj, Hoeflich1995apj, Boos2021apj}, which produces a temperature profile that increases towards the core, though second-order effects such as over-densities of radioactive elements near the ejecta surface \citep[e.g., in KSN-2018ku;][]{Ni2022natas} and non-spherically symmetric ejecta components \citep[e.g., in KSN-2021V;][]{Ni2023bapj} are possible depending on the explosion mechanism.
Under this framework, we examine whether the color evolutions of the three \tas\ sub-populations
in Figure~\ref{fig:gpphys} can be explained as a result of evolving thermal emission sources and highlight
implications for their origins as follows:
\begin{itemize}
    \item \textbf{Early-yellow:} Without the addition of other effects, the evolution of color temperature can be driven by the expected evolution of the photosphere which expands in radius but retreats into deeper ejecta layers with higher temperature \citep{Piro&Nakar2013apj}. This will produce blueward evolution in all colors.
    From the bottom to the top panel of Figure~\ref{fig:gpphys} the colors of the early-yellow events appear to evolve blueward along the blackbody line, consistent with increasing temperature.
    The color evolution can be interpreted as the result of a change in temperature from $\sim$ 8,000 to 13,000~K between $\sim$ 2 to 10 days (bottom to top panel) since the epoch of first light, which matches excellently with the inferred color temperature evolutions from \tas\ bolometric light curves with radioactive \ni56-powered models \citep[e.g.,][]{Piro&Nakar2014apj}.

    \item \textbf{Early-blue:} In contrast, the early-blue events in the figure show little change in color clustered around a much higher mean temperature $\sim$ 17,000~K.
    Their early high temperatures likely require additional energy sources that heat the shallower ejecta layers, such as \ni56\ clumps/shells in the outer ejecta \citep{Magee2020aab} or shock interaction with a companion star \citep{Kasen2010apj}.
    This is consistent with their early light curves which are found to accommodate short-timescale excess emission within the first 1--2 days in all cases.
    Note that the amounts of energy attributable to the observed early excess emissions---estimated in the range of $\sim$ (3.5--5.8) $\times$ 10$^{46}$~ergs in the optical bands (Appendix~\ref{sec:reggaus})---are also within expectations for some of these energy sources which are predicted to be present in \tase\ by some models \citep[e.g., collision with a subgiant companion in the single-degenerate scenario;][]{Kasen2010apj, Maoz2014araa}.

    \item \textbf{Early-red}: Finally, all of the early-red events exhibit colors that are inconsistent with thermal SEDs of any temperature when observed at sufficiently early phases ($\lesssim$ 3 days; bottom and middle panels of Figure~\ref{fig:gpphys}). 
    In particular, except for one event with a large uncertainty (KSN-2019bl), 
    the colors of the early-red events take on an approximately linear distribution in \bv\ versus \vi\ at $\sim$ 3 days (middle panel) which is nearly orthogonal to blackbody evolution (dotted line).
    Their evolution during 3--10 days (middle to top panel) appears to be consistent with the direction 
    of this linear distribution, indicating that the evolution is primarily due to non-thermal effects. We discuss the possible nature of these non-thermal effects below (Section~{\ref{sec:lines-red}}).
\end{itemize}

\subsection{Origin of Color Evolution II: Broad Line Features and Additional Power Sources} \label{sec:lines}

As shown above, non-thermal effects are required to explain the color evolution of early-red events, and we examine whether broad line features can explain their observed colors and color evolutions in this section.
We also discuss the color evolution of the two earliest events in our sample, KSNe-2018ku and 2021V, highlighting their similarities to early-red events and possible contributions from additional power sources.

\subsubsection{Broad Line Features for Early-Red Events} \label{sec:lines-red}

Based on a number of well observed \tase, two prominent spectral features that can substantially affect the evolution of optical colors at early times are the \fex\ complex located in the $B$-band and the \caii\ near-infrared (NIR) triplet located in the $i$-band \citep[for example, see Appendix~\ref{sec:synph} for the spectral evolution of SN~2011fe in KSP filters, and see spectral models in][]{Parrent2014apss}.
The former feature affects the entire $B$-band, and may also include contributions from \mgx\ and sometimes \tiii\ in the case of 91bg-like events \citep{Doull2011pasp}.
The latter feature can affect up to 1/3 of the $i$-band wavelength range, especially in \tase\ that have ``High-Velocity Features'' \citep[HVFs; e.g., of both \caii\ and \siii\ in SN 2021aefx; see][and Figures A1 and 6 therein]{Ni2023bapj}.
While SET18 argue that a flux difference of $\sim$ 50\% based on $\sim$ 0.5-mag differences in the \bv\ colors of the red and blue groups is far too large to be attributable to line features, suppression of as much as $\sim$ 70--80\% of early $B$-band flux by \fex\ compared to the expectation based on $V$ and $i$ bands was recently seen in normal \tase\ \citep[e.g., SN~2018aoz;][]{Ni2022natas}.

The opposing blueward and redward evolutions of \bv\ and \vi\ colors, respectively, in early-red events (see Figure~\ref{fig:gpphys} and Section~\ref{subsec:popid}) may be attributed 
to the weakening of \caii\ (in $i$ band) and \fex\ (in $B$ band) features over time.
Such line evolution is clearly seen in the spectrophotometric time series of the normal 
\tas~2011fe \citep[see Figure~\ref{fig:synph};][identified as a red object in SET18]{Pereira2013aa},
which produces similar opposing color evolutions seen in our early-red events (Figures~\ref{fig:BVsplit} and \ref{fig:VIsplit}).
Note that a range of colors that evolve in this way can in principle arise from variations in the steepness of the distribution of nuclear burning products in the ejecta \citep[such variation has been inferred in several normal \tase\ with early bolometric light curves using \ni56-powered models; e.g.,][]{Piro&Nakar2014apj, Moon2021apj, Ni2022natas}.
We show below that the spread in the observed colors of the early-red events in our sample,
including their aforementioned approximately linear distribution in \bv\ and \vi\ at 3.2 days,
is consistent with what can be expected from potential differences in their distributions of nuclear burning products.

To investigate this hypothesis, we adopt SN~2011fe as a typical early-red case \citep{Stritzinger2018apj} 
and estimate the ``line coloration'' parameter ($\Delta \Vec{C}$)
which we define to represent the color change from the continuum 
expected from the \fex\ and \caii\ features.
We conduct synthetic photometry on the spectrum of SN~2011fe before and after excising the features (see Appendix~\ref{sec:synph} for the details).
The results are scaled by 0--1.5, illustrating the line coloration ``direction'' for a range of suppression strengths relative to SN~2011fe, while holding fixed the ratio of the suppression due to \fex\ and \caii.
The grey arrow in the middle and bottom panels of Figure~\ref{fig:gpphys} shows the line coloration direction from 3.5 days, while the grey shaded uncertainty region includes the results from various phases in 2.6--9.5 days which approximately accounts for variations in the steepness of the distribution of nuclear burning products in the ejecta.
The slope of the arrow appears consistent with that of the distribution of the colors of early-red events in the nearby 3.2-day phase (middle panel). 
This suggests that the colors of early-red events could be explained by a $\sim$ 13,000~K 
continuum SED affected by \fex\ and \caii\ suppression of various strengths, but with a similar ratio as found in SN~2011fe.
Since this result is based on a single object as a template, we point to the need for further exploration of the effect using a broader sample of early-red \tase\ with early spectroscopic coverage.

\subsubsection{Broad Line Features for SNe~2018aoz and 2021aefx} \label{sec:lines-cases}

The effects of broad line features on early \tas\ colors are also seen in the color evolutions of KSNe-2018ku and 2021V (= SNe 2018aoz and 2021aefx) from $<$ 1 hour since first light until 3.2 days (solid and dashed green curves, respectively, beginning with stars in the bottom panel of Figure~\ref{fig:gpphys}).
These are two SNe in our KSP sample that were detected earliest and show 
color evolution that is not generally seen in the other events (Section~\ref{subsec:popspec}). 
We discuss each in more detail here in the context of the physical processes driving the color evolutions of the early populations.

First, the observed features of KSN-2018ku show consistency with early-red events from as early as $\sim$ 0.5 days 
since first light (Section~\ref{subsec:popspec}) 
as seen in Figure~\ref{fig:gpphys} (the right green star from one of the two green stars in the top and middle panels), 
attesting to its similarity to the early-red population. 
The SN is featured with an extremely red color in \bv\ ($\sim$ 1.8 mag) at $\sim$ 0.5 days \citep{Ni2022natas} 
due to the effects by \fex\ lines
whose subsequent evolution leads to the approximately vertical color change of KSN-2018ku from 
its reddest \bv\ color to $\sim$ 0.1 mag between 0.5 and 3.2 days in Figure~\ref{fig:gpphys} (solid green curve in the bottom panel).
The strong similarities between KSN-2018ku and the early-red population 
support the interpretation that the former is an extreme case of the latter,
suggesting that the line-blanketing effects by \fex\ lines may be present 
in early-red events starting from as early as within $\lesssim$ 1 day since first light
(i.e., preceding their earliest color detections; Table~\ref{tab:snparam}).

KSN-2021V also shares similarities in its color
evolution to those of the early-red population 
as evidenced by the apparent consistency of its colors (see Figure~\ref{fig:gpphys}
and the locations of the left green star) with the linear (middle panel) and clustered (top panel) distributions 
of red circles. 
However, its initial color is consistent with a $\sim$ 15,000~K blackbody (bottom panel)---when its 
SED was dominated by the additional heating from its early excess emission \citep[see][and Figure 4 therein]{Ni2023bapj}---before the excess emission fades in relative strength compared to the underlying power-law rise by $\sim$ 2--3 days since first light (dashed green curve in the bottom panel).
In addition, considering that SN~2009ig, a \tas\ showing extremely similar early spectroscopic features as KSN-2021V \citep[see Figure 5 in][]{Ni2023bapj}, was also identified as a red object in SET18, 
we associate KSN-2021V with the early-red population.
We note, however, the early excess emission and blue thermal colors of the source visible until $\sim$ 2--3 days \citep[][]{Ni2023bapj}, which may be attributable to the presence of an additional power source,  
is unique among the early-red population.

\subsubsection{Additional Power Sources} \label{sec:pow-sources}

Both the early-blue events in our sample and KSN-2021V accommodate early excess 
emission in their light curves, indicating the presence of additional power sources.
However, their colors undergo distinctly different evolutions where 
the former show slowly-evolving hot thermal colors; in contrast, the latter shows a brief appearance of a hot thermal color that fades quickly over 2--3 days to reveal similar colors to the early-red population.
These differences may point to different origins for their excess emissions,
such as mixing of radioactive material towards the outer ejecta by sub-sonic burning \citep{Maeda2010apj} 
to produce a stratified distribution of excess heat for the former;
and collision with a companion star to produce an ejecta hole briefly revealing the hotter interior \citep{Kasen2010apj} for the latter. 
We note that comprehensive comparisons between their observed early excess emissions in our
sample to model predictions will be presented in our forthcoming paper (Ni et al. 2024 in prep.).

\section{Summary and Conclusion}\label{sec:conc}

The KSP has obtained early multi-color light curves of 19 infant/early \tase, including 7 of them detected within $\lesssim$ 1 day since 
the estimated epoch of 
first light and the rest within $\lesssim$ 3 days.
We have identified at least 3 distinct populations in this sample based on the early colors in the first $\sim$ 1--3 days and their subsequent evolutions. The properties of the populations as well as their host galaxy environments are summarized as follows:

\begin{enumerate}
    \item ``Early-blue'' events show a slowly-evolving blue color within 1--10 days;
    accommodate early excess emission in addition to the power-law rise of their light curves in this phase;
    have similar peak $B$-band magnitudes ($-$19.45 $\pm$ 0.03 mag) that are consistent with peculiar 91T-like events or the bright end of normal \tase; have shallower \siii\ absorption features, preferentially found in the SS subtype (which includes 91T-like events) and on the boundary between SS and CN; are slower-evolving with \dm15\ $<$ 1.2 mag; and are found in low-mass and/or late-type galaxies harboring active star-formation and younger, more metal-poor stellar populations.
    Their early colors are approximately consistent with thermal SEDs at a relatively high blackbody temperature of $\sim$ 17,000~K, likely due to the presence of additional power sources responsible for their early excess emission.
    
    \item ``Early-red'' events evolve blueward in \bv\ and redward in \vi\ within 1--10 days; show pure power-law rise in the early light curves; are consistent with the Phillips relation across the full range of normal \tas\ brightnesses and \dm15 values; exhibit \siii\ absorption features with strengths and velocities consistent with CN/NV or BL/HV subtypes; are found in host galaxies of various morphologies and masses, with preference for massive early-type hosts; and appear to be the majority population, comprising (60 $\pm$ 15)~\% of early-red and early-blue events out to at least $z \sim$ 0.08. Their SEDs are incompatible with blackbodies in early phases ($\lesssim$ 3 days), most likely due to the suppression of $B$- and $i$-band flux by \fex\ and \caii, respectively, which weaken over time to produce the opposing color evolutions in \bv\ and \vi.

    \item ``Early-yellow'' events evolve blueward in both \bv\ and \vi\ color; show pure power-law rise in the early light curves, similar to early-red events; have near-peak light curve properties consistent with early-red events, but lower \siii~velocities; are found in massive early-type galaxies with old, metal-rich stellar populations, consistent with quiescent star-formation environments; and may be relatively rare, estimated to comprise $\sim$ (18 $\pm$ 11)\% of the three populations based on only one event found within $z \lesssim$ 0.08.
    Their colors during $\sim$ 2--10 days are compatible with blackbody SEDs evolving from $\sim$ 8,000 to 13,000~K caused by the typical expansion of \tas\ ejecta with a centrally-concentrated \ni56\ distribution.

    \item KSNe-2018ku and 2021V (= SNe~2018aoz and 2021aefx, respectively) resemble early-red events during $\sim$ 2--10 days, though in earlier phases, SN~2018aoz exhibits a rapid redward color evolution in \bv\ with non-thermal SED while SN~2021aefx has a prominent blue thermal excess emission that reddens and fades.
    Both of these reddening behaviors beginning from an extremely early phase have not yet been identified in any other normal \tase\ to date.
\end{enumerate}

We find that the early-red and early-blue populations are marginally separable based on the intersection of their multi-color evolution with an oblique slice of the 3-D space of \bv, \vi, and time, and provide a decision boundary for classifying early \tase\ that intersect with the slice.
The distribution of their colors, however, are nearly equally compatible with them being two distinct populations \citep{Stritzinger2018apj, Han2020apj} whose distributions of colors are $\sim$ 1.5-$\sigma$ separated in our sample, or extreme ends of a continuous population \citep{Bulla2020apj}.
Further detections of early \tase\ within $\lesssim$ 2 days since first light in at least three bands including $B$ (containing \fex) and $i$ (or $I$; containing the \caii\ NIR triplet) are required to fully test the distinction between the two populations.

\citet{Bulla2020apj} suggested that the lower degree of separability 
between red and blue objects in the ZTF sample may be attributable to the 
the broader (4000--5500~\AA) $g$-band filter of the ZTF being less sensitive 
than $B$ band (4000--5000~\AA) to the spectral differences between the two groups (which primarily comes from \fex\ in $B$ band). 
As we have shown, it may also be attributable to the presence of early-yellow events straddling their boundaries. 
We also note that since early-yellow events appear to prefer early-type hosts, compared to the red and blue events of SET18 and their extension in \citet{Han2020apj}---which all come from late-type hosts---the ZTF untargetted sample of red/blue events is more likely to be contaminated by early-yellow events.

The distinct early multi-color evolution of the three populations points to differences in their explosion process and/or explosion asymmetry.
For instance, the different evolution of early-red and early-yellow events may be due to how nuclear burning products are distributed along the line of sight.
The explosions of early-red events may produce a more extensive distribution of nuclear burning products with radioactive \ni56\ and intermediate mass elements located closer to the ejecta surface, leading to strong \fex\ and \caii\ suppression and a bluer color temperature compared to the early-yellow events whose multi-color evolution could explained by a centrally-concentrated radioactive heat source.
For early-blue events, 
the presence of an additional heat source responsible for their early excess emission and slowly-evolving blue colors likely differentiates their explosions and/or progenitor systems from those of early-red and early-yellow ones. 
These differences may be attributable to viewing angle effects due to asymmetry---such as off-center carbon ignition producing an asymmetric \ni56\ distribution \citep{Maeda2010apj, Boos2021apj}, or collision with the companion star producing angle-dependent excess emission \citep{Kasen2010apj}---or the existence of multiple \tas\ production channels.

The observed preference of early-blue events in our sample for different host galaxy types and stellar population properties compared to the other populations (especially early-yellow) may point to different progenitor channels for them based on stellar population age.
Considering the differences between the near-peak light curve properties of early-blue events and the other populations after standardization with normal \tas\ templates, their respective preferences for low- and high-mass hosts indicates a potential connection with the ``mass step'' of \tas\ Hubble residuals.
On the other hand, given that the colors of early-red and early-blue events are still compatible with a single continuous distribution, it is possible that those two populations share a common origin.
For instance, sub-Chandrasekhar-mass explosions are predicted to produce the full range of \tas\ brightnesses with brighter ones preferentially from younger host environments \citep{vanKerkwijk2010apj, Shen2021apjl}, which appears to be consistent with the observed peak brightnesses and host galaxy properties of the early-red and early-blue populations.
We emphasize, however, that these trends for the early multi-color populations are currently based on a very small number of observed members for each population (5 early-blue, 8 early-red, and 4 early-yellow), while more statistically robust conclusions will require continued observations focused on obtaining early multi-color light curves of \tase\ along with systematic efforts to constrain their near-peak and host galaxy properties.

Two extreme cases of KSNe-2018ku and 2021V (= SNe~2018aoz and 2021aefx, respectively) show unique early behavior among the 19 \tase\ of our sample, though the distinction of KSN-2018ku may be due to an observational effect rather than an intrinsic difference.
Considering the much earlier and deeper detection of KSN-2018ku within $\lesssim$ 0.5 days since first light compared to all of the other early-red events, and its overall consistency with early-red power-law rise, color evolution, and near-peak light curve and spectroscopic characteristics after $\sim$ 0.5 days, it is possible that additional members of the early-red population would reveal infant-phase excess emission and/or rapid redward color evolution with earlier and deeper multi-color observations.

On the other hand, the observed multi-color evolution of KSN-2021V before $\sim$ 2 days is apparently incompatible with any of the three early multi-color populations. However, the later consistency of its light curves after 2 days with early-red power-law rise and color evolution, and its intermediate classification between CN and BL consistent with early-red events suggests that it may be a rare case of an early-red event that experienced a briefly visible heating of the ejecta surface from some localized power source, such as ejecta collision with a companion star or the presence of an isolated over-density of radioactive material near the ejecta surface.

\nopagebreak

\vskip 5.8mm plus 1mm minus 1mm
\vskip1sp
\section*{Acknowledgments}
\vskip4pt

Y.Q.N. thanks Mark M. Phillips and Marten H. van Kerkwijk for helpful comments and Ernest Chang for the SNID results of KSP-SN-2019dz.
This research has made use of the KMTNet system operated by the Korea Astronomy and Space Science Institute (KASI) and the data were obtained at three host sites of CTIO in Chile, SAAO in South Africa, and SSO in Australia.
Data transfer from the host site to KASI was supported by the Korea Research Environment Open NETwork (KREONET).
This research was supported by KASI under the R\&D program (Project No. 2023-1-868-03) supervised by the Ministry of Science and ICT.
This research is also based on observations obtained at the international Gemini-S Observatory, a program of National Science Foundation’s (NSF) NOIRLab, which is managed by the Association of Universities for Research in Astronomy (AURA) under a cooperative agreement with the NSF on behalf of the Gemini Observatory partnership: the NSF (United States), National Research Council (NRC; Canada), Agencia Nacional de Investigaci\'{o}n y Desarrollo (Chile), Ministerio de Ciencia, Tecnolog\'{i}a e Innovaci\'{o}n (Argentina), Minist\'{e}rio da Ci\^{e}ncia, Tecnologia, Inova\c{c}\~{o}es e Comunica\c{c}\~{o}es (MCTI; Brazil), and KASI (Republic of Korea).
The Gemini-S observations were obtained under the Canadian Gemini Office (PID: GS-2016B-Q-1, GS-2017B-Q-10, GS-2018A-Q-137, GS-2019A-Q-237, GS-2021A-Q-114, GS-2021B-Q-114, and GS-2023B-Q-207) of the NRC and the K-GMT Science Program (PID: GS-2017B-Q-21 and GS-2018A-Q-117) of KASI and accessed through the Gemini Observatory Archive at NSF’s NOIRLab.
This paper includes data gathered with the 2.5m du Pont telescope and 6.5m Magellan telescopes located at Las Campanas Observatory, Chile.
Some of the data presented herein were obtained at SOAR telescope, which is a joint project of the MCTI do Brasil, the US NSF’s NOIRLab, the University of North Carolina at Chapel Hill, and Michigan State University.
The SOAR observations were obtained under NOIRLab (PID: 2023B-394234).
using the Las Cumbres Observatory (LCO) Observation Portal and accessed through the LCO Science Archive.
D.-S.M. and M.R.D. are supported by Discovery Grants from the Natural Sciences and Engineering Research Council of Canada (NSERC; Nos. RGPIN-2019-06524 and RGPIN-2019-06186, respectively).
D.-S.M. was supported in part by a Leading Edge Fund from the Canadian Foundation for Innovation (CFI; project No. 30951).
M.R.D. was supported in part by the Canada Research Chairs Program and the Dunlap Institute at the University of Toronto.

\vspace{1mm}

\facilities{KMTNet, du Pont (WFCCD), Magellan-Clay (LDSS-3), Gemini-S (GMOS), SOAR (Goodman)}

\software{Astropy \citep{Astropy2013aa}, \texttt{emcee} \citep{Foreman-Mackey2013pasp}, \texttt{george} \citep{Ambikasaran2015itpam}, \texttt{HOTPANTS} \citep{Becker2015ascl}, \texttt{IRAF} \citep{Tody1993aspc}, pPXF \citep{Cappellari2017mnras}, \texttt{scikit-learn} \citep{scikit-learn}, SciPy \citep{Virtanen2020natme}, SNAP \citep{Ni2022zndo}}, SNID \citep{Blondin2007apj}, SNooPy \citep{Burns2011aj}, Source Extractor \citep{Bertin&Arnouts1996aas}, \texttt{synphot} \citep{synphot}

\appendix
\restartappendixnumbering

\section{KSP Multi-Object Photometry} \label{sec:mop}

Here, we describe the ``Multi-Object Photometry'' (MOP) which is used to measure the fluxes of the KSP \tase\ that are located within a non-linear flux background dominated by a small number of nearby sources.
The method measures the fluxes of the SN and nearby sources by applying ``forced-photometry'', where the profiles of each source are fitted simultaneously.
In the following subsections, we describe the model and its fitting process (Appendix~\ref{subsec:mopmod});
how MOP compares to other photometry methods (Appendix~\ref{subsec:cophot});
and the mitigation of systematic uncertainties for our photometry methods (Appendix~\ref{subsec:debias}).

\subsection{Model Description and Implementation}\label{subsec:mopmod}

The profiles of sources are approximated using either the image PSFs for point sources---i.e., the SN, stars, and galaxy cores---or S\'ersic functions for galaxy wings \citep{Caon1993mnras} in the MOP model.
The model instrumental intensities $I_{\rm MOP}$ are described by the equations:
\begin{align}
  I_{\rm MOP}(\Vec{x}) &= \sum_i^{N_p} A_i\,PSF(\Vec{x} - \Vec{x}_i) + \sum_i^{N_s} A_i\,\Tilde{SER}_i(\Vec{x} - \Vec{x}_i)
  \label{eq:mop}
  \\
  \Tilde{SER}_i(\Vec{x}) &= \frac{1}{k^2} \sum_{\Vec{x}\,^{\prime} \in\,K_r} SER_i(\Vec{x} - \Vec{x}\,^{\prime}) \times PSF(\Vec{x}\,^{\prime})
  \label{eq:conv}
\end{align}
where $\Vec{x}$ is pixel position on the 2-D image, and $A_i$ and $\Vec{x}_i$ are the heights and positions of the sources in the model, respectively, comprised of $N_p$ PSFs and $N_s$ S\'ersics. 
$PSF$ and $SER_i$ are the normalized distribution functions for the image PSF and the S\'ersics, respectively---i.e., both sum to unit flux over pixel-space (thus, the heights $A_i$ in Equation~\ref{eq:mop} also correspond to the integrated fluxes of each source).
$\Tilde{SER}_i$ is the result of a discrete convolution between $SER_i$ and $PSF$ evaluated over a square kernel stamp $K$, whose size and sampling are scalings of the $PSF$ major-axis FWHM and image pixel sampling, respectively, by factors of $r_K$ and $n_K$.
In practice, $r_K$ = $n_K$ = 4 is sufficiently large, and larger values affect an insignificant ($\lesssim$ 1\%) change in measured flux.

We adopt PSF and S\'ersic distributions with normalized flux and elliptical shape, described as follows:
\begin{align}
    PSF(\Vec{x},\ [a_{\rm maj}, a_{\rm min}, \theta, b]\ ) &= \frac{b-1}{\pi a_{\rm maj}a_{\rm min}}\left(1+||W\Vec{x}\,||^2\right)^{-b}
    \label{eq:moff}
    \\
    SER(\Vec{x},\ [a_{\rm maj}, a_{\rm min}, \theta, n]\ ) &= \frac{b_n^{2n} \exp(-b_n||W\Vec{x}\,||^{1/n})}{\Gamma(2n+1)\,\pi a_{\rm maj} a_{\rm min}}
    \label{eq:sers}
    \\
    W(a_{\rm maj}, a_{\rm min}, \theta) &= \begin{bmatrix}
    \cos\theta/a_{\rm maj} & -\sin\theta/a_{\rm maj} \\
    \sin\theta/a_{\rm min} & \cos\theta/a_{\rm min}
    \end{bmatrix}
    \label{eq:whit}
\end{align}
where Equation~\ref{eq:moff} is a normalized Moffat function with FWHM = $2a\sqrt{2^{1/b} - 1}$ \citep{Trujillo2001mnras}; $b_n$ is defined such that $a_{\rm maj}$ and $a_{\rm min}$ are the semi-major and semi-minor axes of the half-light ellipse over which the normalized S\'ersic distribution integrates to 1/2, respectively, obtained as the solution to $\Gamma(2n) = 2\gamma(2n, b_n)$ \citep{Ciotti1991aa}; and W is the whitening matrix used to obtain Mahalanobis distance $W\Vec{x}$ for input into circular Moffat and S\'ersic functions. 

The shape parameters of the PSF ($a_{\rm maj}, a_{\rm min}, \theta, b$) are measured from a hand-picked set of nearby reference stars (Section~\ref{sec:photflux}), while those of each S\'ersic ($a_{\rm maj}, a_{\rm min}, \theta, n$) are measured by fitting the model in advance to a deep pre-SN stacked image template. The model fitted to the template, therefore, has $3\,N_p + 7\,N_s$ parameters and is obtained by minimizing the following $\chi^2$ loss function:
\begin{align}
    \chi^2 = &\sum_{\Vec{x}\,\in\,R_{\rm fit}}\frac{[I(\Vec{x}) - I_{\rm sky}(\Vec{x}) -  I_{\rm MOP}(\Vec{x})]^2}{|I(\Vec{x}) - I_{\rm sky}(\Vec{x})| + \sigma_{\rm sky}^2}
    \label{eq:chi}\\
    &I_{\rm sky}(\Vec{x}) = [m_x, m_y]^T\cdot\Vec{x} + c
    \label{eq:sky}
\end{align}
where the signal to be fit is the intensity of the image ($I$) in the fit region ($R_{\rm fit}$; see below) after subtracting sky background ($I_{\rm sky}$), and the total error is the sum of Poisson and sky background noise.
We adopt a linear sky background (Equation~\ref{eq:sky}) with fixed slope ($m_x, m_y$) and offset ($c$) measured before the fit (see below) along with the sky background noise ($\sigma_{\rm sky}$).
The loss function is minimized using the Levenberg-Marquardt algorithm\footnote{\texttt{scipy.optimize.curve\_fit}}.
The inverse Hessian of the loss function at the optimum is used to estimate standard errors. 
For the instrumental fluxes, $A_i$, the standard error of the fit is typically smaller than the total error obtained from summing the photon statistics of each pixel independently---i.e., $\sigma^2_{\rm phot} = \sum_{i}|A_i| + N_{\rm 90}\,\sigma_{\rm sky}^2$, where $N_{\rm 90}$ is the number of pixels in an aperture enclosing 90\% of the source flux---in the low S/N regime, while it is larger for very high S/N.

We determine the sources to be included in the MOP model as follows.
First, a few bright objects in the template stacked image that appear to be approximately point or S\'ersic sources are included in an initial fit. 
The fit region $R_{\rm fit}$ is the union of $N_p$ + $N_s$ circular apertures centered on each source, with radii defined as a scaling of the average FWHM of the image PSF by a constant factor $n_{\rm ap}$.
We select $n_{\rm ap}$ between 3 and 9 so that $R_{\rm fit}$ extends $\sim$ 3\,FWHM beyond the physical radius of each source, including the S\'ersic sources.
Second, we search the residual for features that appear to be sources missed by the initial fit, and either include them in the model---if the sources are bright enough to be detected in individual images or near enough to overlap with the SN, and match PSF or S\'ersic profiles---or mask them---if they are faint or not overlapping with the SN.
Finally, this process of fitting followed by inclusion/masking of residual sources is repeated a few times until a good final fit is achieved on the template stacked image. 
The results are then applied to fit the SN and nearby sources on individual images, keeping the same S\'ersic shape parameters and masks that were obtained from the final fit, with the ``forced'' MOP model (see below).
In some cases, a good fit is not achieved because (1) the residual includes non-PSF/non-S\'ersic features that overlap with the SN (e.g., host galaxies with non-S\'ersic shape) and/or (2) the SN is coincident with another source, and we then proceed with other photometry methods (Section~\ref{sec:photflux}).

\begin{figure*}[t!]
\epsscale{\scl}
\gridline{\leftfig{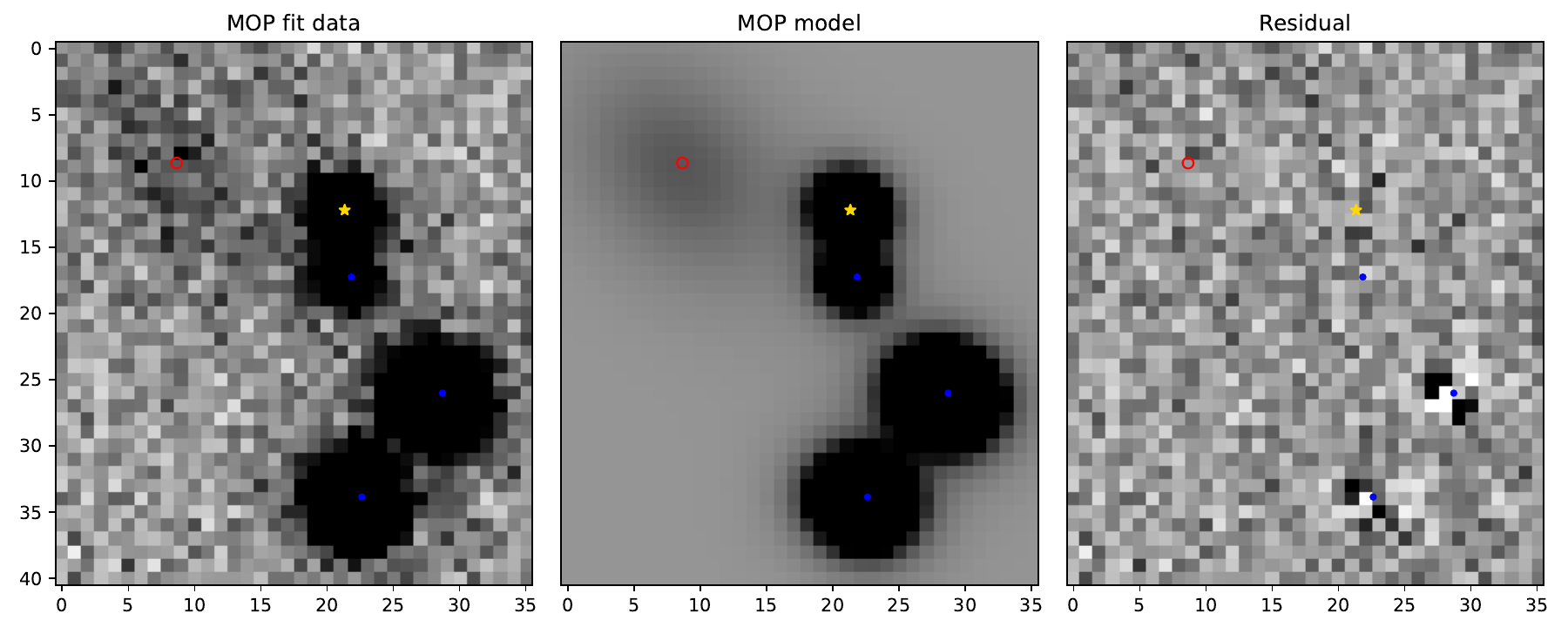}{0.7\textwidth}{}
          \rightfig{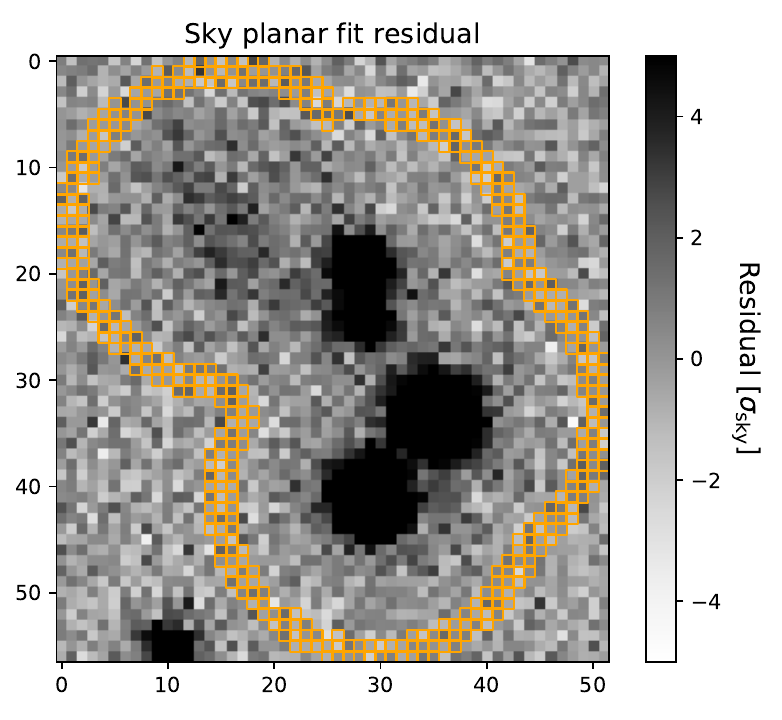}{0.3\textwidth}{}}
\vspace{-6mm}
\caption{Example of forced MOP with 15 fit parameters for a typical $B$-band image of KSN-2016ad near peak including 4 nearby sources ($N_p$ = 4 and $N_s$ = 1), fitting $A_i$ and $\Vec{x}_i$. Panels (left to right) show the (1) sky-subtracted image stamp containing the fit region $R_{\rm fit}$ with $n_{\rm ap}$ = 3; (2) best-fit (\chisqr = 1.41) MOP model; (3) residual; and (4) larger sky-subtracted stamp with the sky annulus pixels in orange outlines ($n_{\rm in}$ = 5, $n_{\rm out}$ = 6). The intensity scale (right colorbar) for all four image stamps spans a linear range of $\pm$5-$\sigma_{\rm sky}$ (= the black-/white-points) and the colored symbols represent the central positions of the SN (gold star), point sources (blue dots), and S\'ersic host galaxy (red dot).}
\label{fig:mop}
\end{figure*}

\begin{figure}[t!]
\epsscale{\scl}
\plotone{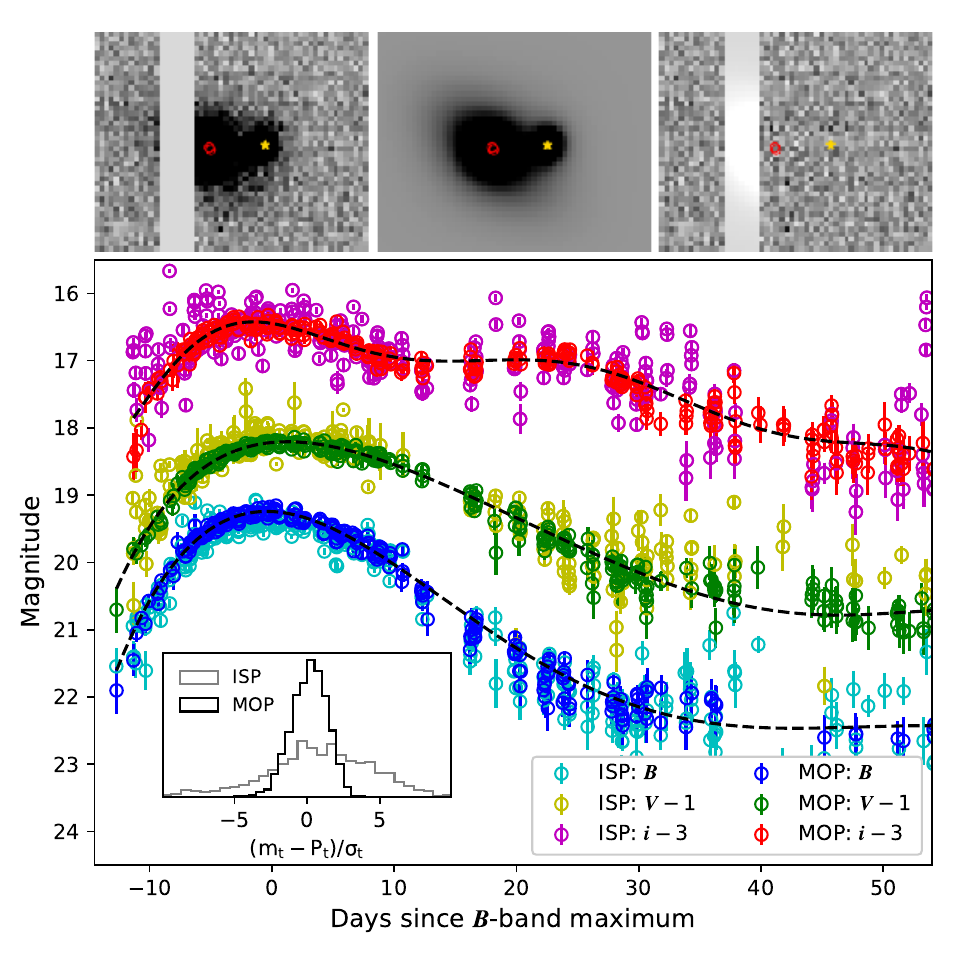}
\caption{(Top) Same as (1)--(3) panels of Figure~\ref{fig:mop}, but for KSN-2017cv
($N_p$ = 1, $N_s$ = 2), a typical SN overlapping with a luminous host galaxy, with \chisqr = 0.87. 
Note that a vertical saturation artefact is masked.
(Bottom) Comparison of light curves for KSN-2017cv obtained using two methods 
of ISP and forced MOP.
Dashed curves represent polynomial fits ($P_t$) to the MOP light curves which are used to measure the light curve ``scatter'' (=~$SD_t[(m_t - P_t)/\sigma_t]$; see text). 
Similar fits were used for the measurement of the ISP light curve scatter.
The inset shows histograms of the uncertainty-scaled residuals, $(m_t - P_t)/\sigma_t$, for the ISP (grey step curve) and MOP (black step curve) light curves.
In this case, MOP outperforms ISP by reducing the light curve scatter by $\sim$ 72\% over ISP.
\label{fig:mopdiff}}
\end{figure}

The ``forced'' MOP model with $(N_p + N_s)$ sources has either 1 or 3 fit parameters per source, depending on if the central positions $\Vec{x}_i$ are fixed or allowed to vary. 
For each image, we usually perform an initial fit with fixed $\Vec{x}_i$---obtained from image subtraction light curves of the SN and the template MOP fit---to estimate the S/N of each source. 
We then refine this fit result by relaxing $\Vec{x}_i$ for sources with high S/N ($>$ 3 for point sources and $>$ 5 for S\'ersics) in a second round fit.
Repeated convolutions in Equation~\ref{eq:mop}---the most computationally expensive step of the MOP model---are avoided by evaluating the sum of the unconvolved S\'ersics, i.e., $\sum_i^{N_s} A_i\,SER_i(\Vec{x} - \Vec{x}_i)$, over a single $n_K$-super-sampled model grid $G_{\rm fit}$, which includes $R_{\rm fit}$ and a half-kernel (see $K$ above) wide border region. 
All of the S\'ersic convolutions (Equation~\ref{eq:conv}) are evaluated in one step by multiplying the sum of unconvolved S\'ersics, i.e., $\Sigma_i^{N_s} SER_i$($G_{\rm fit}$), with the kernel in Fourier space.
Using a fast Fourier transform\footnote{\texttt{scipy.signals.fftconvolve}}, the complexity per model evaluation over $G_{\rm fit}$ of given size ($L\times L$) scales as $\mathcal{O}(L^2\log L)$.
The fit converges quite quickly, typically in fewer than 300 iterations, with KSN-2017iw ($N_s + N_p$ = 7) representing the upper extreme when fitting $\Vec{x}_i$. Applying forced MOP to a set of $\sim$ 2000 images typically takes $\lesssim$ 12 hours.

Prior to fitting the full MOP or forced MOP models, the linear sky background model ($I_{\rm sky}$; Equation~\ref{eq:sky}) is fitted to a ``multi-object annulus'' around the sources, and $\sigma_{\rm sky}$ is estimated with the RMS noise of the residual.
The ``multi-object annulus'' is defined as the union of $N_p$ + $N_s$ circular annuli spanning a radius interval from $n_{\rm in}$ to $n_{\rm out}$ times the average FWHM of the image PSF, where $n_{\rm in}$ and $n_{\rm out}$ are constant scalings for the interior and exterior radii of the annulli.
We select $n_{\rm in}$ and $n_{\rm out}$ to avoid as much flux from the SN or any other sources as possible by taking a series of annuli with adjacent pairs of $n$ in [4, 5, 6, 7, 10, 12] and adopting the pair that yields the smallest average intensity in the annulus as $n_{\rm in}$ and $n_{\rm out}$.
If this process selects the largest annulus (= [10, 12]), then an extended set of $n$ including [14, 16, 18, 20, 22] are considered.
The fit is conducted twice, rejecting 2-$\sigma$ outliers in the residual after the first fit which removes any sources that remain in the multi-object annulus. 

Figure~\ref{fig:mop} shows an example of applying forced MOP with free parameters $A_i$ and $\Vec{x}_i$ to a typical near-peak image of KSN-2016ad---involving a relative large number of sources ($N_p + N_s$ = 5) for a \tas\ in our sample---including (left to right) image stamps of the fit region, the best-fit model, the residual, and the selected sky annulus.
The residual of the fit (with \chisqr\ = 1.4) near the SN position is consistent with noise ($\sim\sigma_{\rm sky}$), indicating the SN flux is fully captured.

\subsection{Comparison with Previous Photometry Methods}\label{subsec:cophot}

Prior to MOP, photometry of KSNe 201509b, 2018ku, and 2021V were done by fitting a single PSF at the SN location---after sky subtraction as described above---when (1) the SN is located far from bright sources and (2) nearby sources are substantially below the noise floor of individual images and incapable of affecting the SN flux (this is especially important in the earliest phases when the SN itself is faint).
When detectable nearby sources affect PSF photometry, image subtraction photometry (ISP) had been performed for some KSP sources to remove the flux backgrounds
\citep[see Section~\ref{sec:photcal} and][]{Afsariardchi2019apj}.
However, the ISP process can potentially lead to non-ideal side effects, including (1) degrading the quality of target images when their seeing is better than that of the subtraction template and (2) addition of noise and non-linear residuals from subtraction, sometimes resulting in extreme photometric outliers.

Recent surveys of \tase\ for precision cosmology have adopted ``Scene Modelling Photometry'' \citep[SMP;][]{Scolnic2018apj, Brout2019apj} approaches to measuring the SN fluxes, where a temporally-varying SN flux and a temporally-fixed 2-D model of the host galaxy background are fitted to the series of target images instead of performing image subtraction. 
SMP avoids the degradation of seeing and depth caused by subtraction and has been shown to decrease the frequency of 5-$\sigma$ photometric outliers by $\sim$ 80\% compared to image subtraction \citep{Brout2019apj}.
The temporally-fixed host galaxy in SMP is modelled as a grid of fluxes in real or Fourier space and the fit is performed over a time series of image stamps.
The host galaxy model grid is usually super-sampled in order to capture flux variations at the sub-pixel level, often requiring several hundred parameters \citep[e.g., 500;][]{Brout2019apj}.
Such a fit typically requires a large number of iterations for convergence, with each iteration involving a large number of convolution steps equal to the number of images, making SMP less feasible for KSP due to the large number of high-cadence images.
MOP with fewer model parameters provides a feasible alternative used to obtain reliable photometry for KSP \tase.

We compare the light curves obtained by MOP and ISP as follows, choosing two events, KSNe-2016bo and 2017cv, for their luminous overlapping host galaxies where ISP would ordinarily be required to measure the light curves.
Typical \chisqr\ for forced MOP fits to the two SNe are $\sim$ 1.3 and $\sim$ 0.9, respectively.
Figure~\ref{fig:mopdiff} shows the results for KSN~2017cv (bottom panel) and image stamps of the fit region from a near-peak epoch (top panel).
As seen in the top panel, two S\'ersic components are used to adequately fit the core and wings of the host galaxy with saturation artifact being masked from the fit. 
The correlation between the detected $BVi$-band magnitudes obtained with MOP and ISP is 90--99\%.
We quantify the ``scatter'' in the $BVi$-band light curves ($m_t$) by fitting polynomials ($P_t$).
We define ``scatter'' as the standard deviation ($SD$) of the fit residuals scaled by the data uncertainties ($\sigma_t$), i.e., scatter~=~$SD_t[(m_t - P_t)/\sigma_t]$.
For KSN~2016bo, we find that scatter improves from 4.4 to 1.7 for MOP compared to ISP, a difference of $\sim$ 61\%, partly attributable to extreme photometric outliers being much less common in MOP.
Similarly, for KSN~2017cv, we find an improvement from 4.4 to 1.2 (see Figure~\ref{fig:mopdiff} inset), a difference of $\sim$ 72\%.

\subsection{Additional Photometric Procedures}\label{subsec:debias}

We include the following procedures in our photometry,
which is largely based on MOP but also ISP and aperture photometry methods, 
in order to have more precise measurements of the early \tas\ light curves.
First, in ISP and aperture photometry, we subtract background flux at the location of the SN arising from image subtraction residuals or background sources (including host galaxies) in the aperture.
For this, we measured the background flux at the location of the SN in pre-SN images,
and subtracted it when the background is largely constant and independent of the limiting magnitude of a given image.
(In the case where the background is correlated with the seeing, we apply the method described below).
The presence of a constant background in pre-SN images is also confirmed by examining for the following
relationship for its S/N ratio
\begin{equation}
    \ln{\rm (S/N)} = \frac{\ln(10)}{2.5}(m_{\rm lim}-m_{\rm bg}) + \ln(3)
\end{equation}
where $m_{\rm lim}$ and $m_{\rm bg}$ are the limiting magnitude and constant background magnitude, respectively.

Secondly, fitting the SN flux distribution in MOP and ISP can introduce biases in flux measurements due to the influence of flux distributions of nearby---but not necessarily coincident (as in the case above)---sources, including galaxies that 
are not perfectly modeled in MOP as well as subtraction residuals.
These imperfectly-modeled flux distributions can overlap with that of the SN, producing a biased flux measurement, with the degree of overlap dependent on the seeing conditions.
This effect (when present) is visible in pre-SN images and decreases as the image quality increases,
and we obtain a strong linear correlation between the magnitude of this artificial flux 
at the SN location and its noise.
In our SN flux measurements, 
we remove this effect by subtracting
a flux value obtained by mapping the noise of a given image to the artificial flux-noise
relationship from the pre-SN images.
We note that, after this correction, the distribution of the ratio of artificial flux-to-noise (F/N) 
at the SN location in pre-SN images is approximately symmetric, bell-curved, and homoscedastic with respect to image quality, 
which is consistent with unbiased flux measurements (see below).
For most SNe in our sample, the $SD$ of F/N is $\sim$ 1 (i.e., standard normal), 
while in a few cases we have $SD$ $\gtrsim$ 1, 
indicating slightly under-estimated photometric uncertainties.

Here, we briefly explore the origin of the F/N distribution in pre-SN images and show how a standard normal one results from unbiased measurements.
We consider the forced MOP model with $m$ = $N_s + N_p$ sources fit to a sky-subtracted region of $n$ pixels.
The model is equivalent to linear regression \citep[see][for a review]{Bishop2007book} with $m$ features (i.e, $\{g_i\}_{i=1}^m$ representing the modelled flux distribution of each $i^{\rm th}$ source) fitting the vector of observed pixel intensities $\Vec{y}$ such that we have $\Vec{y} = G \cdot \Vec{a} + \Vec{\epsilon}$, where $G$ and $\Vec{a}$ are the feature matrix and fitted parameter vector, respectively:
\begin{equation}
    \Vec{y} = \begin{bmatrix}
    I_1 \\
    \vdots \\
    I_n
    \end{bmatrix},\ \ \ G = \begin{bmatrix}
    g_1(\Vec{x}_1) & \dots  & g_{m}(\Vec{x}_1) \\
    \vdots & \ddots & \vdots \\
    g_1(\Vec{x}_n) & \dots  & g_{m}(\Vec{x}_n)
    \end{bmatrix},\ \ \ \Vec{a} = \begin{bmatrix}
    A_1 \\
    \vdots \\
    A_m
    \end{bmatrix}
\end{equation}
Note that $A_i$, representing the fluxes of each source in MOP (Appendix~\ref{subsec:mopmod}), are the only fit parameters of the model.
The noise term $\Vec{\epsilon}$ is i.i.d. white noise $\Vec{\epsilon_0} \sim \mathcal{N}_n(0, I_n)$ scaled by the data uncertainties $\Sigma = \diag(||I_i||+\sigma_{\rm sky}^2)$, i.e., $\Vec{\epsilon} = \Sigma^{1/2} \Vec{\epsilon_0}$.
The estimated fluxes $\hat{a}$ and their covariance matrix $\Sigma_{\hat{a}}$ are
\begin{align}
    \hat{a} &= (G^{\prime T}G^{\prime})^{-1}G^{\prime T}\Vec{y^{\prime}}\\
    \Sigma_{\hat{a}} &= (G^{\prime T}G^{\prime})^{-1}\chi^2_R
\end{align}
where $G^{\prime} = \Sigma^{-1/2}G$ and $\Vec{y^{\prime}} = \Sigma^{-1/2}\Vec{y}$ are whitened features and data and $\mathbb{E}[\chi^2_R] = 1$ \citep[Cochran's theorem;][]{Cochran1934pcps}.
If the model is correct and $\hat{a}_0$ are the true fluxes (i.e., the data is generated by $\Vec{y} \equiv G \cdot \hat{a}_0 + \Vec{\epsilon}$), then
\begin{align}
    \hat{a} &= (G^{\prime T}G^{\prime})^{-1}G^{\prime T} (G^{\prime}\hat{a}_0 + \Vec{\epsilon_0})\\
    &= \hat{a}_0 + (G^{\prime T}G^{\prime})^{-1}G^{\prime T}\Vec{\epsilon_0}
\end{align}
means that the measurement is unbiased on average, since $\mathbb{E}[\hat{a}] = \hat{a}_0$. Further, the artificial F/N is equal to
\begin{align}
    {\rm F/N} = \Sigma_{\hat{a}}^{-1/2}(\hat{a}-\hat{a}_0) &= \frac{1}{\sqrt{\chi^2_R}}\hat{P}^{\prime}\Vec{\epsilon_0} \sim \mathcal{N}_m(0, I_m)\\
    {\rm where}\ \ \mathbb{E}\left[\frac{1}{\sqrt{\chi^2_R}}\right] &= \sqrt{\frac{n-m}{2}}\ \frac{\Gamma(\frac{n-m-1}{2})}{\Gamma(\frac{n-m}{2})}\ \ ;
\end{align}
$\hat{P}^{\prime} = (G^{\prime T}G^{\prime})^{-1/2}G^{\prime T}$ is an orthonormal matrix (i.e., $\hat{P}^{\prime}\hat{P}^{\prime T} = I_m$)---interpreted as a whitened projection onto features; and $\mathbb{E}[1/\sqrt{\chi^2_R}]$ converges to 1$^+$ in the limit of large $n-m$. 
Thus, standard normal F/N (measurable in pre-SN images since $\hat{a}_0$ = 0), is the expected result of unbiased photometry.

\section{Standardization: Template Fitting Distance and S/K--correction}\label{sec:standard}

\begin{figure}[t!]
\epsscale{\scl}
\plotone{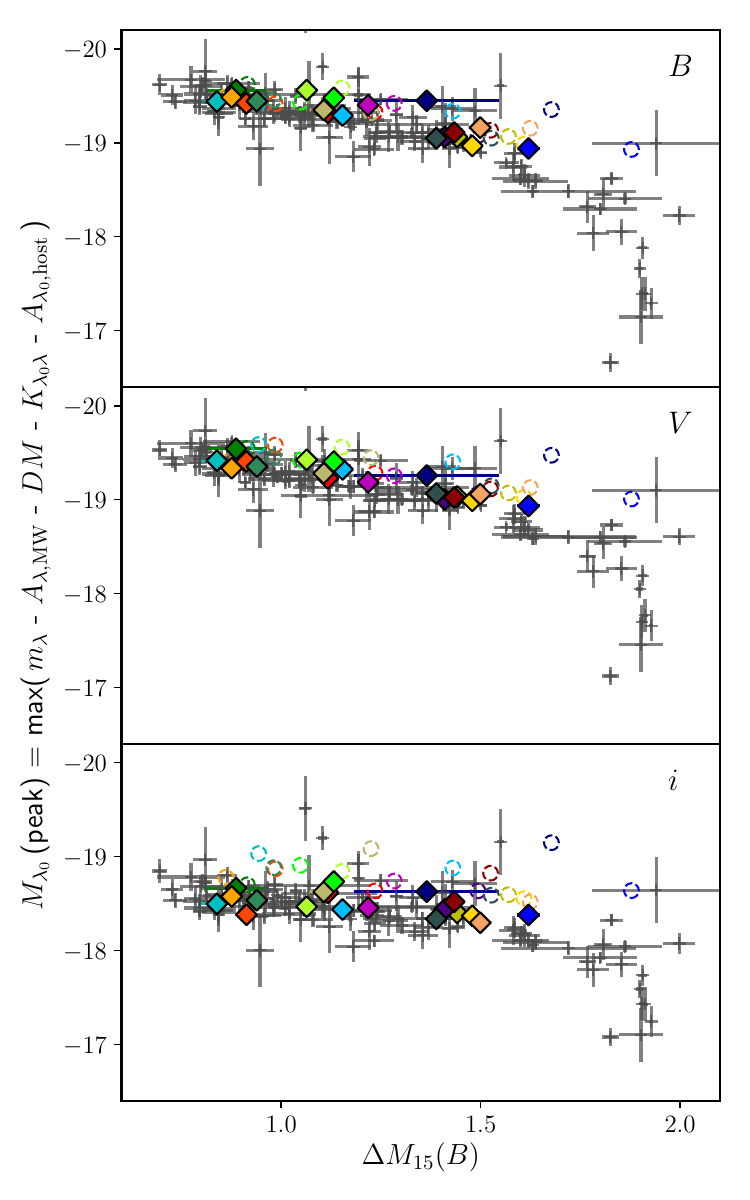}
\caption{Peak absolute magnitudes, $M_{\lambda 0}$\,(peak) in $BVi$ bands (top to bottom) and Phillips parameters (\dm15) of the standardized rest-frame light curves of the KSP \tase\ 
(black-bordered diamonds, colored as in Figure~\ref{fig:ksplc}) obtained with Equation~\ref{eq:standard} compared to the Phillips diagram of a sample of \tase\ \citep{Burns2018apj}.
The 19 KSP \tase\ show apparent consistency with the Phillips relation in all three bands.
The same SNe but without K--corrections, represented by the same-colored dashed open circles, 
show the magnitude of the correction effect.
\label{fig:phillipsband}}
\end{figure}

The light curves of normal \tase\ in the photospheric phase from $-$10 to 15 days since peak are known to form a family of functions parameterized by the Phillips parameter, \dm15, which measures the post-peak decline rate in the $B$-band light curve \citep{Phillips1999aj}.
To effectively include some peculiar Type Ia SNe, such as the rapidly evolving 91bg-like subtype, the color stretch parameter, \sbv, has been used as an alternative to \dm15, defined by \sbv\ = $t_{BV}$/(30 days), where $t_{BV}$ is the time between $B$-band peak to the maximum post-peak \bv\ color \citep{Burns2014apj}.
Our method for light curve standardization is based on determining the best-fit \sbv-parameterized normal \tas\ template in the following sequence of steps.

First, we perform an initial template fit to the $V$ and $i$ band light curves which is used to measure S--corrections for the $B$-band instrumental magnitudes (Section~\ref{sec:photcal}).
For the template model, we used $\texttt{EBV\_model2}$ of \texttt{SNooPy} \citep{Burns2011aj}, which is based on transforming the normal \tas\ template of \citet{Hsiao2007apj} by the fit parameters $DM$, $E(\bv)_{\rm host}$, and $s_{BV}$, adopting $z_{\rm host}$ and $E(\bv)_{\rm MW}$ (Table~\ref{tab:snparam}) as fixed inputs.
For all 16 previously unpublished \tase, we obtain best-fit \chisqr\ in the range of 1--15 \citep[see][for the fit results of the 3 published events]{Moon2021apj, Ni2022natas, Ni2023bapj}.
$s_{BV}$ of the best-fit templates ranged in 0.8--1.1 for the SNe, well within expected values for normal \tase\ \citep{Burns2011aj}, while $DM$ and $E(\bv)_{\rm host}$ were consistent with Hubble flow distances and spectra, respectively (see Section~\ref{sec:restlc}). 
Note that in some cases where the initial fit found $E(\bv)_{\rm host} <$ 0, which confirms that no physical amount of extinction can improve the fit, we performed a second fit where we fixed $E(\bv)_{\rm host}$ = 0.

Second, we perform a template fit to all three $BVi$ filters with the same model after applying the S-correction, obtaining estimates of $DM$ and $E(\bv)_{\rm host}$ for the SNe (Table~\ref{tab:snparam}) and best-fit templates for measuring $BVi$-band K--corrections.
The best-fit \sbv\ for these templates range in 0.8--1.1 with \chisqr\ ranging in 1--12.
We test the K--corrected rest-frame light curves (Equation~\ref{eq:standard}) by measuring the peak absolute magnitudes $M_{\lambda_0}$ and \dm15\ of the SNe before and after applying K--corrections with polynomial fits (see Section~\ref{subsec:nonearly}), comparing the results with the Phillips relation \citep{Phillips1999aj} formed by a large sample of \tase\ \citep{Burns2018apj}.
Figure~\ref{fig:phillipsband} shows that after applying K--correction, all SNe follow the Phillips relation closely in all three $BVi$ bands, while their \dm15\ values ranging in 0.84--1.50 mag also corresponds to an expected \sbv\ range of 0.77--1.07 \citep[based on Equation 4 of][]{Burns2014apj} which shows consistency with the \sbv\ of the best-fit templates, supporting the validity of the K--corrections.

Finally, we attempt to repeat the template fitting steps except (1) using a 91bg-like template for the cases with $s_{BV}\sim$ 0.8; and (2) using a fixed $DM$ = $DM_{\rm cosmo}(z_{\rm host})$ with the normal template.
For the former, we obtained worse fits with much larger \chisqr\ and $s_{BV}$ that were inconsistent with 91bg-like events.
For the latter,
we obtain worse \chisqr\ in all cases and unreasonably large $E(\bv)_{\rm host}$ values for KSNe-2016M, 2017iw, 2018oh, 2019dz, 2021iq and 202112D in the range of 0.1--0.3, inconsistent with either non-detections of their \nai~D spectral features or small $E(\bv)_{\rm host}$ values measured from them (Table~\ref{tab:spec}).
In the cases of KSNe-2017iw and 2018oh, the $E(\bv)_{\rm host}$ were also less compatible with their locations within the halos of their host galaxies (Table~\ref{tab:hosts}).
These results support (1) that the SNe are clearly better matched to normal templates; and (2) the use of the template-fitted $DM$ over the Hubble flow estimates.

\section{Excess Emission Characterization: Regularized Gaussian Method} \label{sec:reggaus}

Our search for excess emission in early \tas\ light curves and characterization of the excess emission properties is conducted as follows.
First, we fit a pure power-law (Equation~\ref{eq:pow}) to the rest-frame early light curve up to $\sim$ 40\% of $B$-band peak flux iteratively---with fit interval start times ($t_{\rm iter}$) ranging from 25 days before $B$-band maximum to $\sim$ 3--4 days before the end of the interval, depending on how many data points are available. 
This method is often used to estimate the onset of the power-law rise ($t_{\rm PL}$) in \tase\ while accommodating the possibility of early excess emission \citep[e.g.,][]{Olling2015nat, Dimitriadis2019apj}.
We also investigate whether the light curve can accommodate a ``forced'' early excess emission by fitting a power-law $+$ Gaussian (Equations~\ref{eq:pow}+\ref{eq:gex}), which has often been used to characterize observed early excess emissions in \tase\ \citep[e.g.,][]{Ni2023bapj, Ni2023apj, Wang2024apj}.
For the power-laws and Gaussians, the observed flux normalized to the peak flux in each band ($\lambda$) is modelled as
\begin{align}
    PL_{\lambda}(t) &= \mathcal{H}(t - t_{\rm PL})\ C_{\lambda}(t - t_{\rm PL})^{\alpha_\lambda} 
    \label{eq:pow}\\
    GE_{\lambda}(t) &= \frac{X_{\lambda}}{\sqrt{2\pi\sigma_X^2}}\exp(-\frac{t-\mu_X}{2\sigma_X^2})
    \label{eq:gex}
\end{align}
where $\mathcal{H}$ is the Heaviside step function, $t$ represents time in rest-frame days since $B$-band maximum, ($t_{\rm PL}$, $\alpha_{\lambda}$, $C_{\lambda}$) represents the power-law onset, indices, and coefficients, respectively, and ($\mu_X$, $\sigma_X$, $X_{\lambda}$) do the Gaussian excess emission mean and standard deviation (days), and normalized time-integrated energy density (days; ergs~Hz$^{-1}$ divided by peak flux in ergs~day$^{-1}$~Hz$^{-1}$).
The pure power-law therefore has 7 fit parameters while the Gaussian adds 5 more.

\begin{figure}[t!]
\epsscale{\scl}
\plotone{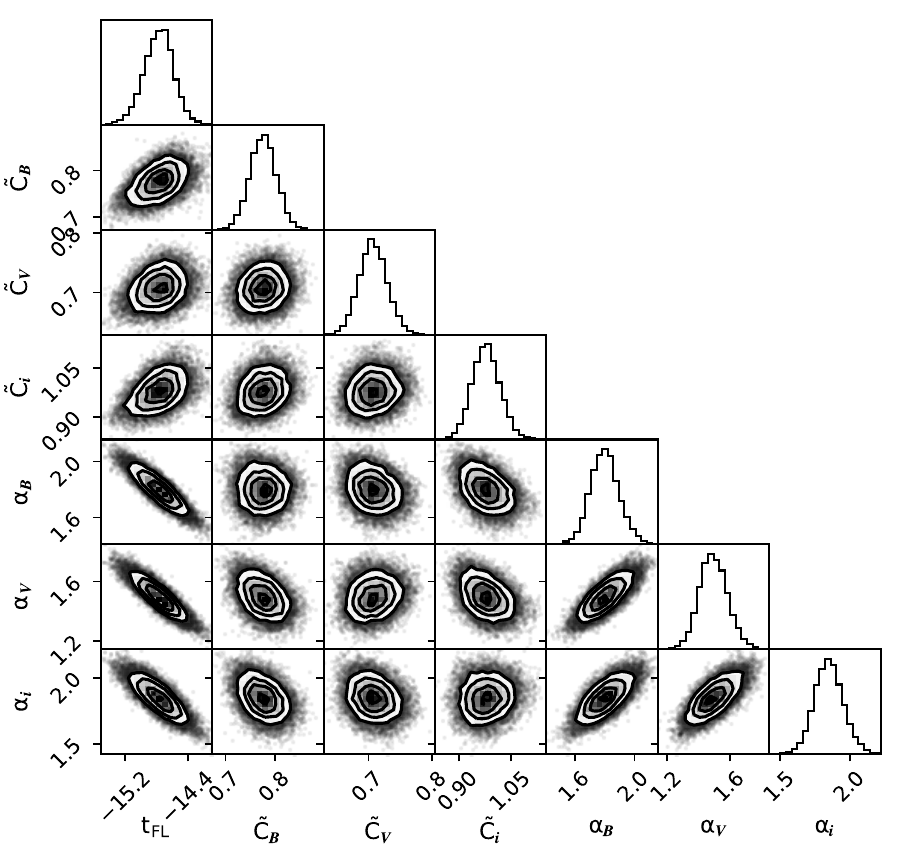}
\caption{Joint probability distributions of the transformed fit parameters of the power-law fit (Equation~\ref{eq:pow}; see text) for a typical SN with no detected excess emission, KSN-2017iw, sampled with MCMC. Comparison of the best-fit power-law model to the early light curves of KSN-2017iw is shown in Figure~\ref{fig:powfit}.
\label{fig:powmc}}
\end{figure}

We obtain the optimal $t_{\rm iter}$ resulting in the lowest \chisqr\ that gives a reasonable $t_{\rm PL}$ for all cases except for KSNe~2017fo and 2019bl, which were less suitable for the iterative method due to having sparse data points and excessive noise, respectively, in their early light curves.
We also establish a ``forced'' detection of excess emission indicating that the light curve accommodates an excess emission if the Gaussian component of the power-law $+$ Gaussian fit (1) has positive S/N---i.e., the errorbar-weighted mean of $X_{\lambda}$ has S/N $>$ 1; and, when $t_{\rm iter}$ is available, (2) consists of data points that indeed deviate from power-law rise as indicated by consistency between the epochs of the Gaussian component and $t_{\rm iter}$---i.e., $|\mu_X - t_{\rm iter}| < 3\,\sigma_X$.
Table~\ref{tab:powfit} (column 2) includes all of the ``forced'' detections established with these criteria, though two of them, KSNe-2016bo and 201901ah, are found to be spurious upon further investigation (see below).
Absent such a ``forced'' detection, or in the case of a spurious one, we proceed to estimate $t_{\rm PL}$ with a pure power-law fit to the entire early light curve, obtaining the best-fit parameters in Table~\ref{tab:powfit} (columns 3 and 4).
For one case of KSN~2017fo with sparse data points in the early light curve, we obtained reliable results by augmenting with Gaussian process interpolated data points every 8 hrs (see Appendix~\ref{sec:gpint}).

The best-fit is obtained by conducting a Levenberg-Marquardt minimization of $\chi^2$, estimating 1-$\sigma$ errorbars on the parameters with bootstrap.
We also explore degeneracies between the parameters by using the affine-invariant, Markov Chain Monte Carlo ensemble sampler \texttt{emcee} \citep{Foreman-Mackey2013pasp} to approximate their joint probability distribution \citep{Goodman2010camcs}.
The fit is shown for KSN-2017iw (Figure~\ref{fig:powfit}), a typical example of a SN that does not admit a ``forced'' detection of excess emission.
Figure~\ref{fig:powmc} presents the probability distribution sampled with MCMC for pairs of parameters in this fit.
All of the parameters are well-constrained, though there is some degeneracy between $\alpha_{\lambda}$ and $t_{\rm PL}$ leading to the relatively large uncertainty noted in Section~\ref{sec:res} associated with estimating the epoch of first light using power-law fits.
Note that we applied the change of variables $\Tilde{C_{\lambda}} = C_{\lambda} 10^{\alpha_{\lambda}}$ such that the effects of $\Tilde{C_{\lambda}}$ and $\alpha_{\lambda}$ decouple towards the end of the fit interval where the S/N of the data is highest, which has been found to reduce the strong degeneracy between $C_{\lambda}$ and $\alpha_{\lambda}$ \citep{Miller2020apj}.

\begin{deluxetable*}{lc|cc|ccc|c|c}
\tabletypesize{\footnotesize}
\tablecolumns{9} 
\tablewidth{0.99\textwidth}
 \tablecaption{Power-law and excess emission fit parameters.}
 \tablehead{
 \colhead{KSN} & \colhead{Forced $GE{\rm ^a}$ [$t_{>X}$]} & \colhead{$t_{\rm PL}$} & \colhead{$\alpha_{(BVi)}$} & \colhead{$\mu_X$} & \colhead{$\sigma_X$} & \colhead{$X_{(BVi)}$} & \colhead{$E_{X, BVi}$} & \colhead{\chisqr}
 } 
\startdata
2017gp & Yes \textbf{[$-$14.0]} & \textbf{$-$16.19 $\pm$ 1.15} & (1.73, 1.47, 1.59) & $-$15.64 & 0.97 & (0.089, 0.058, 0.056) & 5.78 $\times$ 10$^{46}$ ergs & 1.0 \\
2016M & Yes \textbf{[$-$11.2]} & \textbf{$-$15.95 $\pm$ 2.18} & (1.43, 1.15, 1.29) & $-$13.27 & 0.51 & (0.070, 0.065, 0.069) & 5.04 $\times$ 10$^{46}$ ergs & 1.1 \\
2016ad & Yes \textbf{[$-$14.5]} & \textbf{$-$16.76 $\pm$ 0.28} & (1.60, 1.58, 1.57) & $-$15.82 & 0.57 & (0.041, 0.057, 0.074) & 4.29 $\times$ 10$^{46}$ ergs & 1.2 \\
2017iw & No & $-$14.84 $\pm$ 0.20 & (1.83, 1.53, 1.89) & & & & & 2.2 \\
2016bo  & (Yes) & $-$13.25 $\pm$ 0.34 & (1.34, 1.03, 1.36) & & & & & 1.6 \\
2017cv & No & $-$14.55 $\pm$ 1.03 & (1.96, 1.47, 2.21) & & & & & 1.1 \\
2018oh & No & $-$15.97 $\pm$ 0.36 & (2.41, 2.07, 2.37) & & & & & 1.5 \\
2017cz & No & $-$18.14 $\pm$ 1.19 & (2.98, 2.45, 2.66) & & & & & 2.0 \\
2017fo & No & $-$15.61 $\pm$ 0.60 & (1.32, 1.31, 1.04) & & & & & 0.1 \\
2019bl & No & $-$16.36 $\pm$ 0.31 & (2.92, 1.69, 2.49) & & & & & 14.2 \\
2018ng & No & $-$14.46 $\pm$ 1.09 & (1.88, 1.68, 1.22) & & & & & 4.6 \\
201903ah & (Yes) & $-$17.11 $\pm$ 1.0 & (1.83, 1.49, 1.79) & & & & & 1.4 \\
2019dz & No & $-$14.92 $\pm$ 0.92 & (1.63, 1.20, 1.18) & & & & & 1.0 \\
2016bu & No & $-$16.36 $\pm$ 0.32 & (2.24, 1.83, 1.51) & & & & & 1.3 \\
2021iq & Yes \textbf{[$-$13.9]} & \textbf{$-$14.91 $\pm$ 1.12} & (1.22, 1.23, 0.90) & $-$14.11 & 0.23 & (0.045, 0.069, 0.058)
& 4.58 $\times$ 10$^{46}$ ergs & 2.4 \\
202112D & Yes \textbf{[$-$15.6]} & \textbf{$-$16.71 $\pm$ 0.54} & (1.86, 1.58, 1.76) & -16.34 & 0.38 & (0.026, 0.054, 0.046) & 3.47 $\times$ 10$^{46}$ ergs & 4.5\\
\hline
SN 2017cbv${\rm ^b}$ & & $-$19.79 & (1.78, 1.68, 1.58) & $-$17.24 & 1.20 & (0.079, 0.064, 0.073) &
\enddata
\tablenotetext{{\rm a}}{This column is for ``forced'' detection'' of Gaussian excess emission (see Appendix~\ref{sec:reggaus}), where ``Yes'' means a ``forced'' detection from the initial power-law $+$ Gaussian fit while brackets indicate the cases where we found that this ``forced'' detection was likely spurious.}
\tablenotetext{{\rm b}}{Results of fitting an unregularized power-law $+$ Gaussian to the early fluxes of SN~2017cbv until $\sim$ 40\% of peak $B$-band flux.}
\tablecomments{Bold font means $t_{\rm PL}$ (= ``first light'') is obtained from conducting a pure power-law fit to the restricted early light curve starting from $t_{>X}$ while the remaining parameters are from a regularized power-law $+$ Gaussian fit (Appendix~\ref{sec:reggaus}).}
\end{deluxetable*} 
\label{tab:powfit}

For the cases where we obtained a ``forced'' detection of excess emission, we estimate $t_{\rm PL}$ by performing a second pure power-law fit to the early light curve, excluding the data points within 1.5-$\sigma_X$ of $\mu_X$---which is expected to contain $\sim$ 90\% of the possible excess emission flux.
This results in the early data points within a $<$ 3.4-day interval being excised from the fit.
In two cases, KSNe-2016bo and 201903ah, this second fit resulted in extreme parameters for normal \tase\ \citep[e.g., see][]{Miller2020apj}, including small power-law indices $\lesssim$ 1.0 and a very short rise time of 12.3 days for 2016bo.
By not excluding the early data points, we found much more reasonable parameters for these two SNe (Table~\ref{tab:powfit}, columns 3 and 4), indicating the data points were consistent with pure power-law rise and a spurious detection of excess emission as noted above.
The estimated $t_{\rm PL}$ for the SNe with non-spurious ``forced'' detections are presented in Table~\ref{tab:powfit} (columns 3) with bold font to differentiate them from the pure power-law fit results above.

The five SNe with non-spurious ``forced'' detections above accommodate early excess emission in their light curves.
Visual inspection confirms the reality of the excess emissions
in the five SNe based on the presence of a decline, plateau, or change in the slope of the light curve rise spanning multiple epochs and/or bands which deviates from pure power-law rise (see Figures~\ref{fig:Bsplit}--\ref{fig:Isplit}), though all of the excess emissions are clearly under-sampled with only 1--2 data points in each band.
We recover the power-law indices along with an estimate for the excess emission brightness of each SN by fitting a regularized power-law $+$ Gaussian to the entire early light curve as follows, adopting $t_{\rm PL}$ obtained above as a fixed parameter.
The loss function for the observed data consisting of time, normalized fluxes, and normalized flux uncertainties ($t_i$, $F_{\lambda, i}$, $\sigma_{\lambda, i}$) in each band ($\lambda$) for this model is
\begin{multline}
    \mathcal{L}_{\lambda} = \frac{1}{2}\sum_i\left(\frac{F_{\lambda, i} - (PL_{\lambda}+GE_{\lambda})(t_i)}{\sigma_{\lambda, i}}\right)^2 \\ - (a-1)\ln(X_{\lambda}) + (b-1)\ln(1-X_{\lambda})
    \label{eq:reggaus}
\end{multline}
equal to the negative log posterior probability of the power-law $+$ Gaussian model with a regularization term corresponding to setting a beta distribution prior on the time-integrated excess emission energy density, i.e., $X_{\lambda} \sim {\rm Beta}(a,b)$, with shape parameters $a$ and $b$.
The prior mitigates (1) degeneracy between $\mu_X$ and $\sigma_X$ when there is a data gap before the first detection; and (2) overfitting a small number of excess emission data points by applying prior information from \tase\ with very well-sampled excess emission light curves (see below).
The choice of prior is based on considering the detection of an excess emission photon in any \tas\ for every photon detected at peak as a Bernoulli trial, such that the excess emission photon count in a given band is a binomial distributed random variable, i.e., ${\rm Bin}(n_p, X_{\lambda})$.
The parameters of the distribution are $n_p$, representing the total number of peak photons, and $X_{\lambda}$, which associates with the unknown probability of detecting an excess emission photon per photon detected at peak in the same band.
The natural choice of prior for $X_{\lambda}$ under this distribution is ${\rm Beta}(a,b)$ \citep{Gelman2014book}.
Note that ${\rm Bin}(n_p, X_{\lambda})$ converges to the assumption of Poisson statistics for the excess emission photon count in the limit of large $n_p$ and small $X_{\lambda}$.

\begin{figure}[t!]
\epsscale{\scl}
\plotone{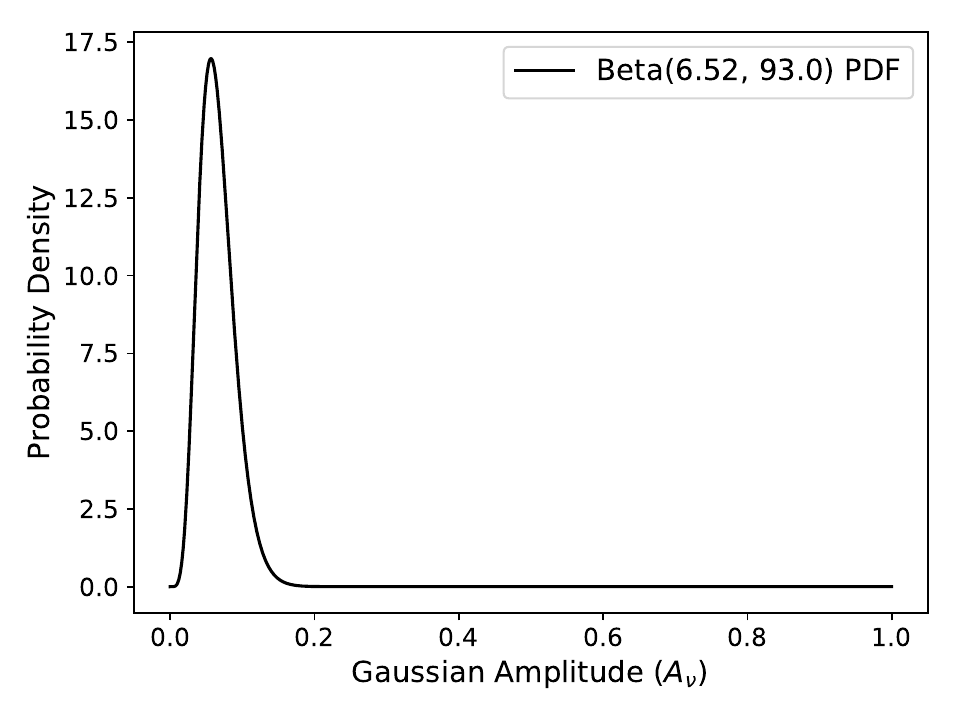}
\caption{Beta($a, b$) distribution PDF used as the prior for Gaussian amplitudes $X_{\lambda}$ in the regularized power-law $+$ Gaussian model (Equation~\ref{eq:reggaus}) for SNe with excess emission.
\label{fig:exbeta}}
\end{figure}

\begin{figure}[t!]
\epsscale{\scl}
\plotone{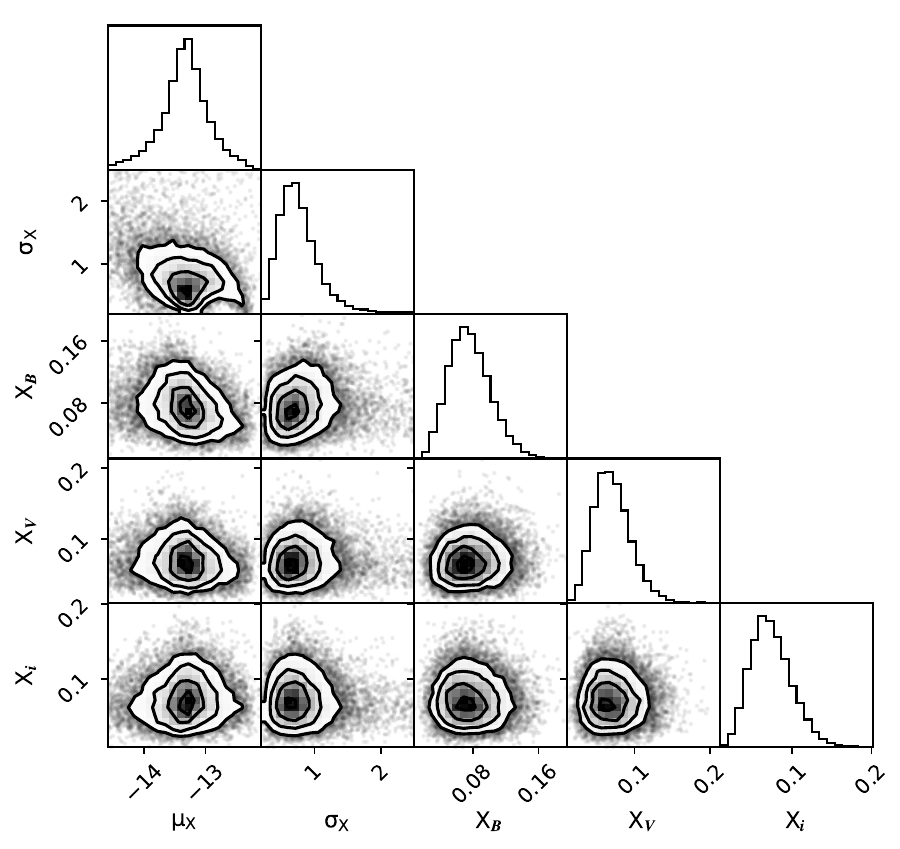}
\caption{MCMC-sampled joint posterior probability distributions of the subset of fit parameters of Equation~\ref{eq:gex} describing the Gaussian component for a typical SN with detected excess emission, KSN-2016M.
Comparison of the best-fit power-law $+$ Gaussian model to the early light curves is shown in Figure~\ref{fig:reggausfit}.
\label{fig:reggausmc}}
\end{figure}

We estimate $a$ and $b$ based on well-sampled light curves of observed excess emissions in normal \tase\ whose properties can conceivably explain the ``forced'' detections.
SNe~2021aefx \citep{Ni2023bapj}, 2017cbv \citep{Hosseinzadeh2017apj}, 2023bee \citep{Wang2024apj}, and 2018oh \citep{Dimitriadis2019apj} provide 4 measurements of $\langle X_{\lambda}\rangle_{BVi}$ for which we minimize the negative log likelihood
\begin{equation}
    \mathcal{L}_{\beta} = -\sum_{\rm SNe}(a-1)\ln\langle X_{\lambda}\rangle_{BVi} + (b-1)\ln(1-\langle X_{\lambda}\rangle_{BVi})
\end{equation}
yielding $a$ = 6.52 and $b$ = 93.0. 
Note that we obtained the best-fit $X_{\lambda}$ for SN~2017cbv, which are provided in Table~\ref{tab:powfit} (row 17), by fitting its early $BVi$-band fluxes until $\sim$ 40\% of its peak $B$-band flux with an unregularized power-law $+$ Gaussian (= Equation~\ref{eq:pow}+\ref{eq:gex}).
The prior PDF for $X_{\lambda}$ with these parameters is shown in Figure~\ref{fig:exbeta}.

Table~\ref{tab:powfit} (columns 4-9) presents the results of fitting the regularized Gaussian $+$ power-law model, along with the total energy attributable to the excess emission in optical ($BVi$) bands, $E_{X, BVi}$, estimated with a trapezoidal sum of the $BVi$-band time-integrated luminosities with distance moduli from Table~\ref{tab:snparam}, assuming spherical symmetry.
The $E_{X, BVi}$ thus obtained range in (3.5--5.8) $\times$ 10$^{46}$ ergs, corresponding to $\sim$ 0.1\% of typical total radioactive energy output for a \tas.
For one case of KSN-2016M, the 1.5-$\sigma$ cutoff applied to the early light curve for fitting $t_{\rm PL}$ above resulted $\mu_X - t_{\rm PL} > 3\sigma_X$, which is less compatible with what has been observed in other normal Type Ia SNe, such as those used to derive the prior PDF above as well as KSN~2018ku \citep{Ni2023bapj}, all of whose excess emissions overlap $t_{\rm PL}$ to within $3\sigma_X$. 
Thus, for this SN, a 2.0-$\sigma$ cutoff was applied instead, resulting in better consistency of $\mu_X - t_{\rm PL} = 2.6\sigma_X$ (Table~\ref{tab:powfit}) for the recovered Gaussian excess emission with the other \tase.
The regularized power-law $+$ Gaussian fit is shown for KSN-2016M (Figure~\ref{fig:reggausfit}) as a typical case of a SN with excess emission.
The joint distributions of the Gaussian parameters (Equation~\ref{eq:gex}) for this fit are shown in Figure~\ref{fig:reggausmc}, where all parameters appear to be well-constrained and non-degenerate as a result of the regularization.

\section{Synthetic KSP Photometry of SN~2011fe} \label{sec:synph}

We perform synthetic photometry on the flux calibrated spectral time series of SN 2011fe \citep{Pereira2013aa} to obtain early $BVi$-band light curves mimicking KSP observations.
We used \texttt{synphot} \citep{synphot} to integrate the product of each spectrum with the corresponding KSP filter transmission curves, obtaining $BVI$-band magnitudes.
We mimick the KSP $I$-band calibration to Sloan $i$' magnitudes by adding the S--correction $\Delta m = m_{i^{\prime}} - m_{I}$, based on the difference in synthetic magnitudes between the two filters for the Hubble Space Telescope Vega spectrum \citep{Bohlin2014aj} available from the CALSPEC database\footnote{\url{http://ssb.stsci.edu/cdbs/calspec/alpha_lyr_stis_008.fits}}.
We validate the results by comparing them to real observations of SN~2011fe \citep{Guillochon2017apj}, as shown in Figures~\ref{fig:Bsplit}--\ref{fig:Isplit} (bottom left panels), finding an excellent agreement between the synthetic and observed photometry.
Note that $i$-band magnitudes in the figure are derived from $I$-band observations, applying S--corrections from the spectral series with $<$ 0.5-day extrapolations by a polynomial fit.

\begin{figure}[t!]
\epsscale{\scl}
\plotone{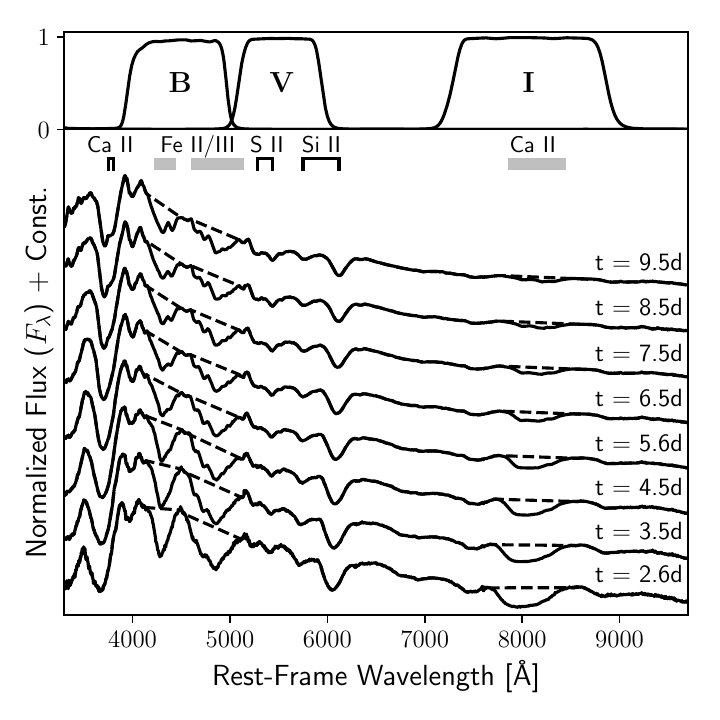}
\caption{(Top) The KSP $BVI$ filter transmission curves \citep{Kim2016jkas}. (Bottom) The spectrophotometric time series of SN~2011fe \citep[black solid curves;][]{Pereira2013aa} until 10 rest-frame days since first light \citep[= MJD 55796.687;][]{Nugent2011nat} with spectral epochs on the right. The locations of prominent spectral features are marked at the top of the panel.
The dashed lines show the removal and interpolation of three absorption features affecting early \tas\ color evolution in the KSP filters: 
$B$-band features mainly attributable to \fex\ with contribution from \mgx, and the \caii\ NIR triplet in the $I$ band \citep{Parrent2014apss}.
\label{fig:synph}}
\end{figure}

The spectra of SN~2011fe from before 10 days since its epoch of first light---MJD 55796.687 based on a similar pure power-law fit \citep{Nugent2011nat} as our method (Appendix~\ref{sec:reggaus})---are shown in Figure~\ref{fig:synph} (bottom panel) compared to the KSP filter transmission curves (top panel).
We examine how \fex\ and \caii\ lines alter the continuum colors of early-red \tase\ in the KSP filters by masking the corresponding absorption features of SN~2011fe, obtaining pseudo-continuum spectra. 
Taking differences between the synthetic colors obtained from the original and ``pseudo-continuum'' spectra, we derive the color change produced by the lines ($\Delta \Vec{C}_{\rm 11fe}$; called ``line coloration'' in Section~\ref{sec:lines-red}) at a range of epochs, which corresponds to various photosphere depths relative to the distribution of nuclear burning products in the ejecta.
The resulting $\Delta \Vec{C}_{\rm 11fe}$ for the 3.5-day spectrum, in particular, appears to match the direction of the approximately linear distribution formed by the observed colors of early-red events at the similar phase of 3.2 days since first light excellently (see Section~\ref{sec:lines}), indicating that the observed colors could be due to suppression by \fex\ and \caii\ lines in a similar ratio as in SN~2011fe.

\section{Hierarchical Gaussian Class-Conditional Model} \label{sec:hgaus}

The distribution of early \tas\ color observations (\bv\ and \vi) in Section~\ref{subsec:popsep} and their covariance matrices ($\Sigma$; see Appendix~\ref{sec:gpint}) is modelled as the random vector $X = (\vi, \bv)$ with the hierarchical distribution
\begin{align}
    X|_{\mu, \Sigma} &\sim \mathcal{N}(\mu, \Sigma)\\
    \mu &\sim \mathcal{N}(\mu_0, \Sigma_\mu) \label{eq:mugaus}\\
    \Sigma &\sim \mathcal{W}^{-1}(\nu, \Psi) \label{eq:iw}
\end{align}
where $\mu$ represents the intrinsic color of a SN; $\mu_0$ does the mean of the intrinsic colors; $\Sigma_\mu$ and $\Sigma$ represent the covariance matrices of the intrinsic SN color and the measurement process, respectively, as independent Gaussian sources of randomness; and $\mathcal{W}^{-1}(\nu, \Psi)$ in Equation~\ref{eq:iw} is the inverse Wishart distribution \citep{Wishart1928} with $\nu$ degrees of freedom and scale matrix $\Psi$.
Note that under this model, $X$ is marginally normal given $\Sigma$, with
\begin{equation}
    X|_{\Sigma} \sim \mathcal{N}(\mu_0, \Sigma_\mu) + \mathcal{N}(0, \Sigma) = \mathcal{N}(\mu_0, \Sigma_\mu + \Sigma) \label{eq:hgaus}
\end{equation}
while the full marginal distribution of $X$ is the sum of a normal distribution and a bi-variate (i.e., $k = 2$) t-distribution \citep{Murphy2012} with the following parameters:
\begin{equation}
    X \sim \mathcal{N}(\mu_0, \Sigma_\mu) + {\it t}_{\nu-k+1}(0, \frac{\Psi}{\nu - k + 1})
\end{equation}

The choice of models for $\mu$ and $\Sigma$ are based on the following assumptions.
For $\mu$, Equation~\ref{eq:mugaus} models the intrinsic colors of early \tase\ as being approximately normally distributed around a similar color $\mu_0$.
Since the distribution of \tas\ colors is seen to be potentially separable into distinct populations (Section~\ref{subsec:popid}), the validity of Equation~\ref{eq:mugaus} is conditional on the SNe being from a single population.
For $\Sigma$, the inverse Wishart distribution (Equation~\ref{eq:iw}) is the usual conjugate prior for the covariance of a multivariate normal distribution. 
Note that the Wishart distribution generalizes the scaled $\chi^2$ distribution to random matrices, and the choice of prior is also equivalent to assuming that the precision $P = \Sigma^{-1}$ is Wishart distributed. 
The $\chi^2$ distribution for precision arises from considering the observation as originating from a measurement process averaging $\nu$ independent sources of error $\Sigma_i$, for which the variance of the maximum likelihood estimator is $\Sigma = [\Sigma_{i=1}^\nu (\Sigma_i^{-1})]^{-1}$.
Therefore, the assumption of the precision $\Sigma^{-1} = P = \Sigma_{i=1}^\nu (P_i)$ being $\chi^2$ distributed is the same as assuming that the positive semi-definite matrices $P_i$ are squared normal.

\begin{figure}[t!]
\epsscale{\scl}
\plotone{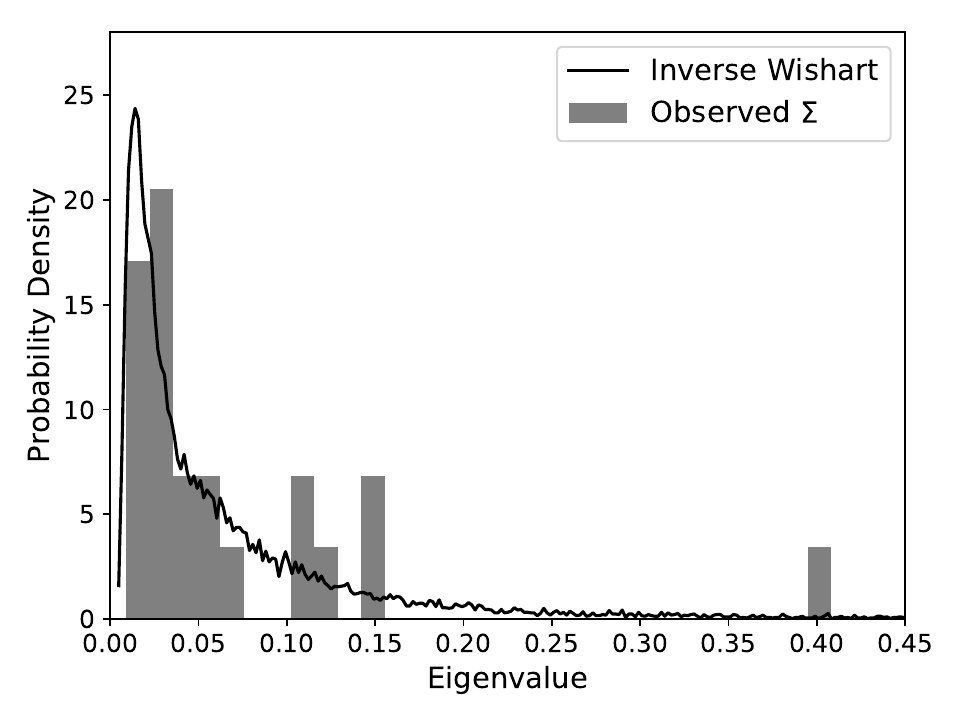}
\caption{Eigenvalue PDF of the observed color covariance matrices (grey histogram) 
of all the 11 early-blue and early-red SNe in Figure~\ref{fig:slice} 
compared to the eigenvalue PDF estimated from 10000 Monte Carlo samples (black solid curve) drawn from the best-fit inverse Wishart distribution (Equation~\ref{eq:iw}).
\label{fig:iw}}
\end{figure}

The parameters of the model are estimated in two steps, minimizing the negative log likelihood of the observations ($X$, $\Sigma$) under Equations~\ref{eq:hgaus} and \ref{eq:iw} to estimate ($\mu_0, \Sigma_\mu$) and ($\nu, \Psi$), respectively. 
We minimize the negative log likelihoods using Constrained Optimization By Linear Approximation (COBYLA\footnote{\texttt{scipy.optimize.minimize}}).
Under this minimization scheme, the matrices ($\Sigma_\mu, \Psi$) are required to be positive definite by constraint, with each of their three fit parameters $\sigma_x$, $\sigma_y$, and $\rho_{x,y}$---representing $SD$ in $\vi$ and $\bv$ directions and correlation between the two, respectively---following $\sigma_x > 10^{-3}$, $\sigma_y > 10^{-3}$, and $\rho_{x,y} \in (-1+10^{-3}, 1-10^{-3})$.
In addition, $\nu > k + 1$ is required to ensure the inverse Wishart mean $\Psi/(\nu-k-1)$ is also positive definite.
We note that in all cases (including in Sections~\ref{subsec:popsep} and \ref{subsec:philcomp}) the optimal parameters are well within the boundaries of the constraints.
We validate the inverse Wishart fit to the color covariance matrices by comparing the eigenvalues of observed matrices to those drawn from the best-fit inverse Wishart distribution.
For instance, Figure~\ref{fig:iw} compares the PDF of the observed color covariance eigenvalues for the 11 \tase\ in Figure~\ref{fig:slice} to that of the best-fit inverse Wishart distribution, approximated by drawing 10000 Monte Carlo samples, showing that the model provides an adequate fit to the observations.

\section{Gaussian process Interpolation} \label{sec:gpint}

Here, we describe our augmentation of the early $BVi$-band light curves with Gaussian process (GP) interpolation. 
The interpolant is obtained by fitting the GP model \texttt{george} \citep{Ambikasaran2015itpam} with the stationary 2-D Mat\'ern-3/2 kernel \citep[Equation~\ref{eq:matern};][]{Rasmussen2006} to the time series of 
observed multi-band fluxes ($F_i$) and errors ($\sigma_i$), {[$\Vec{x_i} = (t_i, \lambda_i)$, $F_i$, $\sigma_i$]\}$_{i=1}^{n}$, where $\lambda_i$ is the isophotal wavelength of each filter.
Our choice of kernel results in model predictions that are only once differentiable, and thus, our interpolation occurs under the least restrictive smoothness assumption for the intrinsic time-evolution of SN flux.
In the model, the data $\Vec{y} = \{F_i\}_{i=1}^n$ are considered to be distributed accoring to the $n$-variate Gaussian distribution $\mathcal{N}(0, C)$ with covariance matrix $C_{ij} = K_{3/2}(r_{ij}) + (r_{\sigma}\sigma_i)^2\delta_{ij}$, where $r_{ij}^2 = (\Vec{x}_i-\Vec{x}_j)^T M^{-1} (\Vec{x}_i-\Vec{x}_j)$ is the distance between data points under the metric $M = \diag(r_t^2, r_{\lambda}^2)$ and $K_{3/2}$ is the stationary kernel
\begin{equation}
    K_{3/2}(r_{ij}) = \alpha (1-\sqrt{3}r_{ij})\exp(-\sqrt{3}r_{ij})
    \label{eq:matern}
\end{equation}
The model parameters represent the non-random flux variance ($\alpha$), the metric scales in the time and wavelength directions ($r_t$, $r_{\lambda}$), and an errorbar scaling factor ($r_{\sigma}$) accounting for the underestimation of photometric error (see Appendix~\ref{subsec:debias}). 
We fix $r_{\lambda}$ = 6000~\AA\ due to the small number of filters used for observations \citep[also see][]{Boone2019aj}, which introduces correlation of data across the entire $BVi$ waverange and produces good interpolation results for our sample.
The model fit is performed via a Levenberg-Marquardt minimization of the negative log likelihood of the data
\begin{equation}
    \mathcal{L}_{\rm GP} = \frac{1}{2}\Vec{y}\,^{T}C^{-1}\Vec{y} + \frac{1}{2}\ln|C|
    \label{}
\end{equation}
with all parameters being optimized in log space to restrict their values to the positive real axis.

We fit our GP model to the observed $BVi$-band light curves of each \tas\ from $-$30 days before the first detection to $\sim$ 15 days after first light, obtaining the best-fit metric timescales $r_t$ in the range of 10.5--28.5 days.
We also find the photometric uncertainties are underestimated
with $r_{\sigma}$ in the range of 1.5--3.1 for the light curves of six SNe obtained
from image subtraction (see Appendix~\ref{subsec:cophot}),
while the rest obtained from PSF, multi-object, and aperture photometry
have $r_{\sigma}$ in the range of 0.9--1.6, showing no significant underestimation.
The largest underestimation factor of $r_{\sigma}$ = 3.1 is from KSN-2019bl, 
while the rest show $r_{\sigma}$ $<$ 2.5.

The expected color $c_{ij}$ between two filters $i$ and $j$ is obtained by taking log ratio of the Gaussian process predicted fluxes $f_i$ and $f_j$ at their isophotal wavelengths $\lambda_i$ and $\lambda_j$, i.e., $c_{ij} = -2.5\log_{10}(f_i/f_j)$. 
Treating color as a random variable $C_{ij}$ with mean $c_{ij}$, we estimate the variance $Var[C_{i,j}]$ by considering the fluxes as non-central Gaussian random variables $F_i = f_i + \epsilon_i$ where the noise terms $\epsilon_i$ are distributed according to the the Gaussian process predicted covariance matrix $\Sigma_{i,j}$, as in $(\epsilon_i, \epsilon_j) \sim \mathcal{N}(0, \Sigma_{i,j})$.
The color follows a log ratio distribution \citep{Katz1978}, i.e., $C_{i,j} = -2.5\log_{10}(F_i/F_j)$, which can be approximated with a first order Taylor expansion resulting in a sum of Gaussian random variables: $\ln{F_i/F_j} \sim \ln{f_i/f_j} + (\epsilon_i/f_i) - (\epsilon_j/f_j)$.
Thus, the following equations provide estimates for the variances and covariance of the Gaussian process predicted colors:
\begin{align}
    Var[C_{i,j}] =\ &Var_i + Var_j- 2\,Cov_{i,j}\\
    Cov[C_{i,j}, C_{j,k}] =\ &Cov_{i, j} - Var_j - Cov_{i, k} + Cov_{j, k}\\
    {\rm where}\ \begin{bmatrix}
    Var_i & Cov_{i,j} \\
    Cov_{i,j} & Var_j
    \end{bmatrix} &\equiv \left(\frac{2.5}{\ln(10)}\right)^2 (\Vec{u}\,^T\Sigma_{i,j}\Vec{u})
\end{align}
and $\Vec{u}= [f_i^{-1}, f_j^{-1}]^T$.
Note that the non-centrality of $F_i$ required for the Taylor approximation to be good is ensured by the fact that the Gaussian process results are only used to interpolate detected epochs where S/N $\sim$ $f_i/SD(\epsilon_i)$ $>$ 3.

Figures~\ref{fig:Bsplit}--\ref{fig:VIsplit} compare the observed early light curves and color 
evolution the best-fit GP-predicted ones and their 1-$\sigma$ uncertainty regions, where
we find an excellent agreement between them overall.
(Note that, in the case of KSN-2019bl, the large underestimation of its photometric uncertainties that we found above accommodates the apparent discrepancy between the model and observations seen in the figure).
Variations in only one band on short ($<$ 1 day) timescales consistent with typical noise variation in the light curve are smoothed by the model, though 1-2 of the earliest data points in KSNe-2021iq and 202112D, which may be attributable to multi-band excess emissions (Appendix~\ref{sec:reggaus}), are also somewhat smoothed in some bands.
Since all of the data gaps are $\lesssim$ 1 day---an order of magnitude smaller than any of the metric timescales obtained above---the interpolations across gaps are reasonable in all cases.
Overall, the GP interpolants appear to capture the evolution of the \tase\ on $>$ 1-day timescales excellently while smoothing noise and data gaps on shorter timescales, supporting the use of these models as proxies for their intrinsic color evolution.

\begin{figure*}[t!]
\epsscale{\scl}
\begin{center}
\includegraphics[width=1.0\textwidth]{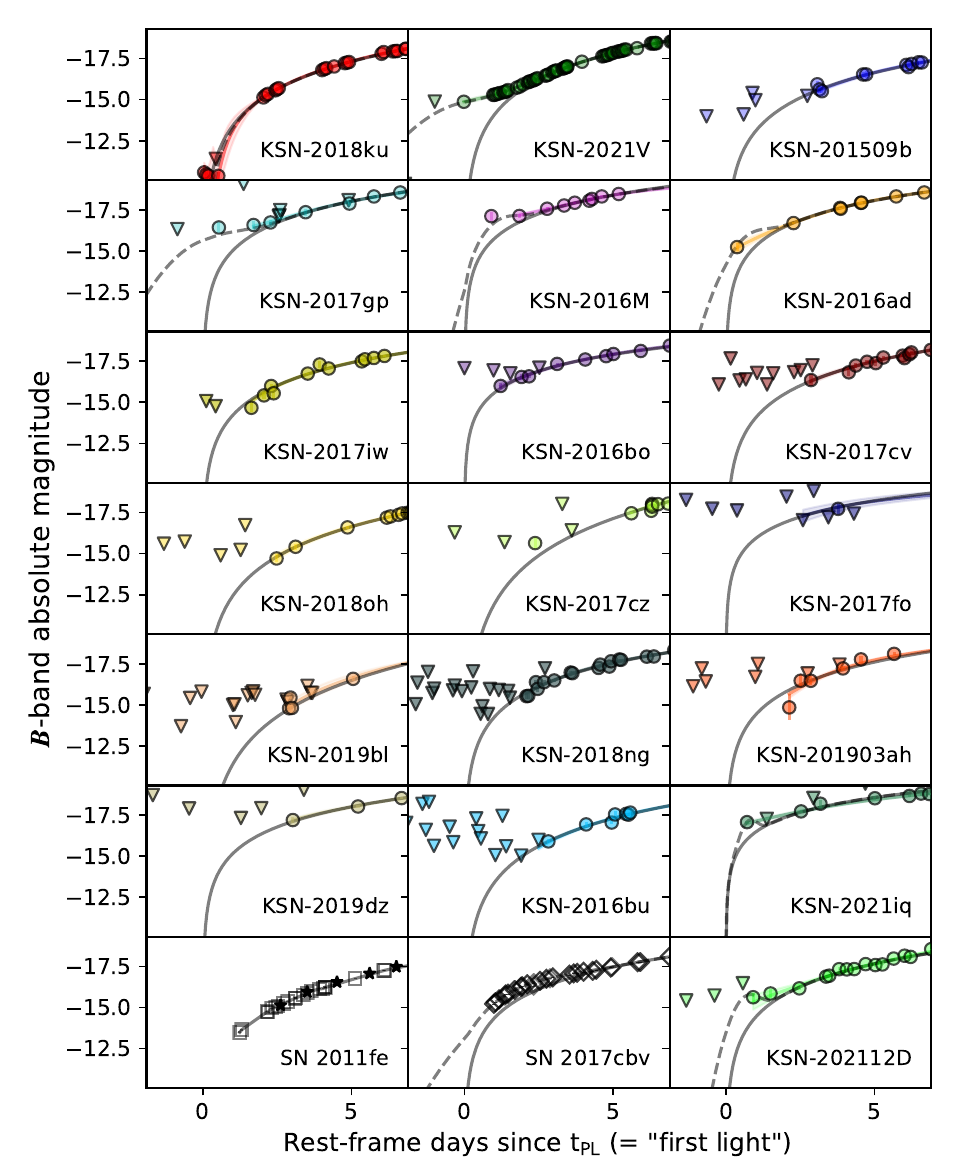}
\end{center}
\vspace{-6mm}
\caption{Early $B$-band light curves of all 19 KSP \tase\ (circles are detections and triangles are upper limits at S/N = 3) compared to the following models: (1) Gaussian process interpolant (colored curves and shaded 1-$\sigma$ regions); (2) pure power-law fits (solid grey curves) used to estimate $t_{\rm PL}$ (often called the epoch of ``first light''); and (3) regularized power-law $+$ Gaussian fits (dashed grey curves) used to characterize excess emission. 
(1) and (2) are for the entire sample of 19 SNe and they often overlap almost perfectly,
whereas (3) is for only the 7 SNe with excess emission. 
Also shown for comparison are SNe~2011fe \citep{Guillochon2017apj} with solid lines connecting data points (open squares) and filled stars from synthetic photometry (Appendix~\ref{sec:synph}), 
and 2017cbv \citep{Hosseinzadeh2017apj} with the best-fit
power-law $+$ Gaussian model (dashed curve) to the observed light curve (open diamonds) and its power-law component (solid curve).
These two non-KSP events provide additional prototypical examples of normal \tase\ exhibiting pure power-law rise and excess emission, respectively.}
\label{fig:Bsplit}
\end{figure*}

\begin{figure*}[t!]
\epsscale{\scl}
\begin{center}
\includegraphics[width=1.0\textwidth]{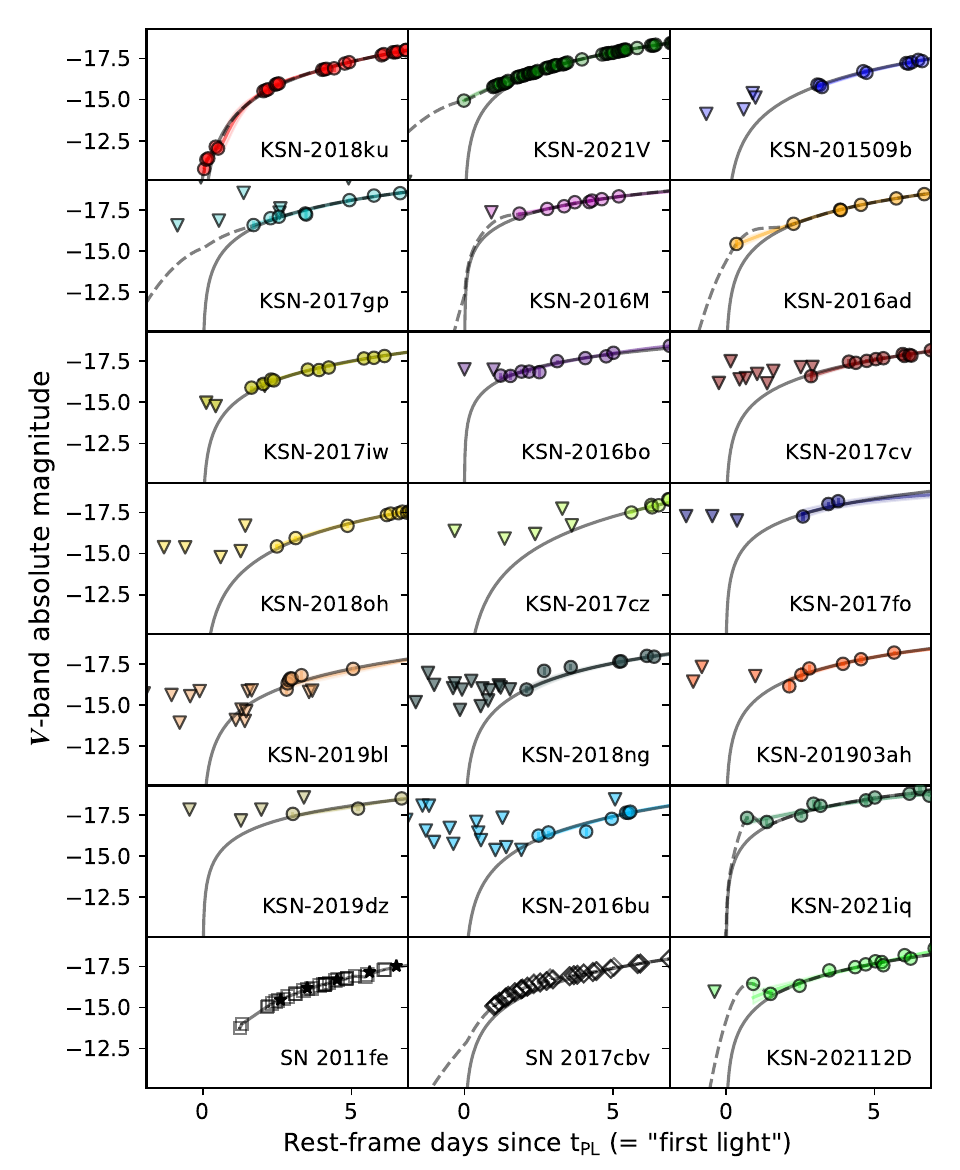}
\end{center}
\caption{Same as Figure~\ref{fig:Bsplit}, but comparing the $V$-band light curves of the same SNe and models.}
\label{fig:Vsplit}
\end{figure*}

\begin{figure*}[t!]
\epsscale{\scl}
\begin{center}
\includegraphics[width=1.0\textwidth]{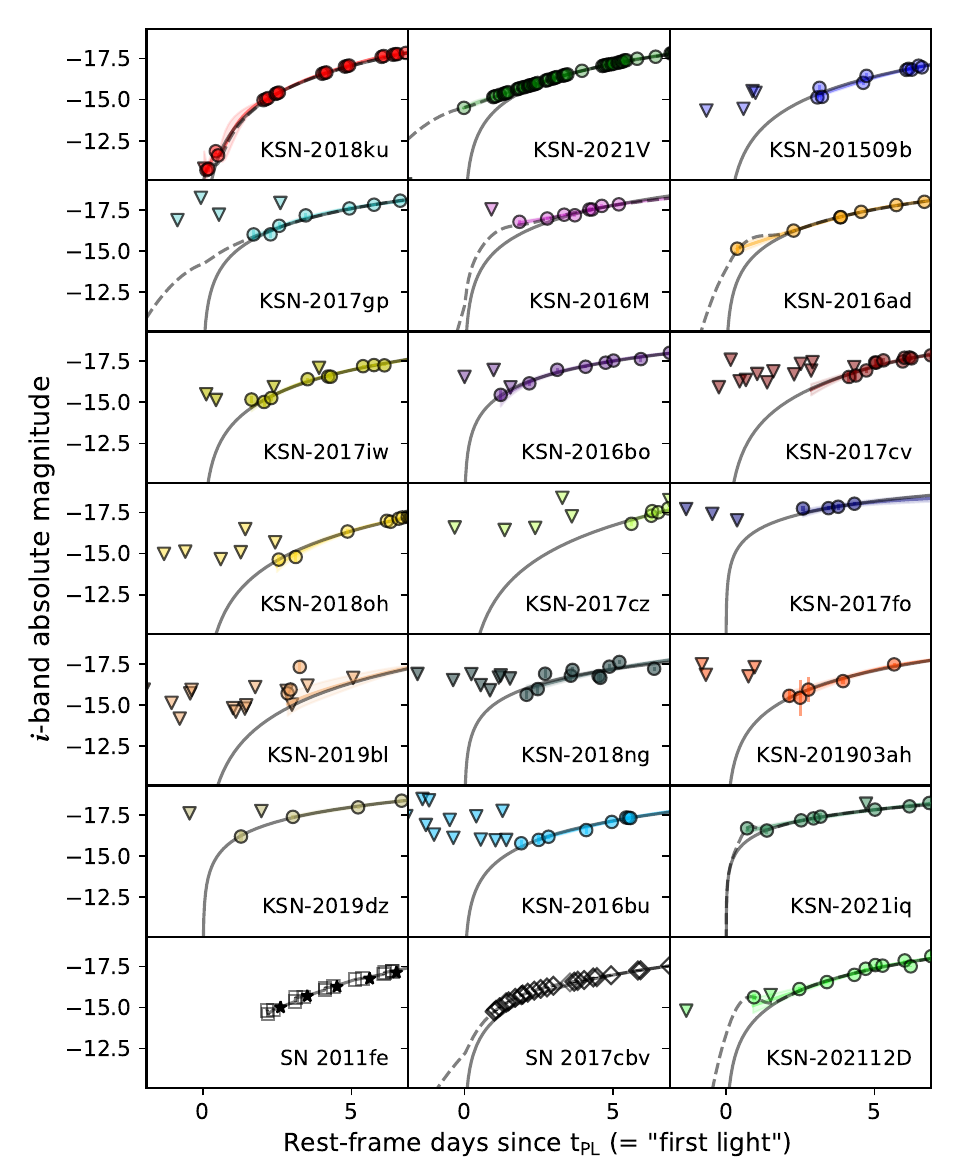}
\end{center}
\caption{Same as Figure~\ref{fig:Bsplit}, but comparing the $i$-band light curves of the same SNe and models.}
\label{fig:Isplit}
\end{figure*}

\begin{figure*}[t!]
\epsscale{\scl}
\begin{center}
\includegraphics[width=1.0\textwidth]{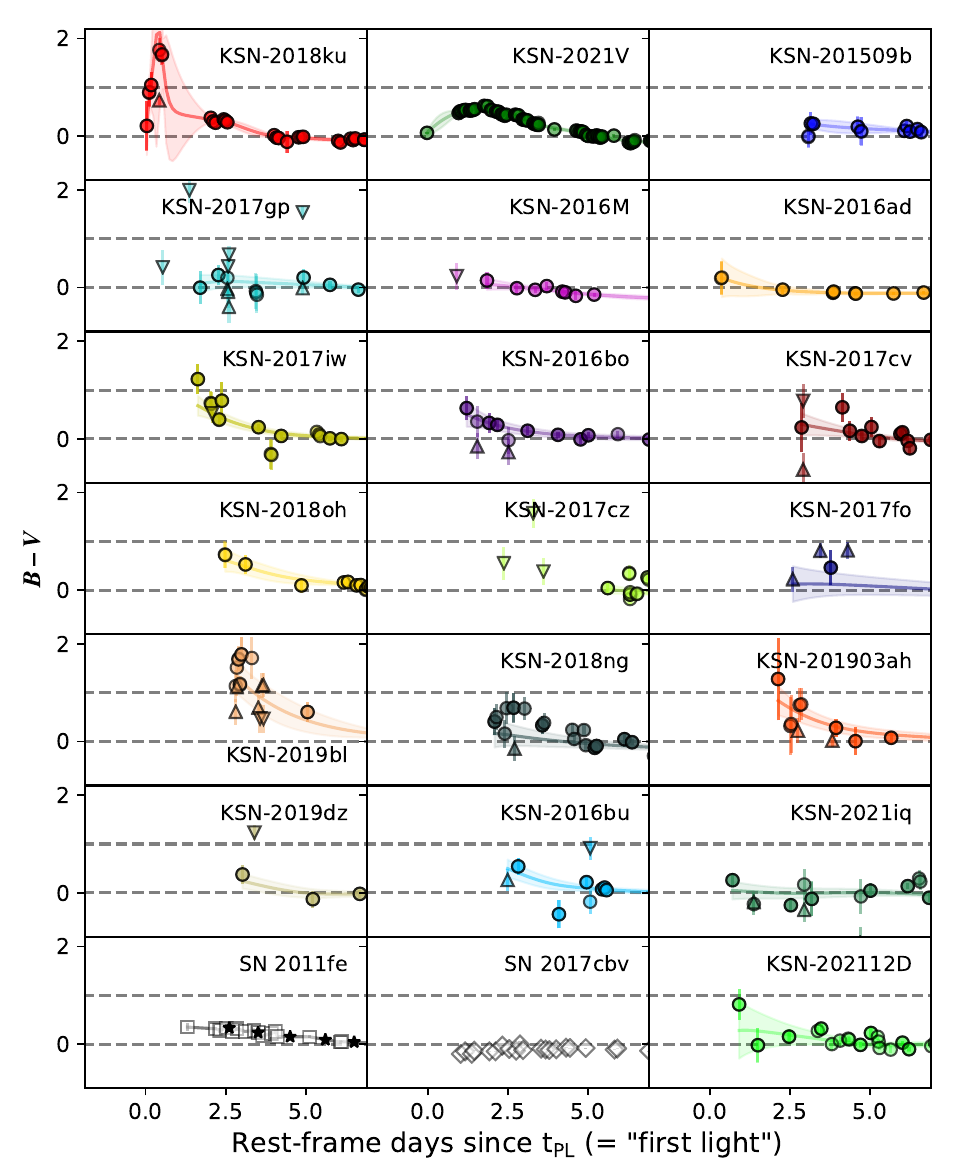}
\end{center}
\vspace{-5mm}
\caption{Same as Figure~\ref{fig:Bsplit}, but comparing the \bv\ color evolution of the same SNe to the Gaussian process interpolants and two dashed horizontal lines representing \bv\ = 0 and 1 mag.
Note that ``early-blue'' color is characterized by slowly-evolving \bv\ within $\pm$ 0.3 mag, while ``early-red'' colors start from \bv\ $\gtrsim$ 0.5 mag at 1 day and monotonically decrease. KSNe-2018ku and 2021V exhibit more complex evolutions featured with increasing \bv\ within $\sim$ 0.5 and 2 days, respectively, followed by a monotonic decrease similar to early-red.}
\label{fig:BVsplit}
\end{figure*}

\begin{figure*}[t!]
\epsscale{\scl}
\begin{center}
\includegraphics[width=1.0\textwidth]{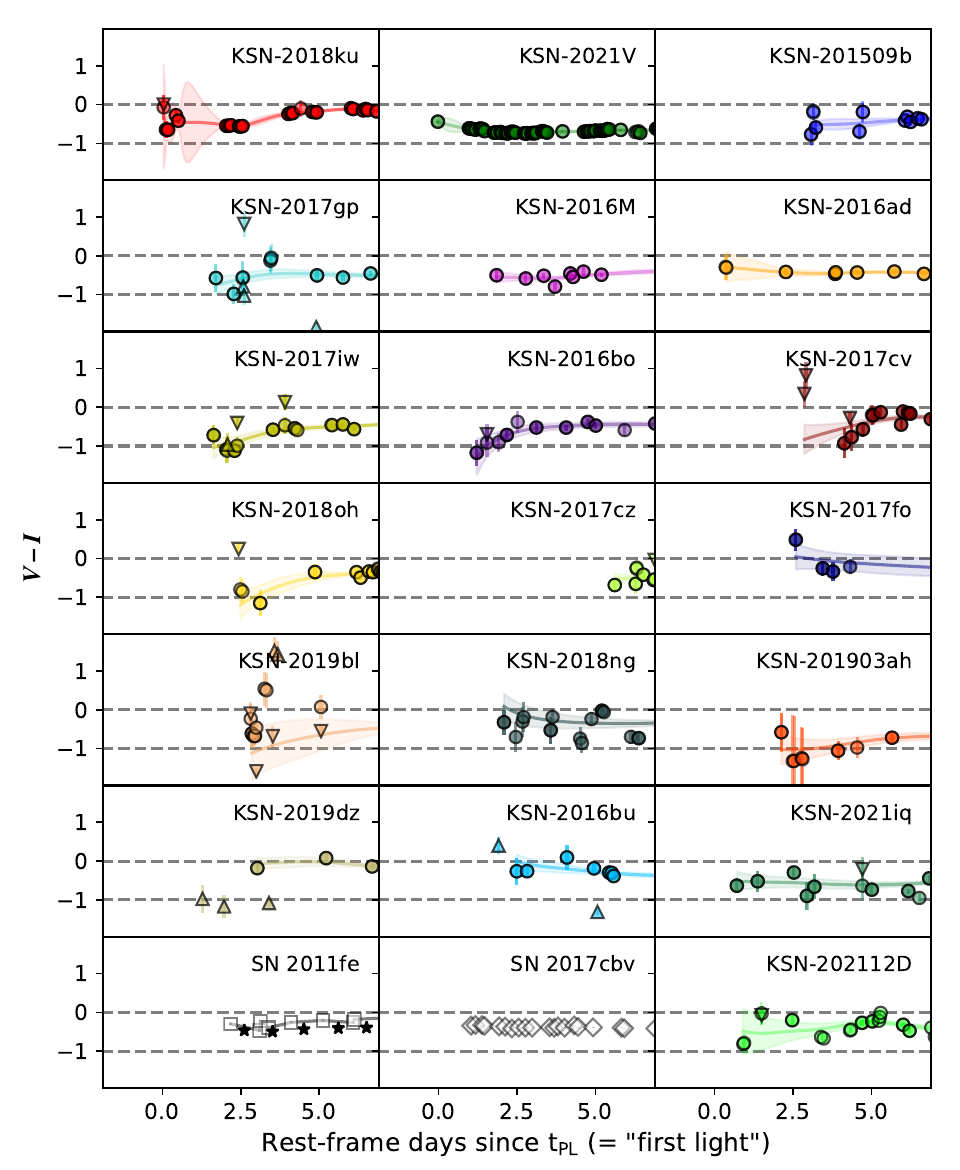}
\end{center}
\vspace{-2mm}
\caption{Same as Figure~\ref{fig:BVsplit}, but comparing the \vi\ color evolution of the same SNe and dashed horizontal lines representing \vi\ = $-$1 and 0 mag. Note, here, that ``early-blue'' color means slowly-evolving \vi\ $\sim$ $-$0.5 mag, while ``early-red'' colors start from \vi\ $\sim$ $-$1.0 at 2 days and monotonically increase. In contrast, ``early-yellows'' colors start from \vi\ $>$ $-$0.2 mag and decrease.}
\label{fig:VIsplit}
\end{figure*}


\begin{thebibliography}{}
\expandafter\ifx\csname natexlab\endcsname\relax\def\natexlab#1{#1}\fi
\providecommand{\url}[1]{\href{#1}{#1}}
\providecommand{\dodoi}[1]{doi:~\href{http://doi.org/#1}{\nolinkurl{#1}}}
\providecommand{\doeprint}[1]{\href{http://ascl.net/#1}{\nolinkurl{http://ascl.net/#1}}}
\providecommand{\doarXiv}[1]{\href{https://arxiv.org/abs/#1}{\nolinkurl{https://arxiv.org/abs/#1}}}

\bibitem[{{Afsariardchi} {et~al.}(2019){Afsariardchi}, {Moon}, {Drout},
  {Gonz{\'a}lez-Gait{\'a}n}, {Ni}, {Matzner}, {Kim}, {Lee}, {Park}, {Gal-Yam},
  {Pignata}, {Koo}, {Ryder}, {Cha}, \& {Lee}}]{Afsariardchi2019apj}
{Afsariardchi}, N., {Moon}, D.-S., {Drout}, M.~R., {et~al.} 2019, \apj, 881,
  22, \dodoi{10.3847/1538-4357/ab2be6}

\bibitem[{{Albareti} {et~al.}(2017){Albareti}, {Allende Prieto}, {Almeida},
  {Anders}, {Anderson}, {Andrews}, {Arag{\'o}n-Salamanca},
  {Argudo-Fern{\'a}ndez}, {Armengaud}, {Aubourg}, {Avila-Reese}, {Badenes},
  {Bailey}, {Barbuy}, {Barger}, {Barrera-Ballesteros}, {Bartosz}, {Basu},
  {Bates}, {Battaglia}, {Baumgarten}, {Baur}, {Bautista}, {Beers}, {Belfiore},
  {Bershady}, {Bertran de Lis}, {Bird}, {Bizyaev}, {Blanc}, {Blanton},
  {Blomqvist}, {Bolton}, {Borissova}, {Bovy}, {Brandt}, {Brinkmann},
  {Brownstein}, {Bundy}, {Burtin}, {Busca}, {Camacho Chavez}, {Cano D{\'\i}az},
  {Cappellari}, {Carrera}, {Chen}, {Cherinka}, {Cheung}, {Chiappini},
  {Chojnowski}, {Chuang}, {Chung}, {Cirolini}, {Clerc}, {Cohen}, {Comerford},
  {Comparat}, {Correa do Nascimento}, {Cousinou}, {Covey}, {Crane}, {Croft},
  {Cunha}, {Darling}, {Davidson}, {Dawson}, {Da Costa}, {Da Silva Ilha},
  {Deconto Machado}, {Delubac}, {De Lee}, {De la Macorra}, {De la Torre},
  {Diamond-Stanic}, {Donor}, {Downes}, {Drory}, {Du}, {Du Mas des Bourboux},
  {Dwelly}, {Ebelke}, {Eigenbrot}, {Eisenstein}, {Elsworth}, {Emsellem},
  {Eracleous}, {Escoffier}, {Evans}, {Falc{\'o}n-Barroso}, {Fan}, {Favole},
  {Fernandez-Alvar}, {Fernandez-Trincado}, {Feuillet}, {Fleming},
  {Font-Ribera}, {Freischlad}, {Frinchaboy}, {Fu}, {Gao}, {Garcia},
  {Garcia-Dias}, {Garcia-Hern{\'a}ndez}, {Garcia P{\'e}rez}, {Gaulme}, {Ge},
  {Geisler}, {Gillespie}, {Gil Marin}, {Girardi}, {Goddard}, {Gomez Maqueo
  Chew}, {Gonzalez-Perez}, {Grabowski}, {Green}, {Grier}, {Grier}, {Guo},
  {Guy}, {Hagen}, {Hall}, {Harding}, {Harley}, {Hasselquist}, {Hawley},
  {Hayes}, {Hearty}, {Hekker}, {Hernandez Toledo}, {Ho}, {Hogg},
  {Holley-Bockelmann}, {Holtzman}, {Holzer}, {Hu}, {Huber}, {Hutchinson},
  {Hwang}, {Ibarra-Medel}, {Ivans}, {Ivory}, {Jaehnig}, {Jensen}, {Johnson},
  {Jones}, {Jullo}, {Kallinger}, {Kinemuchi}, {Kirkby}, {Klaene}, {Kneib},
  {Kollmeier}, {Lacerna}, {Lane}, {Lang}, {Laurent}, {Law}, {Leauthaud}, {Le
  Goff}, {Li}, {Li}, {Li}, {Li}, {Liang}, {Liang}, {Lima}, {Lin}, {Lin}, {Lin},
  {Liu}, {Long}, {Lucatello}, {MacDonald}, {MacLeod}, {Mackereth}, {Mahadevan},
  {Maia}, {Maiolino}, {Majewski}, {Malanushenko}, {Malanushenko}, {Mallmann},
  {Manchado}, {Maraston}, {Marques-Chaves}, {Martinez Valpuesta}, {Masters},
  {Mathur}, {McGreer}, {Merloni}, {Merrifield}, {M{\'e}sz{\'a}ros}, {Meza},
  {Miglio}, {Minchev}, {Molaverdikhani}, {Montero-Dorta}, {Mosser}, {Muna},
  {Myers}, {Nair}, {Nandra}, {Ness}, {Newman}, {Nichol}, {Nidever},
  {Nitschelm}, {O'Connell}, {Oravetz}, {Oravetz}, {Pace}, {Padilla},
  {Palanque-Delabrouille}, {Pan}, {Parejko}, {Paris}, {Park}, {Peacock},
  {Peirani}, {Pellejero-Ibanez}, {Penny}, {Percival}, {Percival},
  {Perez-Fournon}, {Petitjean}, {Pieri}, {Pinsonneault}, {Pisani}, {Prada},
  {Prakash}, {Price-Jones}, {Raddick}, {Rahman}, {Raichoor}, {Barboza Rembold},
  {Reyna}, {Rich}, {Richstein}, {Ridl}, {Riffel}, {Riffel}, {Rix}, {Robin},
  {Rockosi}, {Rodr{\'\i}guez-Torres}, {Rodrigues}, {Roe}, {Roman Lopes},
  {Rom{\'a}n-Z{\'u}{\~n}iga}, {Ross}, {Rossi}, {Ruan}, {Ruggeri}, {Runnoe},
  {Salazar-Albornoz}, {Salvato}, {Sanchez}, {Sanchez}, {Sanchez-Gallego},
  {Santiago}, {Schiavon}, {Schimoia}, {Schlafly}, {Schlegel}, {Schneider},
  {Sch{\"o}nrich}, {Schultheis}, {Schwope}, {Seo}, {Serenelli}, {Sesar},
  {Shao}, {Shetrone}, {Shull}, {Silva Aguirre}, {Skrutskie}, {Slosar}, {Smith},
  {Smith}, {Sobeck}, {Somers}, {Souto}, {Stark}, {Stassun}, {Steinmetz},
  {Stello}, {Storchi Bergmann}, {Strauss}, {Streblyanska}, {Stringfellow},
  {Suarez}, {Sun}, {Taghizadeh-Popp}, {Tang}, {Tao}, {Tayar}, {Tembe},
  {Thomas}, {Tinker}, {Tojeiro}, {Tremonti}, {Troup}, {Trump}, {Unda-Sanzana},
  {Valenzuela}, {Van den Bosch}, {Vargas-Maga{\~n}a}, {Vazquez}, {Villanova},
  {Vivek}, {Vogt}, {Wake}, {Walterbos}, {Wang}, {Wang}, {Weaver}, {Weijmans},
  {Weinberg}, {Westfall}, {Whelan}, {Wilcots}, {Wild}, {Williams}, {Wilson},
  {Wood-Vasey}, {Wylezalek}, {Xiao}, {Yan}, {Yang}, {Ybarra}, {Yeche}, {Yuan},
  {Zakamska}, {Zamora}, {Zasowski}, {Zhang}, {Zhao}, {Zhao}, {Zheng}, {Zheng},
  {Zhou}, {Zhu}, {Zinn}, \& {Zou}}]{Albareti2017apjs}
{Albareti}, F.~D., {Allende Prieto}, C., {Almeida}, A., {et~al.} 2017, \apjs,
  233, 25, \dodoi{10.3847/1538-4365/aa8992}

\bibitem[{{Albrecht} {et~al.}(2006){Albrecht}, {Bernstein}, {Cahn}, {Freedman},
  {Hewitt}, {Hu}, {Huth}, {Kamionkowski}, {Kolb}, {Knox}, {Mather}, {Staggs},
  \& {Suntzeff}}]{Albrecht2006}
{Albrecht}, A., {Bernstein}, G., {Cahn}, R., {et~al.} 2006, arXiv e-prints,
  astro, \dodoi{10.48550/arXiv.astro-ph/0609591}

\bibitem[{{Allington-Smith} {et~al.}(2002){Allington-Smith}, {Murray},
  {Content}, {Dodsworth}, {Davies}, {Miller}, {Jorgensen}, {Hook}, {Crampton},
  \& {Murowinski}}]{Allington-Smith2002pasp}
{Allington-Smith}, J., {Murray}, G., {Content}, R., {et~al.} 2002, \pasa, 114,
  892, \dodoi{10.1086/341712}

\bibitem[{{Ambikasaran} {et~al.}(2015){Ambikasaran}, {Foreman-Mackey},
  {Greengard}, {Hogg}, \& {O'Neil}}]{Ambikasaran2015itpam}
{Ambikasaran}, S., {Foreman-Mackey}, D., {Greengard}, L., {Hogg}, D.~W., \&
  {O'Neil}, M. 2015, IEEE Transactions on Pattern Analysis and Machine
  Intelligence, 38, 252, \dodoi{10.1109/TPAMI.2015.2448083}

\bibitem[{{Armstrong}(1967)}]{Armstrong1967}
{Armstrong}, B. 1967, \jqsrt, 7, 61, \dodoi{10.1016/0022-4073(67)90057-X}

\bibitem[{{Arnett}(1982)}]{Arnett1982apj}
{Arnett}, W.~D. 1982, \apj, 253, 785, \dodoi{10.1086/159681}

\bibitem[{{Ashall} {et~al.}(2016){Ashall}, {Mazzali}, {Sasdelli}, \&
  {Prentice}}]{Ashall2016mnras}
{Ashall}, C., {Mazzali}, P., {Sasdelli}, M., \& {Prentice}, S.~J. 2016, \mnras,
  460, 3529, \dodoi{10.1093/mnras/stw1214}

\bibitem[{{Astropy Collaboration} {et~al.}(2013){Astropy Collaboration},
  {Robitaille}, {Tollerud}, {Greenfield}, {Droettboom}, {Bray}, {Aldcroft},
  {Davis}, {Ginsburg}, {Price-Whelan}, {Kerzendorf}, {Conley}, {Crighton},
  {Barbary}, {Muna}, {Ferguson}, {Grollier}, {Parikh}, {Nair}, {Unther},
  {Deil}, {Woillez}, {Conseil}, {Kramer}, {Turner}, {Singer}, {Fox}, {Weaver},
  {Zabalza}, {Edwards}, {Azalee Bostroem}, {Burke}, {Casey}, {Crawford},
  {Dencheva}, {Ely}, {Jenness}, {Labrie}, {Lim}, {Pierfederici}, {Pontzen},
  {Ptak}, {Refsdal}, {Servillat}, \& {Streicher}}]{Astropy2013aa}
{Astropy Collaboration}, {Robitaille}, T.~P., {Tollerud}, E.~J., {et~al.} 2013,
  \aap, 558, A33, \dodoi{10.1051/0004-6361/201322068}

\bibitem[{{Becker}(2015)}]{Becker2015ascl}
{Becker}, A. 2015, {HOTPANTS: High Order Transform of PSF ANd Template
  Subtraction}, Astrophysics Source Code Library, record ascl:1504.004

\bibitem[{{Bell} {et~al.}(2003){Bell}, {McIntosh}, {Katz}, \&
  {Weinberg}}]{Bell2003apjs}
{Bell}, E.~F., {McIntosh}, D.~H., {Katz}, N., \& {Weinberg}, M.~D. 2003, \apjs,
  149, 289, \dodoi{10.1086/378847}

\bibitem[{{Bertin}(2006)}]{Bertin2006aspc}
{Bertin}, E. 2006, in Astronomical Society of the Pacific Conference Series,
  Vol. 351, Astronomical Data Analysis Software and Systems XV, ed.
  C.~{Gabriel}, C.~{Arviset}, D.~{Ponz}, \& S.~{Enrique}, 112

\bibitem[{{Bertin} \& {Arnouts}(1996)}]{Bertin&Arnouts1996aas}
{Bertin}, E., \& {Arnouts}, S. 1996, \aaps, 117, 393,
  \dodoi{10.1051/aas:1996164}

\bibitem[{{Bertin} {et~al.}(2002){Bertin}, {Mellier}, {Radovich}, {Missonnier},
  {Didelon}, \& {Morin}}]{Bertin2002aspc}
{Bertin}, E., {Mellier}, Y., {Radovich}, M., {et~al.} 2002, in Astronomical
  Society of the Pacific Conference Series, Vol. 281, Astronomical Data
  Analysis Software and Systems XI, ed. D.~A. {Bohlender}, D.~{Durand}, \&
  T.~H. {Handley}, 228

\bibitem[{{Bianchi} {et~al.}(2017){Bianchi}, {Shiao}, \&
  {Thilker}}]{Bianchi2017apjs}
{Bianchi}, L., {Shiao}, B., \& {Thilker}, D. 2017, \apjs, 230, 24,
  \dodoi{10.3847/1538-4365/aa7053}

\bibitem[{{Bilicki} {et~al.}(2014){Bilicki}, {Jarrett}, {Peacock}, {Cluver}, \&
  {Steward}}]{Bilicki2014apjs}
{Bilicki}, M., {Jarrett}, T.~H., {Peacock}, J.~A., {Cluver}, M.~E., \&
  {Steward}, L. 2014, \apjs, 210, 9, \dodoi{10.1088/0067-0049/210/1/9}

\bibitem[{{Bishop}(2007)}]{Bishop2007book}
{Bishop}, C.~M. 2007, {Pattern Recognition and Machine Learning (Information
  Science and Statistics)}

\bibitem[{{Blondin} \& {Tonry}(2007)}]{Blondin2007apj}
{Blondin}, S., \& {Tonry}, J.~L. 2007, \apj, 666, 1024, \dodoi{10.1086/520494}

\bibitem[{{Blondin} {et~al.}(2012){Blondin}, {Matheson}, {Kirshner}, {Mand el},
  {Berlind}, {Calkins}, {Challis}, {Garnavich}, {Jha}, {Modjaz}, {Riess}, \&
  {Schmidt}}]{Blondin2012aj}
{Blondin}, S., {Matheson}, T., {Kirshner}, R.~P., {et~al.} 2012, \aj, 143, 126,
  \dodoi{10.1088/0004-6256/143/5/126}

\bibitem[{{Bohlin}(2014)}]{Bohlin2014aj}
{Bohlin}, R.~C. 2014, \aj, 147, 127, \dodoi{10.1088/0004-6256/147/6/127}

\bibitem[{{Boone}(2019)}]{Boone2019aj}
{Boone}, K. 2019, \aj, 158, 257, \dodoi{10.3847/1538-3881/ab5182}

\bibitem[{{Boos} {et~al.}(2021){Boos}, {Townsley}, {Shen}, {Caldwell}, \&
  {Miles}}]{Boos2021apj}
{Boos}, S.~J., {Townsley}, D.~M., {Shen}, K.~J., {Caldwell}, S., \& {Miles},
  B.~J. 2021, \apj, 919, 126, \dodoi{10.3847/1538-4357/ac07a2}

\bibitem[{{Branch} {et~al.}(2006){Branch}, {Dang}, {Hall}, {Ketchum},
  {Melakayil}, {Parrent}, {Troxel}, {Casebeer}, {Jeffery}, \&
  {Baron}}]{Branch2006pasp}
{Branch}, D., {Dang}, L.~C., {Hall}, N., {et~al.} 2006, \pasa, 118, 560,
  \dodoi{10.1086/502778}

\bibitem[{{Brout} \& {Scolnic}(2021)}]{Brout2021apj}
{Brout}, D., \& {Scolnic}, D. 2021, \apj, 909, 26,
  \dodoi{10.3847/1538-4357/abd69b}

\bibitem[{{Brout} {et~al.}(2019){Brout}, {Sako}, {Scolnic}, {Kessler},
  {D'Andrea}, {Davis}, {Hinton}, {Kim}, {Lasker}, {Macaulay}, {M{\"o}ller},
  {Nichol}, {Smith}, {Sullivan}, {Wolf}, {Allam}, {Bassett}, {Brown},
  {Castander}, {Childress}, {Foley}, {Galbany}, {Herner}, {Kasai}, {March},
  {Morganson}, {Nugent}, {Pan}, {Thomas}, {Tucker}, {Wester}, {Abbott},
  {Annis}, {Avila}, {Bertin}, {Brooks}, {Burke}, {Carnero Rosell}, {Carrasco
  Kind}, {Carretero}, {Crocce}, {Cunha}, {da Costa}, {Davis}, {De Vicente},
  {Desai}, {Diehl}, {Doel}, {Eifler}, {Flaugher}, {Fosalba}, {Frieman},
  {Garc{\'\i}a-Bellido}, {Gaztanaga}, {Gerdes}, {Goldstein}, {Gruen},
  {Gruendl}, {Gschwend}, {Gutierrez}, {Hartley}, {Hollowood}, {Honscheid},
  {James}, {Kuehn}, {Kuropatkin}, {Lahav}, {Li}, {Lima}, {Marshall}, {Martini},
  {Miquel}, {Nord}, {Plazas}, {Roodman}, {Rykoff}, {Sanchez}, {Scarpine},
  {Schindler}, {Schubnell}, {Serrano}, {Sevilla-Noarbe}, {Soares-Santos},
  {Sobreira}, {Suchyta}, {Swanson}, {Tarle}, {Thomas}, {Tucker}, {Walker},
  {Yanny}, {Zhang}, \& {DES COLLABORATION}}]{Brout2019apj}
{Brout}, D., {Sako}, M., {Scolnic}, D., {et~al.} 2019, \apj, 874, 106,
  \dodoi{10.3847/1538-4357/ab06c1}

\bibitem[{{Bulla} {et~al.}(2020){Bulla}, {Miller}, {Yao}, {Dessart}, {Dhawan},
  {Papadogiannakis}, {Biswas}, {Goobar}, {Kulkarni}, {Nordin}, {Nugent},
  {Polin}, {Sollerman}, {Bellm}, {Coughlin}, {Dekany}, {Golkhou}, {Graham},
  {Kasliwal}, {Kupfer}, {Laher}, {Masci}, {Porter}, {Rusholme}, \&
  {Shupe}}]{Bulla2020apj}
{Bulla}, M., {Miller}, A.~A., {Yao}, Y., {et~al.} 2020, \apj, 902, 48,
  \dodoi{10.3847/1538-4357/abb13c}

\bibitem[{{Burns} {et~al.}(2011){Burns}, {Stritzinger}, {Phillips}, {Kattner},
  {Persson}, {Madore}, {Freedman}, {Boldt}, {Campillay}, {Contreras},
  {Folatelli}, {Gonzalez}, {Krzeminski}, {Morrell}, {Salgado}, \&
  {Suntzeff}}]{Burns2011aj}
{Burns}, C.~R., {Stritzinger}, M., {Phillips}, M.~M., {et~al.} 2011, \aj, 141,
  19, \dodoi{10.1088/0004-6256/141/1/19}

\bibitem[{{Burns} {et~al.}(2014){Burns}, {Stritzinger}, {Phillips}, {Hsiao},
  {Contreras}, {Persson}, {Folatelli}, {Boldt}, {Campillay}, {Castell{\'o}n},
  {Freedman}, {Madore}, {Morrell}, {Salgado}, \& {Suntzeff}}]{Burns2014apj}
---. 2014, \apj, 789, 32, \dodoi{10.1088/0004-637X/789/1/32}

\bibitem[{{Burns} {et~al.}(2018){Burns}, {Parent}, {Phillips}, {Stritzinger},
  {Krisciunas}, {Suntzeff}, {Hsiao}, {Contreras}, {Anais}, {Boldt}, {Busta},
  {Campillay}, {Castell{\'o}n}, {Folatelli}, {Freedman}, {Gonz{\'a}lez},
  {Hamuy}, {Heoflich}, {Krzeminski}, {Madore}, {Morrell}, {Persson}, {Roth},
  {Salgado}, {Ser{\'o}n}, \& {Torres}}]{Burns2018apj}
{Burns}, C.~R., {Parent}, E., {Phillips}, M.~M., {et~al.} 2018, \apj, 869, 56,
  \dodoi{10.3847/1538-4357/aae51c}

\bibitem[{{Caon} {et~al.}(1993){Caon}, {Capaccioli}, \&
  {D'Onofrio}}]{Caon1993mnras}
{Caon}, N., {Capaccioli}, M., \& {D'Onofrio}, M. 1993, \mnras, 265, 1013,
  \dodoi{10.1093/mnras/265.4.1013}

\bibitem[{{Cappellari}(2017)}]{Cappellari2017mnras}
{Cappellari}, M. 2017, \mnras, 466, 798, \dodoi{10.1093/mnras/stw3020}

\bibitem[{{Cartier} {et~al.}(2017){Cartier}, {Sullivan}, {Firth}, {Pignata},
  {Mazzali}, {Maguire}, {Childress}, {Arcavi}, {Ashall}, {Bassett}, {Crawford},
  {Frohmaier}, {Galbany}, {Gal-Yam}, {Hosseinzadeh}, {Howell}, {Inserra},
  {Johansson}, {Kasai}, {McCully}, {Prajs}, {Prentice}, {Schulze}, {Smartt},
  {Smith}, {Smith}, {Valenti}, \& {Young}}]{Cartier2017mnras}
{Cartier}, R., {Sullivan}, M., {Firth}, R.~E., {et~al.} 2017, \mnras, 464,
  4476, \dodoi{10.1093/mnras/stw2678}

\bibitem[{{Chambers} {et~al.}(2016){Chambers}, {Huber}, {Flewelling},
  {Magnier}, {Primak}, {Schultz}, {Smartt}, {Smith}, {Tonry}, {Waters},
  {Wright}, \& {Young}}]{Chambers2016tns}
{Chambers}, K.~C., {Huber}, M.~E., {Flewelling}, H., {et~al.} 2016, Transient
  Name Server Discovery Report, 2016-931, 1

\bibitem[{{Childress} {et~al.}(2014){Childress}, {Wolf}, \&
  {Zahid}}]{Childress2014mnras}
{Childress}, M.~J., {Wolf}, C., \& {Zahid}, H.~J. 2014, \mnras, 445, 1898,
  \dodoi{10.1093/mnras/stu1892}

\bibitem[{{Childress} {et~al.}(2013){Childress}, {Scalzo}, {Sim}, {Tucker},
  {Yuan}, {Schmidt}, {Cenko}, {Silverman}, {Contreras}, {Hsiao}, {Phillips},
  {Morrell}, {Jha}, {McCully}, {Filippenko}, {Anderson}, {Benetti}, {Bufano},
  {de Jaeger}, {Forster}, {Gal-Yam}, {Le Guillou}, {Maguire}, {Maund},
  {Mazzali}, {Pignata}, {Smartt}, {Spyromilio}, {Sullivan}, {Taddia},
  {Valenti}, {Bayliss}, {Bessell}, {Blanc}, {Carson}, {Clubb}, {de Burgh-Day},
  {Desjardins}, {Fang}, {Fox}, {Gates}, {Ho}, {Keller}, {Kelly}, {Lidman},
  {Loaring}, {Mould}, {Owers}, {Ozbilgen}, {Pei}, {Pickering}, {Pracy}, {Rich},
  {Schaefer}, {Scott}, {Stritzinger}, {Vogt}, \& {Zhou}}]{Childress2013apj}
{Childress}, M.~J., {Scalzo}, R.~A., {Sim}, S.~A., {et~al.} 2013, \apj, 770,
  29, \dodoi{10.1088/0004-637X/770/1/29}

\bibitem[{{Chung} {et~al.}(2023){Chung}, {Yoon}, {Park}, {An}, {Son}, {Cho}, \&
  {Lee}}]{Chung2023apj}
{Chung}, C., {Yoon}, S.-J., {Park}, S., {et~al.} 2023, \apj, 959, 94,
  \dodoi{10.3847/1538-4357/ad0121}

\bibitem[{{Ciotti}(1991)}]{Ciotti1991aa}
{Ciotti}, L. 1991, \aap, 249, 99

\bibitem[{{Clemens} {et~al.}(2004){Clemens}, {Crain}, \&
  {Anderson}}]{Clemens2004spie}
{Clemens}, J.~C., {Crain}, J.~A., \& {Anderson}, R. 2004, in Society of
  Photo-Optical Instrumentation Engineers (SPIE) Conference Series, Vol. 5492,
  Ground-based Instrumentation for Astronomy, ed. A.~F.~M. {Moorwood} \&
  M.~{Iye}, 331--340, \dodoi{10.1117/12.550069}

\bibitem[{{Cochran} \& {Wishart}(1934)}]{Cochran1934pcps}
{Cochran}, W.~G., \& {Wishart}, J. 1934, Proceedings of the Cambridge
  Philosophical Society, 30, 178, \dodoi{10.1017/S0305004100016595}

\bibitem[{{Dhawan} {et~al.}(2017){Dhawan}, {Leibundgut}, {Spyromilio}, \&
  {Blondin}}]{Dhawan2017aa}
{Dhawan}, S., {Leibundgut}, B., {Spyromilio}, J., \& {Blondin}, S. 2017, \aap,
  602, A118, \dodoi{10.1051/0004-6361/201629793}

\bibitem[{{Dimitriadis} {et~al.}(2019){Dimitriadis}, {Foley}, {Rest}, {Kasen},
  {Piro}, {Polin}, {Jones}, {Villar}, {Narayan}, \&
  {Coulter}}]{Dimitriadis2019apj}
{Dimitriadis}, G., {Foley}, R.~J., {Rest}, A., {et~al.} 2019, \apjl, 870, L1,
  \dodoi{10.3847/2041-8213/aaedb0}

\bibitem[{{Doull} \& {Baron}(2011)}]{Doull2011pasp}
{Doull}, B.~A., \& {Baron}, E. 2011, \pasp, 123, 765, \dodoi{10.1086/661023}

\bibitem[{{Fan} {et~al.}(2023){Fan}, {Moon}, {Park}, {Zaritsky}, {Kim}, {Lee},
  {Li}, {Ni}, {Shin}, {Cha}, \& {Lee}}]{Fan2023mnras}
{Fan}, T.~J., {Moon}, D.-S., {Park}, H.~S., {et~al.} 2023, \mnras, 525, 4904,
  \dodoi{10.1093/mnras/stad2470}

\bibitem[{{Fitzpatrick}(1999)}]{Fitzpatrick1999pasp}
{Fitzpatrick}, E.~L. 1999, \pasp, 111, 63, \dodoi{10.1086/316293}

\bibitem[{{Foley} {et~al.}(2012){Foley}, {Challis}, {Filippenko},
  {Ganeshalingam}, {Landsman}, {Li}, {Marion}, {Silverman}, {Beaton},
  {Bennert}, {Cenko}, {Childress}, {Guhathakurta}, {Jiang}, {Kalirai},
  {Kirshner}, {Stockton}, {Tollerud}, {Vink{\'o}}, {Wheeler}, \&
  {Woo}}]{Foley2012apj}
{Foley}, R.~J., {Challis}, P.~J., {Filippenko}, A.~V., {et~al.} 2012, \apj,
  744, 38, \dodoi{10.1088/0004-637X/744/1/38}

\bibitem[{{Foreman-Mackey} {et~al.}(2013){Foreman-Mackey}, {Hogg}, {Lang}, \&
  {Goodman}}]{Foreman-Mackey2013pasp}
{Foreman-Mackey}, D., {Hogg}, D.~W., {Lang}, D., \& {Goodman}, J. 2013, \pasp,
  125, 306, \dodoi{10.1086/670067}

\bibitem[{{Fremling} {et~al.}(2019){Fremling}, {Dugas}, \&
  {Sharma}}]{Fremling2019tns}
{Fremling}, C., {Dugas}, A., \& {Sharma}, Y. 2019, Transient Name Server
  Classification Report, 2019-498, 1

\bibitem[{{Gallazzi} {et~al.}(2005){Gallazzi}, {Charlot}, {Brinchmann},
  {White}, \& {Tremonti}}]{Gallazzi2005mnras}
{Gallazzi}, A., {Charlot}, S., {Brinchmann}, J., {White}, S. D.~M., \&
  {Tremonti}, C.~A. 2005, \mnras, 362, 41,
  \dodoi{10.1111/j.1365-2966.2005.09321.x}

\bibitem[{{Gelman} {et~al.}(2014){Gelman}, {Carlin}, {Stern}, {Dunson},
  {Vehtari}, \& {Rubin}}]{Gelman2014book}
{Gelman}, A., {Carlin}, J.~B., {Stern}, H.~S., {et~al.} 2014, {Bayesian Data
  Analysis}

\bibitem[{{Goodman} \& {Weare}(2010)}]{Goodman2010camcs}
{Goodman}, J., \& {Weare}, J. 2010, Communications in Applied Mathematics and
  Computational Science, 5, 65, \dodoi{10.2140/camcos.2010.5.65}

\bibitem[{{Gromadzki} {et~al.}(2022){Gromadzki}, {Buckley}, {Wyrzykowski}, \&
  {Ihanec}}]{Gromadzki2022tns}
{Gromadzki}, M., {Buckley}, D., {Wyrzykowski}, L., \& {Ihanec}, N. 2022,
  Transient Name Server Classification Report, 2022-203, 1

\bibitem[{{Guillochon} {et~al.}(2017){Guillochon}, {Parrent}, {Kelley}, \&
  {Margutti}}]{Guillochon2017apj}
{Guillochon}, J., {Parrent}, J., {Kelley}, L.~Z., \& {Margutti}, R. 2017, \apj,
  835, 64, \dodoi{10.3847/1538-4357/835/1/64}

\bibitem[{{Hamuy} {et~al.}(1996){Hamuy}, {Phillips}, {Suntzeff}, {Schommer},
  {Maza}, {Smith}, {Lira}, \& {Aviles}}]{Hamuy1996aj}
{Hamuy}, M., {Phillips}, M.~M., {Suntzeff}, N.~B., {et~al.} 1996, \aj, 112,
  2438, \dodoi{10.1086/118193}

\bibitem[{{Han} {et~al.}(2020){Han}, {Zheng}, {Stahl}, {Burke}, {Vinko},
  {Jaeger}, {Arcavi}, {Brink}, {Cseh}, {Hiramatsu}, {Hosseinzadeh}, {Howell},
  {Ignacz}, {Konyves-Toth}, {Krezinger}, {McCully}, {Ordasi}, {Pinter},
  {Sarneczky}, {Szakats}, {Tang}, {Vida}, {Wang}, {Wei}, {Wheeler}, {Xin}, \&
  {Filippenko}}]{Han2020apj}
{Han}, X., {Zheng}, W., {Stahl}, B.~E., {et~al.} 2020, \apj, 892, 142,
  \dodoi{10.3847/1538-4357/ab7a27}

\bibitem[{{Hodgkin} {et~al.}(2022){Hodgkin}, {Breedt}, {Delgado}, {Harrison},
  {Leeuwen}, {Rixon}, {Wevers}, {Yoldas}, {Ihanec}, {Kruszy{\'n}ska},
  {Rybicki}, {Wyrzykowski}, {Kostrzewa-Rutkowska}, {Eappachen}, \&
  {Marton}}]{Hodgkin2022tns}
{Hodgkin}, S.~T., {Breedt}, E., {Delgado}, A., {et~al.} 2022, Transient Name
  Server Discovery Report, 2022-145, 1

\bibitem[{{Hoeflich} {et~al.}(1995){Hoeflich}, {Khokhlov}, \&
  {Wheeler}}]{Hoeflich1995apj}
{Hoeflich}, P., {Khokhlov}, A.~M., \& {Wheeler}, J.~C. 1995, \apj, 444, 831,
  \dodoi{10.1086/175656}

\bibitem[{{Hogg}(1999)}]{Hogg1999}
{Hogg}, D.~W. 1999, arXiv e-prints, astro,
  \dodoi{10.48550/arXiv.astro-ph/9905116}

\bibitem[{{Hogg} {et~al.}(2002){Hogg}, {Baldry}, {Blanton}, \&
  {Eisenstein}}]{Hogg2002}
{Hogg}, D.~W., {Baldry}, I.~K., {Blanton}, M.~R., \& {Eisenstein}, D.~J. 2002,
  arXiv e-prints, astro, \dodoi{10.48550/arXiv.astro-ph/0210394}

\bibitem[{{Holmbo} {et~al.}(2019){Holmbo}, {Stritzinger}, {Shappee}, {Tucker},
  {Zheng}, {Ashall}, {Phillips}, {Contreras}, {Filippenko}, {Hoeflich},
  {Huber}, {Piro}, {Wang}, {Zhang}, {Anais}, {Baron}, {Burns}, {Campillay},
  {Castell{\'o}n}, {Corco}, {Hsiao}, {Krisciunas}, {Morrell}, {Nielsen},
  {Persson}, {Taddia}, {Tomasella}, {Zhang}, \& {Zhao}}]{Holmbo2019aa}
{Holmbo}, S., {Stritzinger}, M.~D., {Shappee}, B.~J., {et~al.} 2019, \aap, 627,
  A174, \dodoi{10.1051/0004-6361/201834389}

\bibitem[{{Hook} {et~al.}(2004){Hook}, {J{\o}rgensen}, {Allington-Smith},
  {Davies}, {Metcalfe}, {Murowinski}, \& {Crampton}}]{Hook2004}
{Hook}, I.~M., {J{\o}rgensen}, I., {Allington-Smith}, J.~R., {et~al.} 2004,
  \pasp, 116, 425, \dodoi{10.1086/383624}

\bibitem[{{Hosseinzadeh} {et~al.}(2017){Hosseinzadeh}, {Sand}, {Valenti},
  {Brown}, {Howell}, {McCully}, {Kasen}, {Arcavi}, {Bostroem}, {Tartaglia},
  {Hsiao}, {Davis}, {Shahbandeh}, \& {Stritzinger}}]{Hosseinzadeh2017apj}
{Hosseinzadeh}, G., {Sand}, D.~J., {Valenti}, S., {et~al.} 2017, \apj, 845,
  L11, \dodoi{10.3847/2041-8213/aa8402}

\bibitem[{{Houston} {et~al.}(2023){Houston}, {Croton}, \&
  {Sinha}}]{Houston2023mnras}
{Houston}, T., {Croton}, D.~J., \& {Sinha}, M. 2023, \mnras, 522, L11,
  \dodoi{10.1093/mnrasl/slad031}

\bibitem[{{Hsiao} {et~al.}(2007){Hsiao}, {Conley}, {Howell}, {Sullivan},
  {Pritchet}, {Carlberg}, {Nugent}, \& {Phillips}}]{Hsiao2007apj}
{Hsiao}, E.~Y., {Conley}, A., {Howell}, D.~A., {et~al.} 2007, \apj, 663, 1187,
  \dodoi{10.1086/518232}

\bibitem[{{Iben} \& {Tutukov}(1984)}]{Iben&Tutukov1984apjs}
{Iben}, I., J., \& {Tutukov}, A.~V. 1984, \apjs, 54, 335,
  \dodoi{10.1086/190932}

\bibitem[{{Jones} {et~al.}(2004){Jones}, {Saunders}, {Colless}, {Read},
  {Parker}, {Watson}, {Campbell}, {Burkey}, {Mauch}, {Moore}, {Hartley},
  {Cass}, {James}, {Russell}, {Fiegert}, {Dawe}, {Huchra}, {Jarrett}, {Lahav},
  {Lucey}, {Mamon}, {Proust}, {Sadler}, \& {Wakamatsu}}]{Jones2004mnras}
{Jones}, D.~H., {Saunders}, W., {Colless}, M., {et~al.} 2004, \mnras, 355, 747,
  \dodoi{10.1111/j.1365-2966.2004.08353.x}

\bibitem[{{Jones} {et~al.}(2023){Jones}, {Kenworthy}, {Dai}, {Foley},
  {Kessler}, {Pierel}, \& {Siebert}}]{Jones2023apj}
{Jones}, D.~O., {Kenworthy}, W.~D., {Dai}, M., {et~al.} 2023, \apj, 951, 22,
  \dodoi{10.3847/1538-4357/acd195}

\bibitem[{{Kasen}(2006)}]{Kasen2006apj}
{Kasen}, D. 2006, \apj, 649, 939, \dodoi{10.1086/506588}

\bibitem[{{Kasen}(2010)}]{Kasen2010apj}
---. 2010, \apj, 708, 1025, \dodoi{10.1088/0004-637X/708/2/1025}

\bibitem[{Katz {et~al.}(1978)Katz, Baptista, Azen, \& Pike}]{Katz1978}
Katz, D., Baptista, J., Azen, S.~P., \& Pike, M.~C. 1978, Biometrics, 34, 469.
\newblock \url{http://www.jstor.org/stable/2530610}

\bibitem[{{Kennicutt}(1998)}]{Kennicutt1998araa}
{Kennicutt}, Robert~C., J. 1998, \araa, 36, 189,
  \dodoi{10.1146/annurev.astro.36.1.189}

\bibitem[{{Kim} {et~al.}(2016){Kim}, {Lee}, {Park}, {Kim}, {Cha}, {Lee}, {Han},
  {Chun}, \& {Yuk}}]{Kim2016jkas}
{Kim}, S.-L., {Lee}, C.-U., {Park}, B.-G., {et~al.} 2016, \jkas, 49, 37,
  \dodoi{10.5303/JKAS.2016.49.1.037}

\bibitem[{{Lasker} {et~al.}(2008){Lasker}, {Lattanzi}, {McLean}, {Bucciarelli},
  {Drimmel}, {Garcia}, {Greene}, {Guglielmetti}, {Hanley}, {Hawkins},
  {Laidler}, {Loomis}, {Meakes}, {Mignani}, {Morbidelli}, {Morrison},
  {Pannunzio}, {Rosenberg}, {Sarasso}, {Smart}, {Spagna}, {Sturch},
  {Volpicelli}, {White}, {Wolfe}, \& {Zacchei}}]{Lasker2008aj}
{Lasker}, B.~M., {Lattanzi}, M.~G., {McLean}, B.~J., {et~al.} 2008, \aj, 136,
  735, \dodoi{10.1088/0004-6256/136/2/735}

\bibitem[{{Lee} {et~al.}(2023){Lee}, {Pak}, {Jeong}, \& {Oh}}]{Lee2023mnras}
{Lee}, J.~H., {Pak}, M., {Jeong}, H., \& {Oh}, S. 2023, \mnras, 521, 4207,
  \dodoi{10.1093/mnras/stad814}

\bibitem[{{Lee} {et~al.}(2024){Lee}, {Moon}, {Kim}, {Park}, \&
  {Ni}}]{Lee2024apj}
{Lee}, Y., {Moon}, D.-S., {Kim}, S.~C., {Park}, H.~S., \& {Ni}, Y.~Q. 2024,
  \apj, 964, 186, \dodoi{10.3847/1538-4357/ad25ff}

\bibitem[{{Levanon} \& {Soker}(2019)}]{Levanon2019apj}
{Levanon}, N., \& {Soker}, N. 2019, \apjl, 872, L7,
  \dodoi{10.3847/2041-8213/ab0285}

\bibitem[{{Li} {et~al.}(2019){Li}, {Wang}, {Vink{\'o}}, {Mo}, {Hosseinzadeh},
  {Sand}, {Zhang}, {Lin}, {PTSS/TNTS}, {Zhang}, {Wang}, {Zhang}, {Chen},
  {Xiang}, {Rui}, {Huang}, {Li}, {Zhang}, {Li}, {Baron}, {Derkacy}, {Zhao},
  {Sai}, {Zhang}, {Wang}, {LCO}, {Howell}, {McCully}, {Arcavi}, {Valenti},
  {Hiramatsu}, {Burke}, {KEGS}, {Rest}, {Garnavich}, {Tucker}, {Narayan},
  {Shaya}, {Margheim}, {Zenteno}, {Villar}, {UCSC}, {Dimitriadis}, {Foley},
  {Pan}, {Coulter}, {Fox}, {Jha}, {Jones}, {Kasen}, {Kilpatrick}, {Piro},
  {Riess}, {Rojas-Bravo}, {ASAS-SN}, {Shappee}, {Holoien}, {Stanek}, {Drout},
  {Auchettl}, {Kochanek}, {Brown}, {Bose}, {Bersier}, {Brimacombe}, {Chen},
  {Dong}, {Holmbo}, {Mu{\~n}oz}, {Mutel}, {Post}, {Prieto}, {Shields},
  {Tallon}, {Thompson}, {Vallely}, {Villanueva}, {Pan-STARRS}, {Smartt},
  {Smith}, {Chambers}, {Flewelling}, {Huber}, {Magnier}, {Waters}, {Schultz},
  {Bulger}, {Lowe}, {Willman}, {Konkoly/Texas}, {S{\'a}rneczky}, {P{\'a}l},
  {Wheeler}, {B{\'o}di}, {Bogn{\'a}r}, {Cs{\'a}k}, {Cseh}, {Cs{\"o}rnyei},
  {Hanyecz}, {Ign{\'a}cz}, {Kalup}, {K{\"o}nyves-T{\'o}th}, {Kriskovics},
  {Ordasi}, {Rajmon}, {S{\'o}dor}, {Szab{\'o}}, {Szak{\'a}ts}, {Zsidi},
  {Arizona}, {Milne}, {Andrews}, {Smith}, {Bilinski}, {Swift}, {Brown},
  {ePESSTO}, {Nordin}, {Williams}, {Galbany}, {Palmerio}, {Hook}, {Inserra},
  {Maguire}, {Cartier}, {Razza}, {Guti{\'e}rrez}, {North Carolina}, {Hermes},
  {Reding}, {Kaiser}, {ATLAS}, {Tonry}, {Heinze}, {Denneau}, {Weiland},
  {Stalder}, {K2 Mission Team}, {Barentsen}, {Dotson}, {Barclay},
  {Gully-Santiago}, {Hedges}, {Cody}, {Howell}, {Kepler Spacecraft Team},
  {Coughlin}, {Van Cleve}, {Cardoso}, {Larson}, {McCalmont-Everton},
  {Peterson}, {Ross}, {Reedy}, {Osborne}, {McGinn}, {Kohnert}, {Migliorini},
  {Wheaton}, {Spencer}, {Labonde}, {Castillo}, {Beerman}, {Steward}, {Hanley},
  {Larsen}, {Gangopadhyay}, {Kloetzel}, {Weschler}, {Nystrom}, {Moffatt},
  {Redick}, {Griest}, {Packard}, {Muszynski}, {Kampmeier}, {Bjella}, {Flynn},
  \& {Elsaesser}}]{Li2019apj}
{Li}, W., {Wang}, X., {Vink{\'o}}, J., {et~al.} 2019, \apj, 870, 12,
  \dodoi{10.3847/1538-4357/aaec74}

\bibitem[{{Maeda} {et~al.}(2010){Maeda}, {R{\"o}pke}, {Fink}, {Hillebrandt},
  {Travaglio}, \& {Thielemann}}]{Maeda2010apj}
{Maeda}, K., {R{\"o}pke}, F.~K., {Fink}, M., {et~al.} 2010, \apj, 712, 624,
  \dodoi{10.1088/0004-637X/712/1/624}

\bibitem[{{Magee} \& {Maguire}(2020)}]{Magee2020aab}
{Magee}, M.~R., \& {Maguire}, K. 2020, \aap, 642, A189,
  \dodoi{10.1051/0004-6361/202037870}

\bibitem[{{Maoz} {et~al.}(2014){Maoz}, {Mannucci}, \&
  {Nelemans}}]{Maoz2014araa}
{Maoz}, D., {Mannucci}, F., \& {Nelemans}, G. 2014, \araa, 52, 107,
  \dodoi{10.1146/annurev-astro-082812-141031}

\bibitem[{{Marion} {et~al.}(2016){Marion}, {Brown}, {Vink{\'o}}, {Silverman},
  {Sand}, {Challis}, {Kirshner}, {Wheeler}, {Berlind}, {Brown}, {Calkins},
  {Camacho}, {Dhungana}, {Foley}, {Friedman}, {Graham}, {Howell}, {Hsiao},
  {Irwin}, {Jha}, {Kehoe}, {Macri}, {Maeda}, {Mandel}, {McCully}, {Pandya},
  {Rines}, {Wilhelmy}, \& {Zheng}}]{Marion2016apj}
{Marion}, G.~H., {Brown}, P.~J., {Vink{\'o}}, J., {et~al.} 2016, \apj, 820, 92,
  \dodoi{10.3847/0004-637X/820/2/92}

\bibitem[{{Matheson} {et~al.}(2008){Matheson}, {Kirshner}, {Challis}, {Jha},
  {Garnavich}, {Berlind}, {Calkins}, {Blondin}, {Balog}, {Bragg}, {Caldwell},
  {Dendy Concannon}, {Falco}, {Graves}, {Huchra}, {Kuraszkiewicz}, {Mader},
  {Mahdavi}, {Phelps}, {Rines}, {Song}, \& {Wilkes}}]{Matheson2008}
{Matheson}, T., {Kirshner}, R.~P., {Challis}, P., {et~al.} 2008, \aj, 135,
  1598, \dodoi{10.1088/0004-6256/135/4/1598}

\bibitem[{{Matteucci}(2012)}]{Matteucci2012book}
{Matteucci}, F. 2012, {Chemical Evolution of Galaxies},
  \dodoi{10.1007/978-3-642-22491-1}

\bibitem[{{Mazzali} {et~al.}(2007){Mazzali}, {R{\"o}pke}, {Benetti}, \&
  {Hillebrandt}}]{Mazzali2007sci}
{Mazzali}, P.~A., {R{\"o}pke}, F.~K., {Benetti}, S., \& {Hillebrandt}, W. 2007,
  Science, 315, 825, \dodoi{10.1126/science.1136259}

\bibitem[{{Miller} {et~al.}(2020){Miller}, {Yao}, {Bulla}, {Pankow}, {Bellm},
  {Cenko}, {Dekany}, {Fremling}, {Graham}, {Kupfer}, {Laher}, {Mahabal},
  {Masci}, {Nugent}, {Riddle}, {Rusholme}, {Smith}, {Shupe}, {van Roestel}, \&
  {Kulkarni}}]{Miller2020apj}
{Miller}, A.~A., {Yao}, Y., {Bulla}, M., {et~al.} 2020, \apj, 902, 47,
  \dodoi{10.3847/1538-4357/abb13b}

\bibitem[{{Moffat}(1969)}]{Moffat1969aap}
{Moffat}, A.~F.~J. 1969, \aap, 3, 455

\bibitem[{{Moon} {et~al.}(2016){Moon}, {Kim}, {Lee}, {Pak}, {Park}, {He},
  {Antoniadis}, {Ni}, {Lee}, {Kim}, {Park}, {Kim}, {Cha}, {Lee}, \&
  {Gonzalez}}]{Moon2016spie}
{Moon}, D.-S., {Kim}, S.~C., {Lee}, J.-J., {et~al.} 2016, in Society of
  Photo-Optical Instrumentation Engineers (SPIE) Conference Series, Vol. 9906,
  Ground-based and Airborne Telescopes VI, 99064I, \dodoi{10.1117/12.2233921}

\bibitem[{{Moon} {et~al.}(2021){Moon}, {Ni}, {Drout},
  {Gonz{\'a}lez-Gait{\'a}n}, {Afsariardchi}, {Park}, {Lee}, {Kim},
  {Antoniadis}, {Kim}, \& {Lee}}]{Moon2021apj}
{Moon}, D.-S., {Ni}, Y.~Q., {Drout}, M.~R., {et~al.} 2021, \apj, 910, 151,
  \dodoi{10.3847/1538-4357/abe466}

\bibitem[{{Morrell} {et~al.}(2024){Morrell}, {Phillips}, {Folatelli},
  {Stritzinger}, {Hamuy}, {Suntzeff}, {Hsiao}, {Taddia}, {Burns}, {Hoeflich},
  {Ashall}, {Contreras}, {Galbany}, {Lu}, {Piro}, {Anais}, {Baron}, {Burrow},
  {Busta}, {Campillay}, {Castell{\'o}n}, {Corco}, {Diamond}, {Freedman},
  {Gonzalez}, {Krisciunas}, {Kumar}, {Persson}, {Ser{\'o}n}, {Shahbandeh},
  {Torres}, {Uddin}, {Anderson}, {Baltay}, {Gall}, {Goobar}, {Hadjiyska},
  {Holmbo}, {Kasliwal}, {Lidman}, {Marion}, {Mazzali}, {Nugent}, {Perlmutter},
  {Pignata}, {Rabinowitz}, {Roth}, {Ryder}, {Shappee}, {Vink{\'o}}, {Wheeler},
  {de Jaeger}, {Lira}, {Ruiz}, {Rich}, {Prieto}, {Di Mille}, {Osip}, {Blanc},
  \& {Palunas}}]{Morrell2024apj}
{Morrell}, N., {Phillips}, M.~M., {Folatelli}, G., {et~al.} 2024, \apj, 967,
  20, \dodoi{10.3847/1538-4357/ad38af}

\bibitem[{{Mould} {et~al.}(2000){Mould}, {Huchra}, {Freedman}, {Kennicutt},
  {Ferrarese}, {Ford}, {Gibson}, {Graham}, {Hughes}, {Illingworth}, {Kelson},
  {Macri}, {Madore}, {Sakai}, {Sebo}, {Silbermann}, \&
  {Stetson}}]{Mould2000apj}
{Mould}, J.~R., {Huchra}, J.~P., {Freedman}, W.~L., {et~al.} 2000, \apj, 529,
  786, \dodoi{10.1086/308304}

\bibitem[{{Murphy}(2012)}]{Murphy2012}
{Murphy}, K.~P. 2012, {Machine Learning: A Probabilistic Perspective}

\bibitem[{{Ni}(2022)}]{Ni2022zndo}
{Ni}, Y.~Q. 2022, {SuperNova Analysis Package (SNAP)}, 221207,  Zenodo,
  \dodoi{10.5281/zenodo.7411663}

\bibitem[{{Ni} {et~al.}(2023{\natexlab{a}}){Ni}, {Moon}, {Drout}, {Matzner},
  {Leong}, {Kim}, {Park}, \& {Lee}}]{Ni2023bapj}
{Ni}, Y.~Q., {Moon}, D.-S., {Drout}, M.~R., {et~al.} 2023{\natexlab{a}}, \apj,
  959, 132, \dodoi{10.3847/1538-4357/ad0640}

\bibitem[{{Ni} {et~al.}(2022){Ni}, {Moon}, {Drout}, {Polin}, {Sand},
  {Gonz{\'a}lez-Gait{\'a}n}, {Kim}, {Lee}, {Park}, {Howell}, {Nugent}, {Piro},
  {Brown}, {Galbany}, {Burke}, {Hiramatsu}, {Hosseinzadeh}, {Valenti},
  {Afsariardchi}, {Andrews}, {Antoniadis}, {Arcavi}, {Beaton}, {Bostroem},
  {Carlberg}, {Cenko}, {Cha}, {Dong}, {Gal-Yam}, {Haislip}, {Holoien},
  {Johnson}, {Kouprianov}, {Lee}, {Matzner}, {Morrell}, {McCully}, {Pignata},
  {Reichart}, {Rich}, {Ryder}, {Smith}, {Wyatt}, \& {Yang}}]{Ni2022natas}
---. 2022, Nature Astronomy, 6, 568, \dodoi{10.1038/s41550-022-01603-4}

\bibitem[{{Ni} {et~al.}(2023{\natexlab{b}}){Ni}, {Moon}, {Drout}, {Polin},
  {Sand}, {Gonz{\'a}lez-Gait{\'a}n}, {Kim}, {Lee}, {Park}, {Howell}, {Nugent},
  {Piro}, {Brown}, {Galbany}, {Burke}, {Hiramatsu}, {Hosseinzadeh}, {Valenti},
  {Afsariardchi}, {Andrews}, {Antoniadis}, {Beaton}, {Bostroem}, {Carlberg},
  {Cenko}, {Cha}, {Dong}, {Gal-Yam}, {Haislip}, {Holoien}, {Johnson},
  {Kouprianov}, {Lee}, {Matzner}, {Morrell}, {McCully}, {Pignata}, {Reichart},
  {Rich}, {Ryder}, {Smith}, {Wyatt}, \& {Yang}}]{Ni2023apj}
---. 2023{\natexlab{b}}, \apj, 946, 7, \dodoi{10.3847/1538-4357/aca9be}

\bibitem[{{Nordin} {et~al.}(2019){Nordin}, {Brinnel}, {Giomi}, {Santen},
  {Gal-yam}, {Yaron}, \& {Schulze}}]{Nordin2019tns}
{Nordin}, J., {Brinnel}, V., {Giomi}, M., {et~al.} 2019, Transient Name Server
  Discovery Report, 2019-404, 1

\bibitem[{{Nugent} {et~al.}(2011){Nugent}, {Sullivan}, {Cenko}, {Thomas},
  {Kasen}, {Howell}, {Bersier}, {Bloom}, {Kulkarni}, {Kand rashoff},
  {Filippenko}, {Silverman}, {Marcy}, {Howard}, {Isaacson}, {Maguire},
  {Suzuki}, {Tarlton}, {Pan}, {Bildsten}, {Fulton}, {Parrent}, {Sand},
  {Podsiadlowski}, {Bianco}, {Dilday}, {Graham}, {Lyman}, {James}, {Kasliwal},
  {Law}, {Quimby}, {Hook}, {Walker}, {Mazzali}, {Pian}, {Ofek}, {Gal-Yam}, \&
  {Poznanski}}]{Nugent2011nat}
{Nugent}, P.~E., {Sullivan}, M., {Cenko}, S.~B., {et~al.} 2011, \nat, 480, 344,
  \dodoi{10.1038/nature10644}

\bibitem[{{Oke} \& {Sandage}(1968)}]{Oke1968apj}
{Oke}, J.~B., \& {Sandage}, A. 1968, \apj, 154, 21, \dodoi{10.1086/149737}

\bibitem[{{Olling} {et~al.}(2015){Olling}, {Mushotzky}, {Shaya}, {Rest},
  {Garnavich}, {Tucker}, {Kasen}, {Margheim}, \& {Filippenko}}]{Olling2015nat}
{Olling}, R.~P., {Mushotzky}, R., {Shaya}, E.~J., {et~al.} 2015, \nat, 521,
  332, \dodoi{10.1038/nature14455}

\bibitem[{{Park} {et~al.}(2019){Park}, {Moon}, {Zaritsky}, {Kim}, {Lee}, {Cha},
  \& {Lee}}]{Park2019apj}
{Park}, H.~S., {Moon}, D.-S., {Zaritsky}, D., {et~al.} 2019, \apj, 885, 88,
  \dodoi{10.3847/1538-4357/ab4794}

\bibitem[{{Park} {et~al.}(2017){Park}, {Moon}, {Zaritsky}, {Pak}, {Lee}, {Kim},
  {Kim}, \& {Cha}}]{Park2017apj}
---. 2017, \apj, 848, 19, \dodoi{10.3847/1538-4357/aa88ab}

\bibitem[{{Parrent} {et~al.}(2014){Parrent}, {Friesen}, \&
  {Parthasarathy}}]{Parrent2014apss}
{Parrent}, J., {Friesen}, B., \& {Parthasarathy}, M. 2014, \apss, 351, 1,
  \dodoi{10.1007/s10509-014-1830-1}

\bibitem[{Pedregosa {et~al.}(2011)Pedregosa, Varoquaux, Gramfort, Michel,
  Thirion, Grisel, Blondel, Prettenhofer, Weiss, Dubourg, Vanderplas, Passos,
  Cournapeau, Brucher, Perrot, \& Duchesnay}]{scikit-learn}
Pedregosa, F., Varoquaux, G., Gramfort, A., {et~al.} 2011, Journal of Machine
  Learning Research, 12, 2825

\bibitem[{{Peng} \& {Nagai}(2009)}]{Peng2009apjl}
{Peng}, F., \& {Nagai}, D. 2009, \apjl, 705, L58,
  \dodoi{10.1088/0004-637X/705/1/L58}

\bibitem[{{Pereira} {et~al.}(2013){Pereira}, {Thomas}, {Aldering}, {Antilogus},
  {Baltay}, {Benitez-Herrera}, {Bongard}, {Buton}, {Canto}, {Cellier-Holzem},
  {Chen}, {Childress}, {Chotard}, {Copin}, {Fakhouri}, {Fink}, {Fouchez},
  {Gangler}, {Guy}, {Hillebrandt}, {Hsiao}, {Kerschhaggl}, {Kowalski},
  {Kromer}, {Nordin}, {Nugent}, {Paech}, {Pain}, {P{\'e}contal}, {Perlmutter},
  {Rabinowitz}, {Rigault}, {Runge}, {Saunders}, {Smadja}, {Tao},
  {Taubenberger}, {Tilquin}, \& {Wu}}]{Pereira2013aa}
{Pereira}, R., {Thomas}, R.~C., {Aldering}, G., {et~al.} 2013, \aap, 554, A27,
  \dodoi{10.1051/0004-6361/201221008}

\bibitem[{{Perlmutter} {et~al.}(1999){Perlmutter}, {Aldering}, {Goldhaber},
  {Knop}, {Nugent}, {Castro}, {Deustua}, {Fabbro}, {Goobar}, {Groom}, {Hook},
  {Kim}, {Kim}, {Lee}, {Nunes}, {Pain}, {Pennypacker}, {Quimby}, {Lidman},
  {Ellis}, {Irwin}, {McMahon}, {Ruiz-Lapuente}, {Walton}, {Schaefer}, {Boyle},
  {Filippenko}, {Matheson}, {Fruchter}, {Panagia}, {Newberg}, {Couch}, \&
  {Project}}]{perlmutter1999apj}
{Perlmutter}, S., {Aldering}, G., {Goldhaber}, G., {et~al.} 1999, \apj, 517,
  565, \dodoi{10.1086/307221}

\bibitem[{{Phillips} {et~al.}(1999){Phillips}, {Lira}, {Suntzeff}, {Schommer},
  {Hamuy}, \& {Maza}}]{Phillips1999aj}
{Phillips}, M.~M., {Lira}, P., {Suntzeff}, N.~B., {et~al.} 1999, \aj, 118,
  1766, \dodoi{10.1086/301032}

\bibitem[{{Phillips} {et~al.}(2013){Phillips}, {Simon}, {Morrell}, {Burns},
  {Cox}, {Foley}, {Karakas}, {Patat}, {Sternberg}, {Williams}, {Gal-Yam},
  {Hsiao}, {Leonard}, {Persson}, {Stritzinger}, {Thompson}, {Campillay},
  {Contreras}, {Folatelli}, {Freedman}, {Hamuy}, {Roth}, {Shields}, {Suntzeff},
  {Chomiuk}, {Ivans}, {Madore}, {Penprase}, {Perley}, {Pignata}, {Preston}, \&
  {Soderberg}}]{Phillips2013apj}
{Phillips}, M.~M., {Simon}, J.~D., {Morrell}, N., {et~al.} 2013, \apj, 779, 38,
  \dodoi{10.1088/0004-637X/779/1/38}

\bibitem[{{Phillips} {et~al.}(2022){Phillips}, {Ashall}, {Burns}, {Contreras},
  {Galbany}, {Hoeflich}, {Hsiao}, {Morrell}, {Nugent}, {Uddin}, {Baron},
  {Freedman}, {Harris}, {Krisciunas}, {Kumar}, {Lu}, {Persson}, {Piro},
  {Polin}, {Shahbandeh}, {Stritzinger}, \& {Suntzeff}}]{Phillips2022apj}
{Phillips}, M.~M., {Ashall}, C., {Burns}, C.~R., {et~al.} 2022, \apj, 938, 47,
  \dodoi{10.3847/1538-4357/ac9305}

\bibitem[{{Piro} \& {Morozova}(2016)}]{Piro&Morozova2016apj}
{Piro}, A.~L., \& {Morozova}, V.~S. 2016, \apj, 826, 96,
  \dodoi{10.3847/0004-637X/826/1/96}

\bibitem[{{Piro} \& {Nakar}(2013)}]{Piro&Nakar2013apj}
{Piro}, A.~L., \& {Nakar}, E. 2013, \apj, 769, 67,
  \dodoi{10.1088/0004-637X/769/1/67}

\bibitem[{{Piro} \& {Nakar}(2014)}]{Piro&Nakar2014apj}
---. 2014, \apj, 784, 85, \dodoi{10.1088/0004-637X/784/1/85}

\bibitem[{{Polin} {et~al.}(2019){Polin}, {Nugent}, \& {Kasen}}]{Polin2019apj}
{Polin}, A., {Nugent}, P., \& {Kasen}, D. 2019, \apj, 873, 84,
  \dodoi{10.3847/1538-4357/aafb6a}

\bibitem[{{Poulain} {et~al.}(2021){Poulain}, {Marleau}, {Habas}, {Duc},
  {S{\'a}nchez-Janssen}, {Durrell}, {Paudel}, {Ahad}, {Chougule}, {M{\"u}ller},
  {Lim}, {B{\'\i}lek}, \& {Fensch}}]{Poulain2021mnras}
{Poulain}, M., {Marleau}, F.~R., {Habas}, R., {et~al.} 2021, \mnras, 506, 5494,
  \dodoi{10.1093/mnras/stab2092}

\bibitem[{{Poznanski} {et~al.}(2012){Poznanski}, {Prochaska}, \&
  {Bloom}}]{Poznanski2012mnras}
{Poznanski}, D., {Prochaska}, J.~X., \& {Bloom}, J.~S. 2012, \mnras, 426, 1465,
  \dodoi{10.1111/j.1365-2966.2012.21796.x}

\bibitem[{{Prieto} {et~al.}(2006){Prieto}, {Rest}, \&
  {Suntzeff}}]{Prieto2006apj}
{Prieto}, J.~L., {Rest}, A., \& {Suntzeff}, N.~B. 2006, \apj, 647, 501,
  \dodoi{10.1086/504307}

\bibitem[{{Rasmussen} \& {Williams}(2006)}]{Rasmussen2006}
{Rasmussen}, C.~E., \& {Williams}, C. K.~I. 2006, {Gaussian Processes for
  Machine Learning}

\bibitem[{{Riess} {et~al.}(1998){Riess}, {Filippenko}, {Challis},
  {Clocchiatti}, {Diercks}, {Garnavich}, {Gilliland}, {Hogan}, {Jha},
  {Kirshner}, {Leibundgut}, {Phillips}, {Reiss}, {Schmidt}, {Schommer},
  {Smith}, {Spyromilio}, {Stubbs}, {Suntzeff}, \& {Tonry}}]{Riess1998aj}
{Riess}, A.~G., {Filippenko}, A.~V., {Challis}, P., {et~al.} 1998, \aj, 116,
  1009, \dodoi{10.1086/300499}

\bibitem[{{Riess} {et~al.}(2016){Riess}, {Macri}, {Hoffmann}, {Scolnic},
  {Casertano}, {Filippenko}, {Tucker}, {Reid}, {Jones}, {Silverman},
  {Chornock}, {Challis}, {Yuan}, {Brown}, \& {Foley}}]{Riess2016apj}
{Riess}, A.~G., {Macri}, L.~M., {Hoffmann}, S.~L., {et~al.} 2016, \apj, 826,
  56, \dodoi{10.3847/0004-637X/826/1/56}

\bibitem[{{Ruiter}(2020)}]{Ruiter2020iaus}
{Ruiter}, A.~J. 2020, in White Dwarfs as Probes of Fundamental Physics: Tracers
  of Planetary, Stellar and Galactic Evolution, ed. M.~A. {Barstow}, S.~J.
  {Kleinman}, J.~L. {Provencal}, \& L.~{Ferrario}, Vol. 357, 1--15,
  \dodoi{10.1017/S1743921320000587}

\bibitem[{{Ruiter} {et~al.}(2009){Ruiter}, {Belczynski}, \&
  {Fryer}}]{Ruiter2009apj}
{Ruiter}, A.~J., {Belczynski}, K., \& {Fryer}, C. 2009, \apj, 699, 2026,
  \dodoi{10.1088/0004-637X/699/2/2026}

\bibitem[{{Ruiter} {et~al.}(2014){Ruiter}, {Belczynski}, {Sim}, {Seitenzahl},
  \& {Kwiatkowski}}]{Ruiter2014mnras}
{Ruiter}, A.~J., {Belczynski}, K., {Sim}, S.~A., {Seitenzahl}, I.~R., \&
  {Kwiatkowski}, D. 2014, \mnras, 440, L101, \dodoi{10.1093/mnrasl/slu030}

\bibitem[{{Sand} {et~al.}(2018){Sand}, {Graham}, {Boty{\'a}nszki}, {Hiramatsu},
  {McCully}, {Valenti}, {Hosseinzadeh}, {Howell}, {Burke}, {Cartier},
  {Diamond}, {Hsiao}, {Jha}, {Kasen}, {Kumar}, {Marion}, {Suntzeff},
  {Tartaglia}, {Wheeler}, \& {Wyatt}}]{Sand2018apj}
{Sand}, D.~J., {Graham}, M.~L., {Boty{\'a}nszki}, J., {et~al.} 2018, \apj, 863,
  24, \dodoi{10.3847/1538-4357/aacde8}

\bibitem[{{Schlafly} \& {Finkbeiner}(2011)}]{Schlafly&Finkbeiner2011apj}
{Schlafly}, E.~F., \& {Finkbeiner}, D.~P. 2011, \apj, 737, 103,
  \dodoi{10.1088/0004-637X/737/2/103}

\bibitem[{{Scolnic} {et~al.}(2018){Scolnic}, {Jones}, {Rest}, {Pan},
  {Chornock}, {Foley}, {Huber}, {Kessler}, {Narayan}, {Riess}, {Rodney},
  {Berger}, {Brout}, {Challis}, {Drout}, {Finkbeiner}, {Lunnan}, {Kirshner},
  {Sanders}, {Schlafly}, {Smartt}, {Stubbs}, {Tonry}, {Wood-Vasey}, {Foley},
  {Hand}, {Johnson}, {Burgett}, {Chambers}, {Draper}, {Hodapp}, {Kaiser},
  {Kudritzki}, {Magnier}, {Metcalfe}, {Bresolin}, {Gall}, {Kotak}, {McCrum}, \&
  {Smith}}]{Scolnic2018apj}
{Scolnic}, D.~M., {Jones}, D.~O., {Rest}, A., {et~al.} 2018, \apj, 859, 101,
  \dodoi{10.3847/1538-4357/aab9bb}

\bibitem[{{Shappee} {et~al.}(2018){Shappee}, {Piro}, {Stanek}, {Patel},
  {Margutti}, {Lipunov}, \& {Pogge}}]{Shappee2018apj}
{Shappee}, B.~J., {Piro}, A.~L., {Stanek}, K.~Z., {et~al.} 2018, \apj, 855, 6,
  \dodoi{10.3847/1538-4357/aaa1e9}

\bibitem[{{Shen} {et~al.}(2021){Shen}, {Blondin}, {Kasen}, {Dessart},
  {Townsley}, {Boos}, \& {Hillier}}]{Shen2021apjl}
{Shen}, K.~J., {Blondin}, S., {Kasen}, D., {et~al.} 2021, \apjl, 909, L18,
  \dodoi{10.3847/2041-8213/abe69b}

\bibitem[{{Stritzinger} {et~al.}(2002){Stritzinger}, {Hamuy}, {Suntzeff},
  {Smith}, {Phillips}, {Maza}, {Strolger}, {Antezana}, {Gonz{\'a}lez},
  {Wischnjewsky}, {Candia}, {Espinoza}, {Gonz{\'a}lez}, {Stubbs}, {Becker},
  {Rubenstein}, \& {Galaz}}]{Stritzinger2002aj}
{Stritzinger}, M., {Hamuy}, M., {Suntzeff}, N.~B., {et~al.} 2002, \aj, 124,
  2100, \dodoi{10.1086/342544}

\bibitem[{{Stritzinger} {et~al.}(2018){Stritzinger}, {Shappee}, {Piro},
  {Ashall}, {Baron}, {Hoeflich}, {Holmbo}, {Holoien}, {Phillips}, {Burns},
  {Contreras}, {Morrell}, \& {Tucker}}]{Stritzinger2018apj}
{Stritzinger}, M.~D., {Shappee}, B.~J., {Piro}, A.~L., {et~al.} 2018, \apj,
  864, L35, \dodoi{10.3847/2041-8213/aadd46}

\bibitem[{{STScI Development Team}(2018)}]{synphot}
{STScI Development Team}. 2018, {synphot: Synthetic photometry using Astropy},
  Astrophysics Source Code Library, record ascl:1811.001

\bibitem[{{Takats} {et~al.}(2016){Takats}, {Rodriguez}, {Galbany}, \&
  {Yaron}}]{Takats2016tns}
{Takats}, K., {Rodriguez}, O., {Galbany}, L., \& {Yaron}, O. 2016, Transient
  Name Server Classification Report, 2016-932, 1

\bibitem[{{Taubenberger} {et~al.}(2013){Taubenberger}, {Kromer}, {Pakmor},
  {Pignata}, {Maeda}, {Hachinger}, {Leibundgut}, \& {Hillebrand
  t}}]{Taubenberger2013apj}
{Taubenberger}, S., {Kromer}, M., {Pakmor}, R., {et~al.} 2013, \apjl, 775, L43,
  \dodoi{10.1088/2041-8205/775/2/L43}

\bibitem[{{Tody}(1993)}]{Tody1993aspc}
{Tody}, D. 1993, in Astronomical Society of the Pacific Conference Series,
  Vol.~52, Astronomical Data Analysis Software and Systems II, ed. R.~J.
  {Hanisch}, R.~J.~V. {Brissenden}, \& J.~{Barnes}, 173

\bibitem[{{Tonry} {et~al.}(2016){Tonry}, {Denneau}, {Stalder}, {Heinze},
  {Sherstyuk}, {Rest}, {Smith}, \& {Smartt}}]{Tonry2016tns}
{Tonry}, J., {Denneau}, L., {Stalder}, B., {et~al.} 2016, Transient Name Server
  Discovery Report, 2016-927, 1

\bibitem[{{Tonry} {et~al.}(2018){Tonry}, {Stalder}, {Denneau}, {Heinze},
  {Weiland}, {Rest}, {Smith}, {Smartt}, {Young}, {Fulton}, {Mcbrien},
  {O'neill}, \& {Clark}}]{Tonry2018tns}
{Tonry}, J., {Stalder}, B., {Denneau}, L., {et~al.} 2018, Transient Name Server
  Discovery Report, 2018-350, 1

\bibitem[{{Townsley} {et~al.}(2019){Townsley}, {Miles}, {Shen}, \&
  {Kasen}}]{Townsley2019apj}
{Townsley}, D.~M., {Miles}, B.~J., {Shen}, K.~J., \& {Kasen}, D. 2019, \apjl,
  878, L38, \dodoi{10.3847/2041-8213/ab27cd}

\bibitem[{{Trujillo} {et~al.}(2001){Trujillo}, {Aguerri}, {Cepa}, \&
  {Guti{\'e}rrez}}]{Trujillo2001mnras}
{Trujillo}, I., {Aguerri}, J.~A.~L., {Cepa}, J., \& {Guti{\'e}rrez}, C.~M.
  2001, \mnras, 328, 977, \dodoi{10.1046/j.1365-8711.2001.04937.x}

\bibitem[{{van Kerkwijk} {et~al.}(2010){van Kerkwijk}, {Chang}, \&
  {Justham}}]{vanKerkwijk2010apj}
{van Kerkwijk}, M.~H., {Chang}, P., \& {Justham}, S. 2010, \apjl, 722, L157,
  \dodoi{10.1088/2041-8205/722/2/L157}

\bibitem[{{Vazdekis} {et~al.}(2016){Vazdekis}, {Koleva}, {Ricciardelli},
  {R{\"o}ck}, \& {Falc{\'o}n-Barroso}}]{Vazdekis2016mnras}
{Vazdekis}, A., {Koleva}, M., {Ricciardelli}, E., {R{\"o}ck}, B., \&
  {Falc{\'o}n-Barroso}, J. 2016, \mnras, 463, 3409,
  \dodoi{10.1093/mnras/stw2231}

\bibitem[{{Virtanen} {et~al.}(2020){Virtanen}, {Gommers}, {Oliphant},
  {Haberland}, {Reddy}, {Cournapeau}, {Burovski}, {Peterson}, {Weckesser},
  {Bright}, {van der Walt}, {Brett}, {Wilson}, {Millman}, {Mayorov}, {Nelson},
  {Jones}, {Kern}, {Larson}, {Carey}, {Polat}, {Feng}, {Moore}, {VanderPlas},
  {Laxalde}, {Perktold}, {Cimrman}, {Henriksen}, {Quintero}, {Harris},
  {Archibald}, {Ribeiro}, {Pedregosa}, {van Mulbregt}, \& {SciPy 1. 0
  Contributors}}]{Virtanen2020natme}
{Virtanen}, P., {Gommers}, R., {Oliphant}, T.~E., {et~al.} 2020, Nature
  Methods, 17, 261, \dodoi{10.1038/s41592-019-0686-2}

\bibitem[{{Wang} {et~al.}(2024){Wang}, {Rest}, {Dimitriadis}, {Ridden-Harper},
  {Siebert}, {Magee}, {Angus}, {Auchettl}, {Davis}, {Foley}, {Fox}, {Gomez},
  {Jencson}, {Jones}, {Kilpatrick}, {Pierel}, {Piro}, {Polin}, {Politsch},
  {Rojas-Bravo}, {Shahbandeh}, {Villar}, {Zenati}, {Ashall}, {Chambers},
  {Coulter}, {de Boer}, {DiLullo}, {Gall}, {Gao}, {Hsiao}, {Huber}, {Izzo},
  {Khetan}, {LeBaron}, {Magnier}, {Mandel}, {McGill}, {Miao}, {Pan}, {Stevens},
  {Swift}, {Taggart}, \& {Yang}}]{Wang2024apj}
{Wang}, Q., {Rest}, A., {Dimitriadis}, G., {et~al.} 2024, \apj, 962, 17,
  \dodoi{10.3847/1538-4357/ad0edb}

\bibitem[{{Wang} {et~al.}(2009){Wang}, {Filippenko}, {Ganeshalingam}, {Li},
  {Silverman}, {Wang}, {Chornock}, {Foley}, {Gates}, \&
  {Macomber}}]{Wang2009apj}
{Wang}, X., {Filippenko}, A.~V., {Ganeshalingam}, M., {et~al.} 2009, \apjl,
  699, L139, \dodoi{10.1088/0004-637X/699/2/L139}

\bibitem[{{Weymann} {et~al.}(2001){Weymann}, {Vogel}, {Veilleux}, \&
  {Epps}}]{Weymann2001apj}
{Weymann}, R.~J., {Vogel}, S.~N., {Veilleux}, S., \& {Epps}, H.~W. 2001, \apj,
  561, 559, \dodoi{10.1086/323205}

\bibitem[{{Whelan} \& {Iben}(1973)}]{Whelan&Iben1973apj}
{Whelan}, J., \& {Iben}, Icko, J. 1973, \apj, 186, 1007, \dodoi{10.1086/152565}

\bibitem[{Wishart(1928)}]{Wishart1928}
Wishart, J. 1928, Biometrika, 20A, 32, \dodoi{10.1093/biomet/20A.1-2.32}

\bibitem[{{Yaron} \& {Gal-Yam}(2012)}]{Yaron2012pasp}
{Yaron}, O., \& {Gal-Yam}, A. 2012, \pasp, 124, 668, \dodoi{10.1086/666656}

\bibitem[{{Yi} {et~al.}(1999){Yi}, {Lee}, {Woo}, {Park}, {Demarque}, \&
  {Oemler}}]{Yi1999apj}
{Yi}, S., {Lee}, Y.-W., {Woo}, J.-H., {et~al.} 1999, \apj, 513, 128,
  \dodoi{10.1086/306856}

\bibitem[{{Zhang} {et~al.}(2021){Zhang}, {Murakami}, {Stahl}, {Patra}, \&
  {Filippenko}}]{Zhang2021mnras}
{Zhang}, K.~D., {Murakami}, Y.~S., {Stahl}, B.~E., {Patra}, K.~C., \&
  {Filippenko}, A.~V. 2021, \mnras, 503, L33, \dodoi{10.1093/mnrasl/slab020}

\bibitem[{{Zhang} {et~al.}(2020){Zhang}, {Zheng}, {de Jaeger}, {Stahl},
  {Brink}, {Han}, {Kasen}, {Shen}, {Tang}, \& {Filippenko}}]{Zhang2020mnras}
{Zhang}, K.~D., {Zheng}, W., {de Jaeger}, T., {et~al.} 2020, \mnras, 499, 5325,
  \dodoi{10.1093/mnras/staa3191}

\end{thebibliography}
\end{document}